\documentclass[pre,aps,a4paper,reprint,showpacs,amsmath,amssymb]{revtex4-1}  % 2colonnes

\usepackage{graphicx}
\usepackage{hyperref}
\usepackage{xcolor}
\newcommand{\RED}{\color{black}}
\newcommand{\GREEN}{\color{black}}
\newcommand{\BLUE}{\color{black}}

\begin{document}

\title{Evidence of reverse and intermediate size segregation in dry granular flows down a rough incline}

\author{Nathalie Thomas}
\affiliation{CNRS, Aix-Marseille Univ.,
IUSTI UMR 7343, 13453, Marseille, France}     
\email{nathalie.thomas@univ-amu.fr}
\author{Umberto D'Ortona}
\affiliation{CNRS, Aix-Marseille Univ., Centrale Marseille, 
M2P2 UMR 7340, 13451, Marseille, France}
%\email{umberto@l3m.univ-mrs.fr}
%\pacs{45.50.-j}{Dynamics and kinematics of a particle and a system of particles}
%\pacs{45.70.-n}{Granular systems}
%\pacs{45.70.Mg}{Granular flow: mixing, segregation and stratification}
%\pacs{62.20.-x}{Mechanical properties of solids}
%\pacs{89.75.Da}{Systems obeying scaling laws}

\date{09 Feb. 2018}
DOI: 10.1103/PhysRevE.97.022903

\vspace{0.5cm}

\begin{abstract}

 In a dry granular flow, size segregation had been shown to behave differently for a mixture containing a few large particles with a size ratio above 5 (N.~Thomas, Phys. Rev. E {\bf 62}, 961 (2000)). For moderately large size ratios, large particles migrate to an intermediate depth in the bed: this is called ``intermediate segregation". For the largest size ratios, large particles migrate down to the bottom of the flow:  this is called ``reverse segregation" -  in contrast with surface segregation. As the reversal and intermediate depth values  depend on the fraction of particles, this numerical study mainly uses one single large tracer. Small fractions of large beads are also computed showing the link between single tracer behavior and collective segregation process. For each device (half-filled rotating tumbler and rough plane), two (2D) and three (3D) dimensional cases are distinguished.

In the tumbler, the trajectories of a large tracer show that it reaches a constant depth during the flowing phase. For large size ratios, this depth is intermediate. A progressive sinking of the depth is obtained when the size ratio is increased. The largest size ratios correspond to tracers being at the bottom of the flowing layer. All 3D simulation results are in quantitative agreement with the experimental surface, intermediate, and reverse segregation results.

In the flow down a rough incline, a large tracer reaches an equilibrium depth during flow. For large size ratios, the depth is inside the bed, at an intermediate position, and for the largest size ratios, this depth is reverse, located near the bottom. Results are slightly different for a thin or a thick flow. For 3D thick flows, the reversal between surface and bottom positions occurs within a short range of size ratios: no tracer stabilizes near half-height and two reachable intermediate depth layers exist, below the surface and above the bottom reverse layer. For 3D thin flows, all intermediate depths are reachable by a tracer, depending on the size ratio. The numerical study of larger fractions of tracers (5 or 10\%) shows the three segregation patterns (surface, intermediate, reverse) corresponding to the three types of equilibrium depth. The reversal is smoother than for a single tracer, and happens around the size ratio 4.5, in good agreement with experiments. 

\end{abstract}

%\PACS{
%{45.50.-j}{Dynamics and kinematics of a particle and a system of particles}\and
%{45.70.-n}{Granular systems}\and
%{45.70.Mg}{Granular flow: mixing, segregation and stratification}\and
%{45.70.Ht}{Avalanches}
%}

\maketitle

\section{Introduction}

Size segregation in dry granular flow {\RED has been extensively studied as it is an important phenomenon occurring} in natural flows or in industrial applications
\cite{Williams76,DuranRajchenbach93,KnightJaeger93,CantelaubeBideau95,HutterVendsen96,OttinoKhaKhar00,GrayThornton06,SchlickFan15}. Recently, {\RED there have been} significant
advances in the modeling of segregation in dense granular
flows. 
Models based on kinetic theory have
been established for segregation in rapid flows, in the case of particles of
different sizes and/or densities \cite{LarcherJenkins13,LarcherJenkins15,KhakharMcCarthy99}. {\RED These models,} based on particle properties and with no adjustable parameter, are able to predict the evolution of the volume fraction of 
two types of particles that do not differ much in size or mass \cite{LarcherJenkins13}.
Alternative models based on mixture theory have been proposed in {\RED which} unequal stress partitioning reflects the mechanisms {\RED that are} responsible for the
segregation: kinetic sieving and squeeze expulsion \cite{SavageLun88}.
In this continuum framework, particle
segregation results from {\RED the lithostatic pressure} gradient induced
by gravity \cite{GrayThornton05,GrayThornton06}.
Several groups have proposed improvements to take into account the effects of shear rate \cite{MayGolick10,TunuguntlaBokhove14,FanSchlick14,SchlickIsner16},
 kinetic stress gradients (derived from vertical chutes)
\cite{FanHill11,FanHill15}, or the polydispersity of flows with particles of different sizes and densities
\cite{MarksRognon12}, leading to further developments for flows on
inclines \cite{StaronPhillips16}. Quantitative agreement with experiments has been
obtained for the stationary concentration profile of a mixture with a size ratio of 2 
\cite{WiederseinerAncey11}.
 A comparative review can be found in \cite{TunuguntlaReview16}.

Most of these studies are concerned with
small size ratios, the large particles being generally 1.5 to 2 times the size of the
small ones. In some studies, size ratio is varied up to 3 \cite{FanSchlick14}, 3.5
\cite{Tunuguntla16}, or 4 \cite{GolickDaniels14}. {\RED This variation remains}
small compared {\RED with the size ratio range} in our present study. {\RED Even so}, it already {\RED induces} a non-monotonic variation of some parameters, {\RED e.g.} the segregation rate \cite{GolickDaniels14}.  {\RED One of the studies concerning the measurement of the force acting} on an object plunged into a granular flow \cite{Ding11, Ding13, Guillard14} {\RED provides interesting information on the segregation phenomenon} because the intruder is free to move with the flow \cite{Guillard16}, instead of being an obstacle {\RED exerting} drag. In these 2D simulations, {\RED the authors} also noticed an extremum for the normalized segregation force obtained at the size ratio 2.
{\RED Some} segregation theory has been extended to large size ratios (up to 10) \cite{MarksRognon12} and predicts a
monotonic decrease {\RED in the segregation time} with the size ratio and without any change in the segregation pattern. Most of the models do not explicitly depend on
the size ratio,
but rather on a segregating velocity determined for each species \cite{TunuguntlaReview16}. In the few models where 
the size ratio {\RED is explicitly mentioned} \cite{MarksRognon12,TunuguntlaBokhove14}, the segregating velocity cannot change sign when the size ratio is increased, for any particle fraction. In these models, only a difference {\RED in} density between particles could induce a reversal of the segregating velocity direction \cite{TunuguntlaBokhove14}.

Nevertheless, the segregation phenomenon is observed to be different when increasing the size ratio above 4 or 5. It has been shown experimentally that {\BLUE large particles do not reach the surface, as they usually do in surface segregation, but move downwards and stabilize} either at an intermediate depth or at the bottom of the flow for the largest size ratios \cite{Thomas00}. Particle stabilization at an intermediate depth has been named ``intermediate segregation". Large particle segregation at the bottom of the flow has been named
``reverse segregation" by analogy {\RED with} the ``reverse Brazil-nut effect" \cite{ShinbrotMuzzio98,BreuEnsner03,EllenbergerVandu06} observed in vibrated {\RED granular} systems. {\BLUE The origin
of this vibrating effect \cite{HongQuinn01} is due to an inertia driven segregation 
process induced by high amplitude vibrations
\cite{ShinbrotMuzzio98, HongQuinn01} as well as to the absence of convective motion \cite{knight93}.} 
The reverse and intermediate segregations of particles of different sizes (and having the same density) have been observed experimentally {\BLUE in various sheared flows}: channel flow, half-filled cylindrical rotating tumbler and 3D heap flow \cite{Thomas00, FelixThomas04}.  
{\BLUE The corresponding segregation patterns take different forms.}
In a rotating tumbler, if large particles are close to the tumbler {\RED center} in the static part, reverse segregation {\RED occurs}  because particles move {\RED to} the bottom of the flowing layer during flow. {\RED By contrast}, tracers having a small size ratio (from 1.5 to 3) end up at the periphery on the solid part, undergoing surface segregation during flow. For a flow down an incline, {\RED the reverse-segregating} large particles disappear from the surface {\RED during flow}, and are present near the bottom of the deposit, while  {\RED the surface-segregating} large particles cover the flow and deposit surface. {\BLUE For} a flow feeding a 
heap, very large beads form {\BLUE a vertical core} (reverse segregation). For a small size ratio, a ring of large beads forms at the bottom periphery of the heap (surface segregation).

\begin{figure}[htbp]
\includegraphics[width=0.516\linewidth]{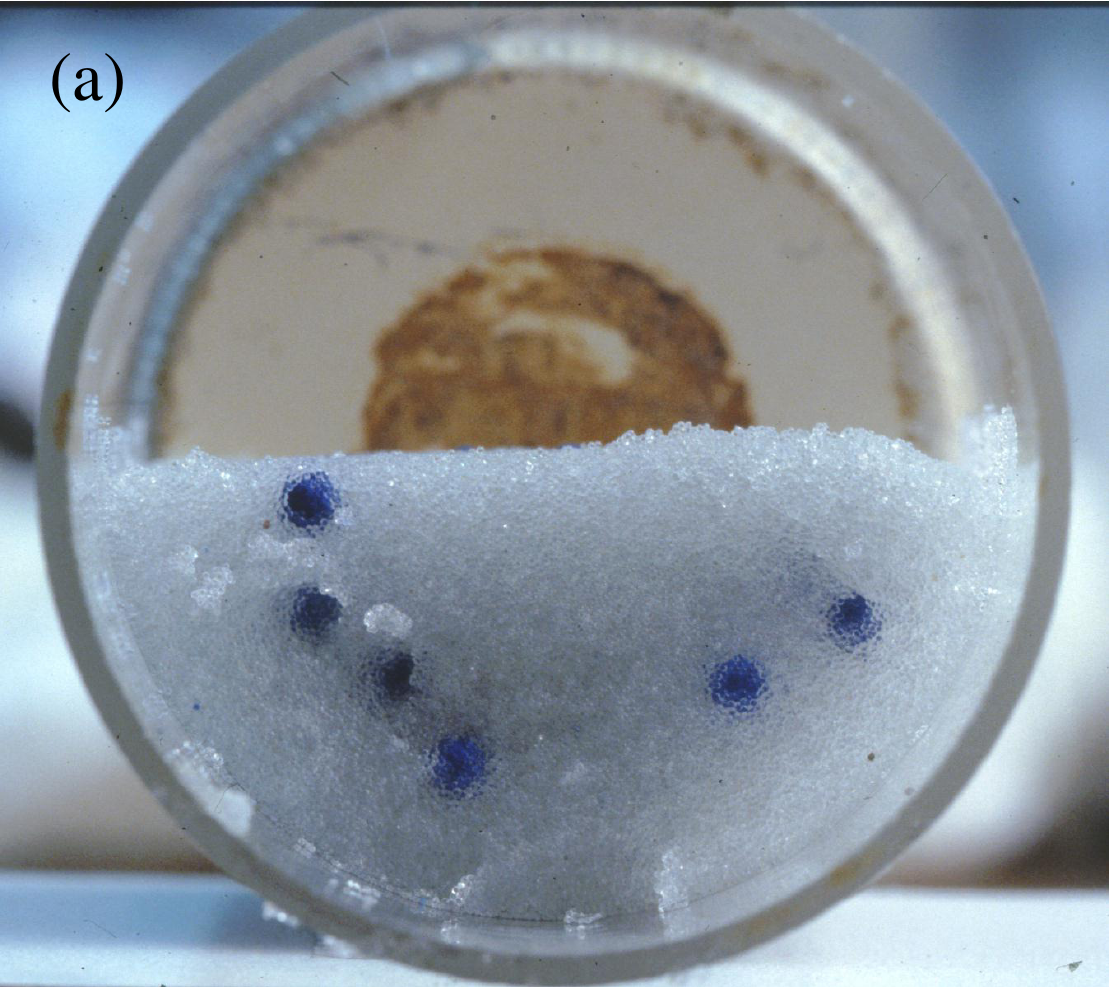}  
\includegraphics[width=0.434\linewidth]{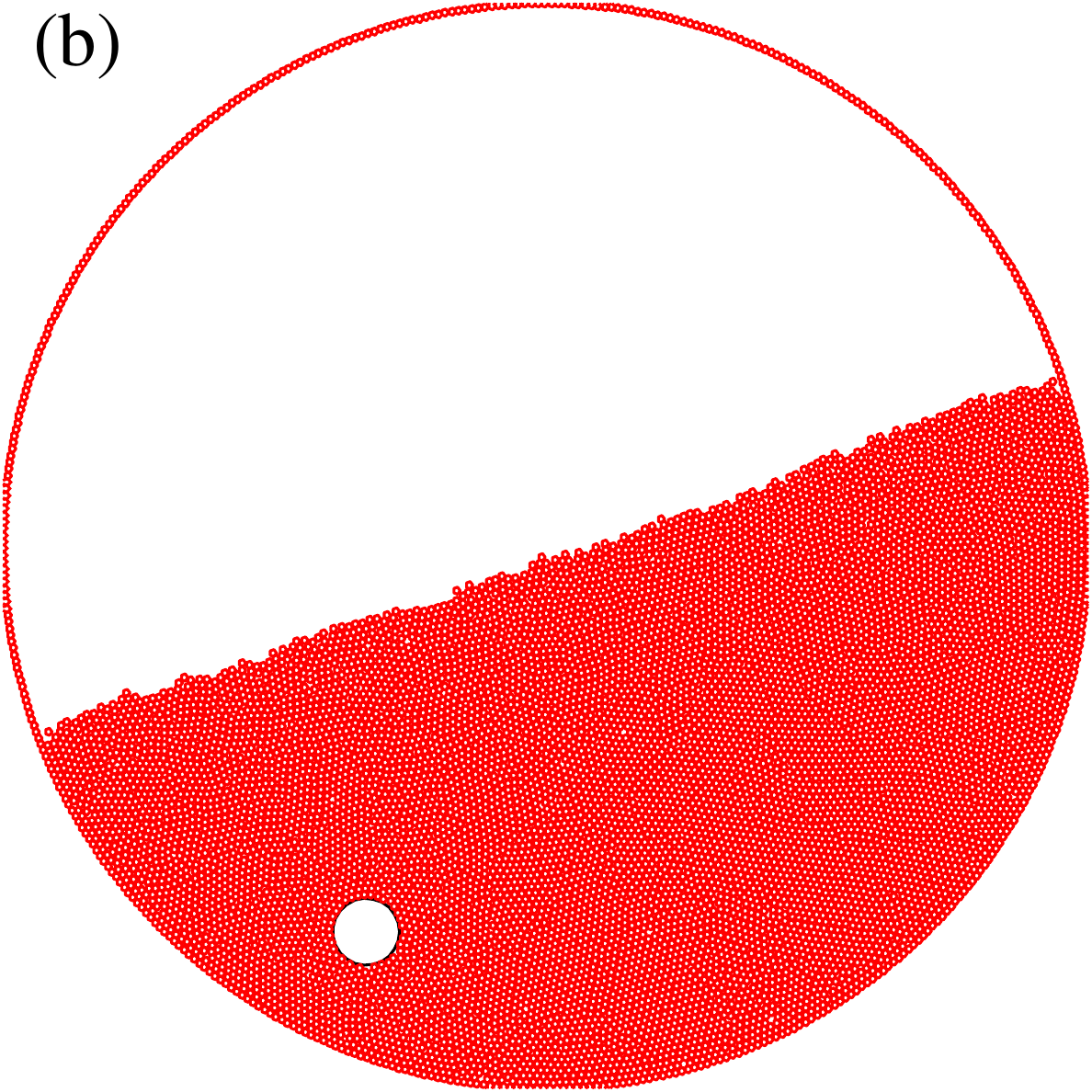}  
\caption{A $D=4.85$~cm rotating cylindrical tumbler with $d= 0.3$~mm
small particles and $d_t=$ 3~mm large particles (tracers): (a) cross-section of an experiment with 3\% blue tracers, slowly impregnated with water after the flow has stopped, then sliced \cite{Thomas00}, (b) 2D simulation with one tracer.} 
\label{tambnath10}
\end{figure}

Intermediate segregation has been precisely observed in the tumbler:  all large particles are found at the same intermediate radial position in the static part (Fig.~\ref{tambnath10}) \cite{Thomas00, FelixThomas04}, forming a segregation {\BLUE half-ring pattern}.  An intermediate ring corresponds to an intermediate depth in the flowing {\BLUE layer} (Fig.~\ref{tambnath16_traj}(a)).  This was measured for size ratios ranging from 4 to approximately 15, for small fractions of large particles (3\%) \cite{Thomas00}. In the experiments, the ring mean radius {\RED decreased} continuously with increasing size ratios, corresponding {\RED in the flowing layer} to a mean depth passing continuously from surface to bottom. 

The reversal of the segregation from the surface toward the bottom depends both on the size ratio and {\RED on} the relative fraction of particles \cite{Thomas00}: {\BLUE a limit between surface and reverse segregations can be defined and it corresponds to a size ratio between 4 and 5 for small fractions (1 to 10\%) of large particles; around 14 for a 30\% fraction}; and no reversal has been observed for a 50\% fraction, for size ratios below 45. As most of the size segregation studies {\RED were} done for equivalent fractions of both species, the reversal was not observed. Moreover, surface and reverse segregations give opposite, {\RED although} not symmetrical, patterns {\BLUE between the two species.} {\RED This asymmetry is} partly due to the use of a smaller fraction of large particles. {\RED However,} when the {\RED tracer} fraction {\RED is increased in an attempt} to reduce the pattern difference, this asymmetry is enhanced: surface segregation leads to a bi-layered {\BLUE(or two concentric zones)} system made {\RED up} of pure components, although reverse segregation progressively leads to an apparent mixing, except near the surface {\BLUE (near the tumbler periphery)} \cite{Thomas00}. Reverse segregation is not {\RED another kind} of surface segregation process, {\RED by which large particles are} placed at the bottom: it is {\RED a different} phenomenon with a different behavior.
 Note that the reversal of the segregation pattern is not due to percolation effects, as suggested by some authors \cite{ThorntonWeinhart12}, because {\BLUE they happen for a large fraction of large particle} \cite{Rahman}. For these reasons, we {\RED limit} our present study {\BLUE to one single large particle or to small fractions (5\% and 10\%)}.

Another series of experiments {\RED involves} particles of different densities and sizes {\BLUE in tumbler flow} \cite{FelixThomas04}. {\RED Similarly}, reverse and intermediate segregations of large particles {\RED are} observed. {\RED The mean segregation} depths are shifted toward the surface for less dense large particles, and shifted toward the bottom for denser large particles. For each {\RED tracer particle} material, the reversal of the segregation is therefore enhanced (resp. reduced) by an increase (resp. a decrease) {\RED in} the density of large tracers.
Only large beads {\RED made} of very light material always segregate {\RED to the} surface, and only very dense beads always segregate {\RED to the bottom (reverse segregation)}, whatever their size.
 These observations suggest {\BLUE that for particles of the same density, the reversal of the size segregation} is due to the increase {\RED in} their mass ratio. {\BLUE Heavy (because large) particles push light (because small)} particles around to make their way down. {\BLUE Moreover,}
the fact that large particles locate {\RED themselves} at a precise intermediate depth shows the existence of vertical gradients of force acting on them. {\BLUE Further studies are needed to extent these results to the case of flow down a solid rough incline.}

In fact, for {\BLUE an incline flow, we have the ``intermediate segregation" if large particles are found at intermediate depths inside the deposit}. Our previous experiments have shown that the mean depth for the large beads in the deposit {\RED varies} continuously with the size ratio from {\RED top to bottom} \cite{Thomas00}. However, these experiments were not precise enough to assess the {\RED occurrence} of intermediate segregation {\BLUE in channel flow:} there was a large spread of the individual positions for these intermediate mean depths. This may be due to the use of 10\% of large beads, but it could also be related to a non-stationary state of the flow, and/or to {\BLUE the modification of the tracer positions during the deposit aggradation}. For {\RED these} reasons, {\BLUE the existence or non-existence} of intermediate equilibrium  depths for a single large tracer 
in a granular flow down an incline is the main {\RED focus} of this article. 
The {\RED case} of several tracers (5\% and 10\% volume fraction) will also be considered for {\BLUE a comparison with a single tracer behavior and with previous experiments of reverse segregation} \cite{Thomas00}.   

This article is organized as follows. In section 2, the 
numerical method is presented. Section 3 {\RED studies} the tracer trajectory and equilibrium radial position in a rotating cylindrical tumbler in two (2D) and three dimensions (3D). The method is validated through {\BLUE quantitative comparison with previous 3D experimental results}. In section 4, the displacement of tracers in a granular 
flow down an incline is studied in 2D, and in 3D. Similarities and discrepancies between 2D and 3D, {\RED as well as the comparison between incline and tumbler flow are discussed}. Then, a study of multiple-tracer {\BLUE flows and a comparison with previous  experiments} are presented.

\section{The numerical method}

The numerical method used is {\RED the} distinct element method (DEM). A linear-spring and 
viscous damper force model  \cite{SchaferDippel96, CundallStrack79} is used to calculate the normal force between contacting particles:
$ \mathbf{F}_{ij}^n=[k_n\,\delta-2\gamma_n m_{\textrm{eff}}(\mathbf{V}_{ij} 
 \cdot \mathbf{\hat{r}}_{ij}  )] \mathbf{\hat{r}}_{ij} $
where $\delta$ and $\mathbf{V}_{ij}=\mathbf{V}_{i}-\mathbf{V}_{j}$ 
are the particle overlap and the relative
velocity of contacting particles{\RED ,} respectively, 
$\mathbf{\hat{r}}_{ij}$ is the unit vector in the direction between two particles
$i$ and $j$, 
$m_{\textrm{eff}}=m_i m_j/(m_i+m_j) $
is the reduced mass of the two particles, 
$ k_n=m_{\textrm{eff}}[(\frac{\pi}{\Delta t})^2+\gamma_n^2]$ is the normal
stiffness and $\gamma_n=\ln e\,/\Delta t$ is the normal damping with
$\Delta t$ the collision time and $e$ the restitution coefficient. 

A standard tangential force with elasticity is implemented: 
$\mathbf{F}_{ij}^t=-\min(| \mu \mathbf{F}_{ij}^n|,|k_s\zeta|)
{\rm sign}(\mathbf{V}^s_{ij}) $
where $\mathbf{V}^s_{ij}$ is the relative tangential velocity of the two 
particles, $k_s$ is the tangential stiffness, and 
$\mathbf{\zeta}(t)=\int_{t_0}^{t}\mathbf{V}^s_{ij}(t')\, {\rm d}t'$
is the net tangential displacement after contact is first established at
time $t=t_0$.  The gravitational acceleration is $g=9.81$\,m\,s$^{-2}$. 
The particle properties correspond to {\RED those of} cellulose acetate: density 
$\rho=1308$\,kg\,m$^{-3}$, restitution coefficient $e=0.87$ and
friction coefficient $\mu=0.7$
\cite{SchaferDippel96,DrakeShreve86,FoersterLouge94,ZamanDOrtona13}.
{\RED To prevent the formation of} a close-packed structure, the particles have a
uniform size distribution ranging from 0.95$d$ to 1.05$d$, with $d$ the particle diameter. The collision time is $\Delta t$ =10$^{-4}$~seconds, consistent with previous
simulations \cite{TaberletNewey06,ChenLueptow11,ZamanDOrtona13} and sufficient for
modeling hard spheres \cite{Ristow00,Campbell02,SilbertGrest07}.  These
parameters correspond to a stiffness coefficient $k_n = 7.32\times10^4$~N~m$^{-1}$ \cite{SchaferDippel96} and a damping coefficient $\gamma_n = 0.206
$~kg~s$^{-1}$.
 The integration time step is $\Delta t/50 = 2\times10^{-6}$~seconds to meet
the requirement of numerical stability \cite{Ristow00}.

{\RED The} rough inclined plane and {\RED the} tumbler walls are modeled as a monolayer of bonded particles of the same size. 
{\RED The} tumbler walls are composed of small particles {\RED in solid body rotation}. In the incline simulations, small beads (or disks) are placed randomly in the simulation domain and{\RED ,} as gravity is set, they
fall on a sticky plane (or line). All small beads touching the bottom of the domain ($z=0$) {\RED stop moving} and form the rough bottom of the inclined plane. The other {\RED beads} constitute the flowing granular material. With this procedure, rough planes {\RED whose} compacity {\RED is} around 0.57 {\RED are} obtained in 3D. A large tracer bead (or disk) is placed usually at
the top of the free surface and at time zero gravity is tilted from 0 to 
23$^\circ$ in 3D (or to $\theta=20$$^\circ$ in 2D), except where {\RED otherwise} stated, and the flow starts. For tumblers, the large tracer is placed first randomly inside the drum,
or at a defined location if needed. The other flowing particles are then
placed randomly inside the tumbler. At time zero, gravity is
switched on, {\RED the} flowing particles fall and {\RED the} wall particles start {\RED a 
rotational} movement. In tumblers and inclined planes, wall particles {\RED are assumed to} have an infinite mass for calculation of the collision force between flowing and fixed particles. 
The velocity-Verlet algorithm is used
to update the position, orientation, and linear and angular momentum of 
each particle. Periodic boundary conditions are applied in the directions $x$ or $x$-$y$ of the box (flow direction or flow - horizontal directions) in the case of an incline, and along the tumbler horizontal axis $y$ in the case of a 3D cylinder.     
In the tumbler case, velocity maps are obtained by binning particles {\RED into boxes}. The simulation domain is divided into 60$\times$60 boxes in the $x-z$ directions. The tumbler having a diameter of 4.85~cm (plus 2 small bead diameters), each box is a square of size around 0.8~mm. From these maps, streamlines and velocity profiles are extracted. Velocity maps are obtained, {\RED the tracer being} either included or excluded in the binning, or by generating a monodisperse flow where the tracer is replaced by exactly the same volume of small particles. All {\RED the} velocity maps {\RED obtained} are identical.

\hspace{0.5cm}

\section{Rotating cylindrical tumblers}     

{\RED In this part,} the aim is to obtain numerical results in 2D and 3D rotating cylindrical tumblers, 
{\RED in order} to compare them precisely with {\GREEN previous} 3D experimental results. This will provide a validation of the numerical method and some insights {\RED into} the processes happening {\RED during flow}.

The {\GREEN previous} experiments used glass beads of different 
diameters and of the same density \cite{Thomas00, FelixThomas04}. In  {\RED those} experiments, the rotating 
cylindrical tumbler (4.2~cm long and 4.85~cm in diameter) {\GREEN was half-filled with small beads and} a few large beads (typically 50) named tracers, initially placed {\RED so} that they {\RED barely interacted}. {\RED The} volume fraction of tracers was 3\%. 
The diameter of the tracers ($d_t=$ 3~mm) was kept constant
while the size of the small particles (diameter $d$) {\RED was} decreased from $d=$ 2.5~mm to $d=$ 90~$\mu$m
to explore size ratios ranging from $d_t/d=1.2$ to 33. 
The cylinder was rotated around its horizontal axis at about 3.6~rpm, {\RED so} that a continuous 
flow with a flat free surface {\RED developed}. After three revolutions, a stationary state 
was reached, with tracers at nearly identical radial positions, leading to a half-ring segregation pattern (Fig.~\ref{tambnath10}(a)). Since each radial position $R_{ti}$ in the solid rotating part corresponds to a depth $h_i$ {\RED during flow}, we interpreted the ring by the fact that all the tracers located {\RED themselves} at the same preferential depth within the flowing layer {\GREEN (Fig.~\ref{tambnath16_traj}(a))}. The radial segregated position $R_t$ was defined as the mean of all radial positions $R_{ti}$.

 {\RED It is important to choose} the same experimental dimensions {\RED for} the simulations,  {\RED so that} experimental and numerical results {\RED can be compared}, because the link between the radial position and the depth within the flowing layer is mainly a function of the tumbler and particle diameters. {\GREEN For instance}, equivalent size or density ratios give different radial positions in different tumbler diameters \cite{FelixThomas04, HillTan14}. 
From a numerical point of view, this experimental protocol is not easy to
reproduce since the number of small particles increases strongly {\RED with} increasing size ratio, already reaching 10$^5$ for 90~$\mu$m small particles in 2D. 
First, we will use dimensions as close as possible to {\RED those used in} the {\GREEN experiments}. {\RED Then}, the tracer size {\GREEN will be increased}  {\RED carefully} to reach larger size ratios.

\subsection{2D simulations of rotating tumbler}     

\subsubsection{Direct comparison with experiments}  

The 2D numerical tumbler of inner diameter ($D=~2R$) 4.85~cm is half-filled with monodisperse small disks and one large tracer (disk) of the same density. 
 {\RED The diameter of} the small particles varies from 2.5~mm to 90~$\mu$m and  {\RED that of} the large tracer is 3 mm. The tumbler rotates at 15~rpm to ensure a continuous flow with a flat free ``1D surface".

\begin{figure}[htbp]
\includegraphics[width=0.464\linewidth]{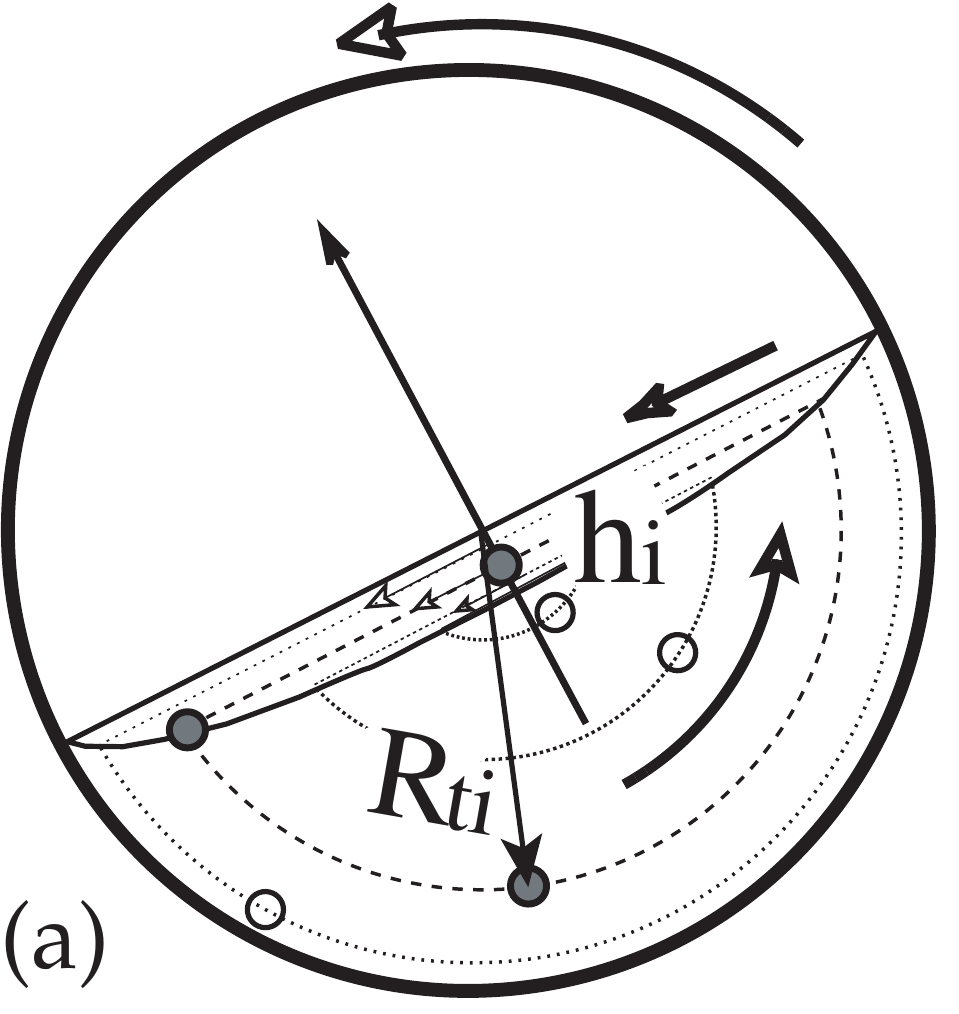}
\includegraphics[width=0.493\linewidth]{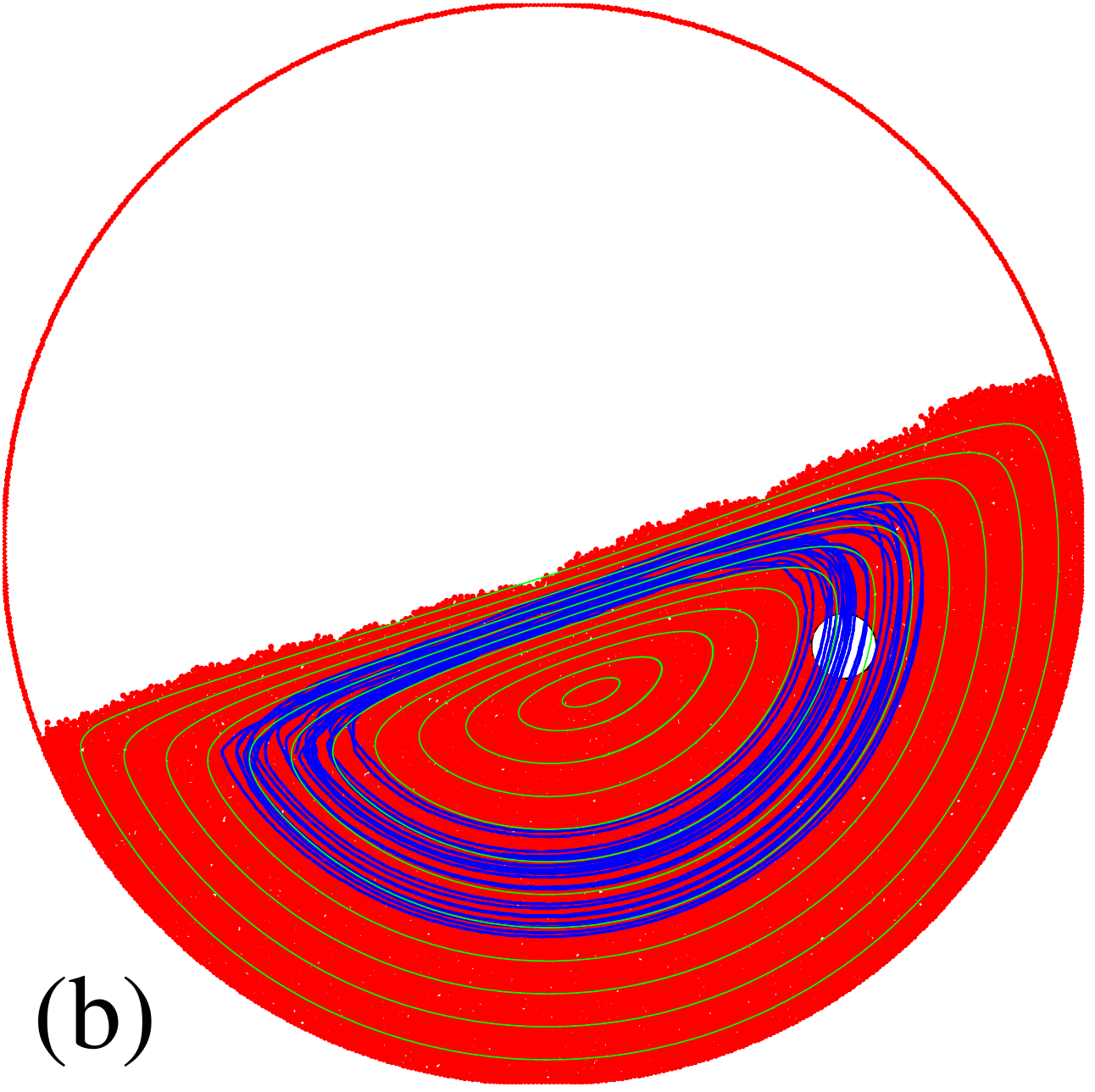} 
\caption{{\GREEN (a) Each depth $h_i$ in the flowing layer corresponds to a radial position $R_{ti}$ in the static part,} (b) A 4.85~cm diameter rotating tumbler with 187~$\mu$m {\GREEN red} small disks and a 3~mm white tracer, $d_t/d=$~16. The blue curve is the tracer trajectory. The tracer {\GREEN stabilizes} at an intermediate depth and {\GREEN radial position}. Green thin lines are the streamlines of the small disks.}
\label{tambnath16_traj}
\end{figure}

Figure~\ref{tambnath16_traj}{\GREEN(b)} shows {\RED the} trajectory of
a large tracer ($d_t/d=16$) passing successively through the flowing layer and the solid rotating zone. After a 
few revolutions (4 to 5, not shown here), the trajectory
converges {\RED to} and fluctuates around an equilibrium radial position: a stationary state is reached.
Each time $i$ the tracer passes through the vertical plane $x=0$ in the static rotating 
zone, the distance from the tracer {\RED center} to the cylinder {\RED center} $R_{ti}$ is measured. 
{\GREEN A} mean position $R_t$ and a standard deviation are computed. Small standard deviations indicate strong {\RED localization} at the same radial position {\GREEN from turn to turn}. {\GREEN This corresponds to stabilization at a constant depth $h$ in the flowing layer. This also} corresponds to segregation {\GREEN rather than to mixing} since several non-interacting tracers would all stabilize at {\GREEN these well-defined depth and radial position}. {\GREEN Consequently, they would regroup, i.e., segregate on this ring, exhibiting this small deviation. We choose to} call the equilibrium radial position $R_t$ a {\GREEN radius} of segregation, because it
corresponds to the experimental segregation half-ring radius obtained with 3\% of tracers {\GREEN (for a comparison between experiment and simulation see Fig.~\ref{tambnath10})}. {\GREEN Averaging and deviation {\RED calculation} are {\RED done} on one tracer during several turns for numerical data, or on several tracers at a given moment for experimental data{\RED ,} thus including trajectory fluctuations, but also potential {\RED tracer} interactions and experimental errors.}

\begin{figure}[htbp]
\center
\includegraphics[width=0.95\linewidth]{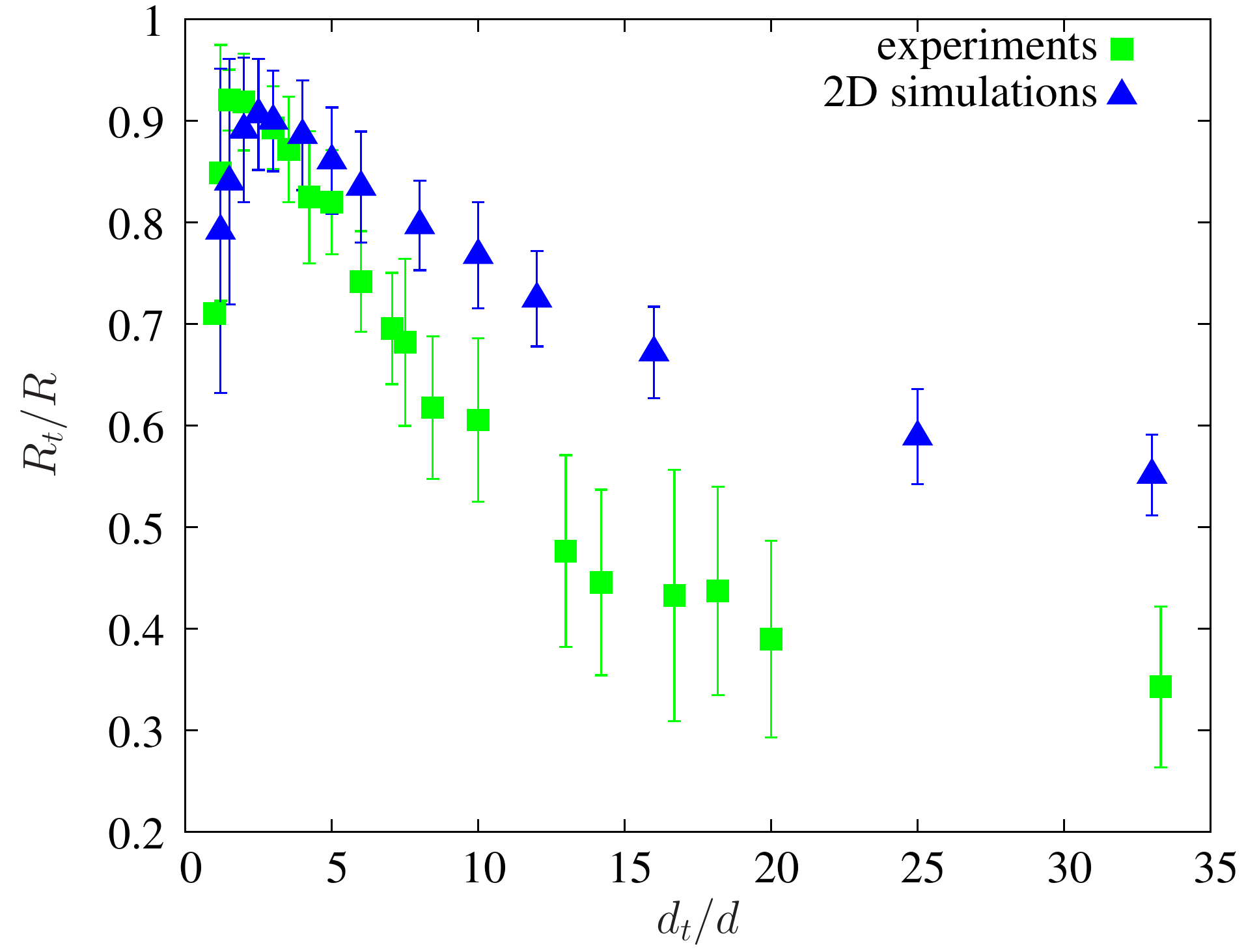}
\caption{Relative {\GREEN tracer radial} positions in the cylindrical tumbler versus size 
ratio $d_t/d$, for : {\GREEN several tracers} in 3D experiments (green~$\blacksquare$) \cite{FelixThomas04} {\GREEN and several  {\RED passages} of one tracer} in 2D simulations (blue~$\blacktriangle$). Error bars {\GREEN show} the standard deviation.}
\label{tambnath}
\end{figure}

{\GREEN In the simulations, a tracer with a size ratio from 1.2 to 3 is at the periphery, {\RED which} corresponds to a surface position {\RED during flow}. Each larger} tracer nearly {\RED remains} at an intermediate radial position $R_t$ inside the drum (Fig.~\ref{tambnath16_traj}(b)), {\RED which corresponds} to an intermediate {\GREEN depth {\RED during flow}. As $R_t$ decreases toward zero {\RED with} increasing size ratio, we deduce that the tracer} position is progressively deeper {\GREEN in the flowing layer. Fig.~\ref{tambnath} represents the evolution of $R_t$ with the diameter ratio $d_t/d$ showing the reversal of the tracer position} {\RED with increasing} size ratio. {\GREEN Each standard deviation value {\RED indicates whether} there is a well-defined position or a dispersed trajectory within the tumbler. {\RED In the event that several tracers are used a well-defined position leads to segregation and the dispersed trajectory leads to mixing}. In the tumbler, {\GREEN the spatial organization} passes from a spread of the instantaneous positions (for size ratio near 1) to well-defined equilibrium mean positions: {\RED at the} surface (maximum of $R_t$ is $R-d_t/2$), then at intermediate depths when $R_t$ decreases, and toward reverse depths for the lowest values of $R_t$.}

{\GREEN  We compare the successive numerical positions of one tracer, and the experimental positions of several tracers, both giving a value for $R_t$ and a standard deviation.} The agreement between experiments and simulations is good, {\RED but only qualitative,} with a similar evolution of the curve. {\GREEN Both simulations and experiments show the reversal of either the equilibrium position or the segregation location} (Fig.~\ref{tambnath}). 
 {\RED There are differences between} 3D experiments and 2D 
simulations: (1) In 3D experiments, the decrease of the curve $R_t/R$ versus $d_t/d$ 
is more rapid than in 2D simulations. 
(2) In 2D simulations, the asymptotic value of the curve is close to 0.55, a larger value than the asymptotic value in the 3D experiments, around $R_t/R= 0.35$. This 2D asymptotic value will  {\RED barely} be reduced for larger size ratios (Fig.~\ref{tambour_2DFin}).
(3) Another difference is observed {\RED regarding} the maximum of the
curve (surface segregation) which occurs for $d_t/d=$ 1.5 or 1.8 in experiments, instead of
$d_t/d=$ 2.5 in the 2D simulations (Fig.~\ref{tambnath}). 
We will see that these differences are due to the 2D nature of these simulations rather than to an experiment-simulation discrepancy. A longer discussion on that point is {\GREEN presented with} the 3D simulations.

\subsubsection{Higher size ratios}    

 {\GREEN To explore the asymptotic value, we need to reach larger size ratios, which would {\RED require} the use of a high number of small particles {\RED in the simulations}. To  {\RED overcome} this disadvantage} several larger tracer sizes {\RED are} tested ($d_t=3$, 
4.85, 6 and 9.7~mm) in the tumbler $D=48.5$~mm, and their equilibrium positions $R_t$ {\RED are} compared  {\GREEN(for size ratios 25 and 40).} 
Up to a diameter of $d_t=6$~mm, $R_t$ are almost identical. For the largest tracer ($d_t=9.7$~mm, {\RED whose} size {\RED is} to
be compared with the drum diameter $D=5d_t$), a small discrepancy 
(relative error of 4\%) {\RED is} observed. We choose to keep the size of the tracer under $d_t=D/10$ to be sure that {\RED there would be} no effect {\RED of} the tracer size.  {\GREEN For the particles and tumbler studied here, the 
thickness of the flowing layer is always larger than one tracer diameter.} 

A tracer diameter $d_t=4.85$~mm is adopted to reach large size ratios up to 60.
\begin{figure}[htbp]
\center
\includegraphics[width=0.95\linewidth]{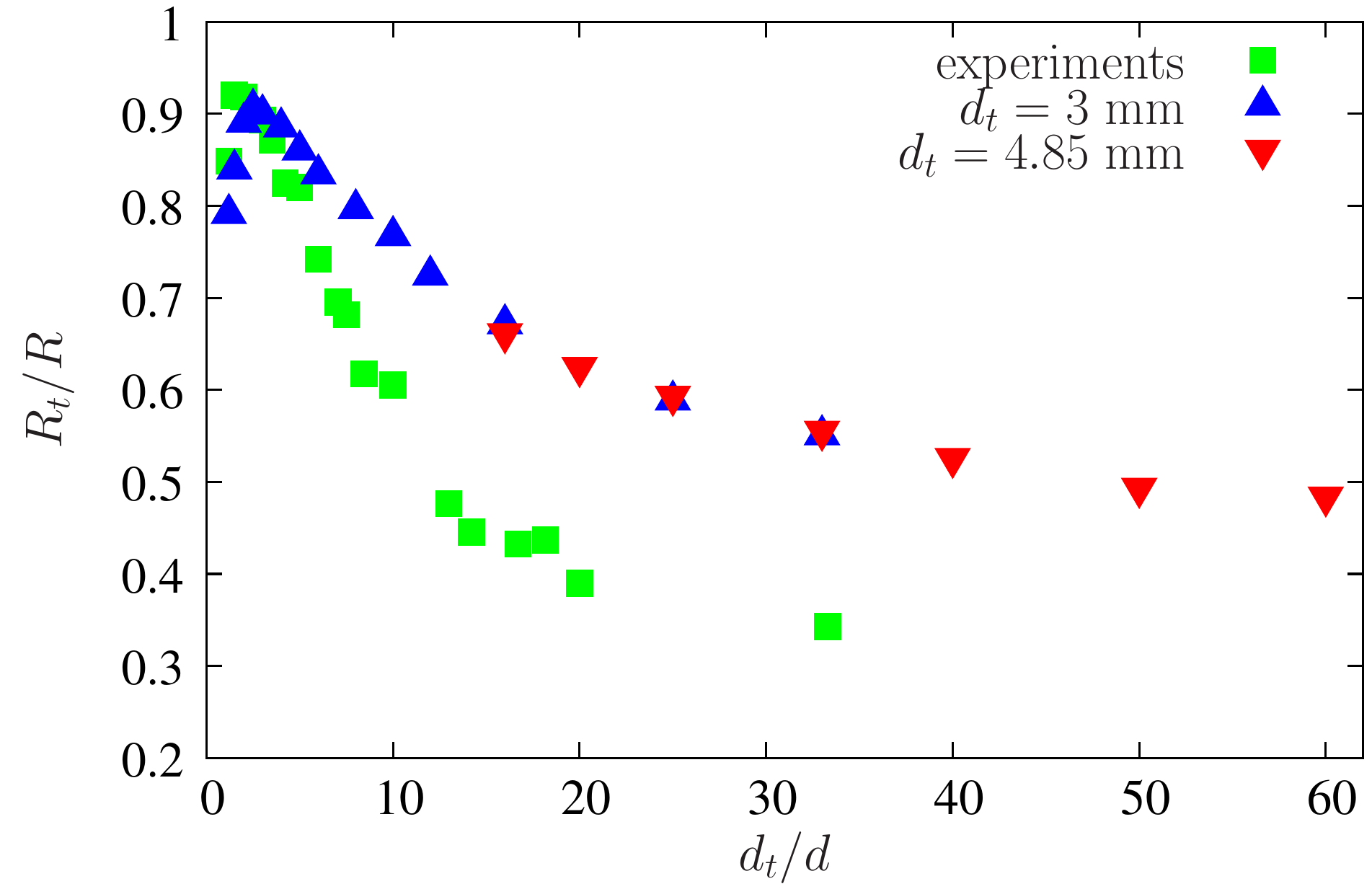}
\caption{Relative {\GREEN tracer} positions in the tumbler versus size ratio $d_t/d$, for 3D
experiments {\GREEN with 3\% of tracers} \cite{FelixThomas04} and 2D simulations with 2 tracer sizes, 3 and 4.85~mm
(no standard deviation here).}
\label{tambour_2DFin}
\end{figure}
Fig.~\ref{tambour_2DFin} 
shows the relative positions $R_t/R$ of the 3~mm and 4.85~mm tracers, compared to 3D experimental results. 
When the 2 different tracers have the same size ratio $d_t/d$, the resulting positions coincide.
For the largest size ratios used in these 2D simulations, the radial position slowly decreases
but remains close to 0.5, and does not reach the experimental value of 0.35.  {\GREEN 2D {\RED simulation} and 3D {\RED experiment}} asymptotic $R_t$ values are different.
Even if there is a qualitative {\GREEN agreement,} 3D simulations are needed for an accurate comparison.

\subsection{3D rotating tumblers}    

\subsubsection{Comparison with experiments}

\begin{figure}[htbp]
\center
\includegraphics[width=0.6\linewidth]{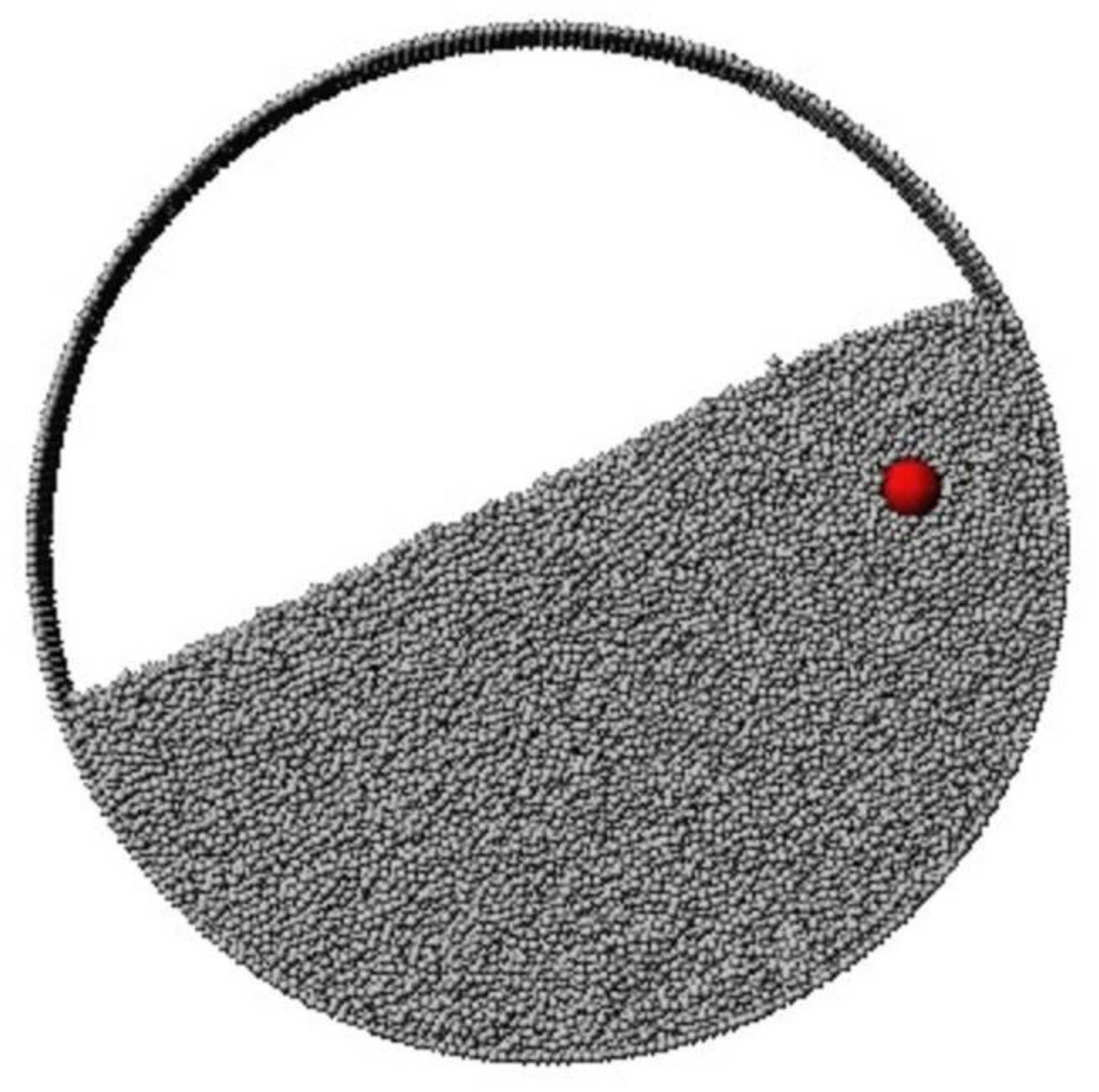}
\caption{3D simulation of a rotating cylinder ({\GREEN diameter} 48.5~mm) with a 3~mm tracer in 0.5~mm small beads.}
\label{tambour3d6}
\end{figure}

To obtain a quantitative agreement, 3D simulations {\RED are} conducted (Fig.~\ref{tambour3d6}). 
The tumbler inner diameter is equal to $D= 48.5$~mm and {\RED it} rotates around the $y$ axis at 15~rpm. 
In a first series (size ratio up to 8), the tracer diameter is set to $d_t = 3$~mm as in experiments, then for larger
size ratios (from 5 to 25) it is set to $d_t=4.8$~mm to reduce the number of small {\RED simulated} beads. 
For size ratios $d_t/d=5$ and 8, both tracer sizes {\RED are} tested. Larger tracers ($d_t=6$ and 9~mm) {\RED are} also used {\GREEN respectively from size ratios 5 to 25 and 12 to 25} to check the {\RED sensitivity} to the tracer size. As in 2D, no differences {\RED are} observed for the 6~mm tracer, and very small discrepancies {\RED are observed} for the 9~mm tracer.

\begin{figure}[htbp]
\center
\includegraphics[width=0.95\linewidth]{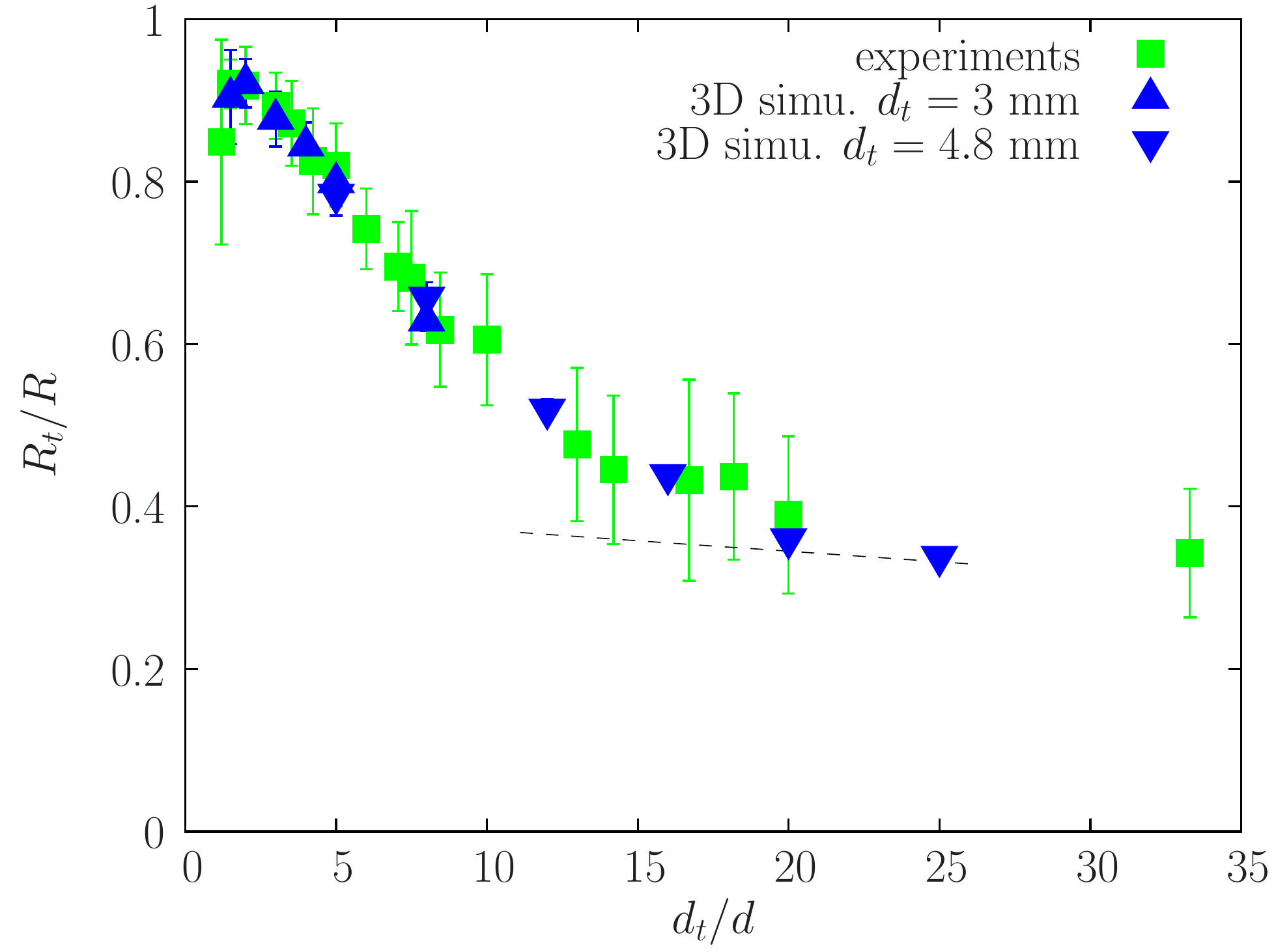}
\caption{Relative radial positions of the tracers versus size ratio, in 3D experiments {\GREEN (3\% of tracers)} \cite{FelixThomas04} and 3D simulations {\GREEN (one tracer, 3 or 4.8~mm). {\RED The} numerical standard deviation is very small (error bars).} {\RED The} dashed line is the position of a tracer touching the bottom of the flow.}
\label{tambour3dfin}
\end{figure}

{\GREEN The 3D numerical results show the evolution of the tracer radial position $R_t$ from the periphery to intermediate positions, toward the reverse position when the size ratio {\RED is increased} (Fig.~\ref{tambour3dfin}). {\RED The} standard deviation is very small{\RED ,} indicating a strong localization on the same radial position from turn to turn.}
{\GREEN The 3D numerical radial} position quantitatively {\RED matches} the 3D experimental {\GREEN radial segregated position of several tracers}. Agreement is very good, even on precise points like: 1) the slope of the curve, 2) the asymptotic value of $R_t/R$ for large
size ratios, and 3) the diameter ratio which corresponds to the maximum of the curve. {\GREEN
The agreement confirms our hypothesis that a few tracers locate {\RED themselves} on a ring which has the same radius {\RED as} the equilibrium radial position of one single tracer. The segregation of several non-interacting tracers can be seen as the regrouping at an identical position because the equilibrium radial position of each tracer depends only on its size ratio. The tracers do not interact much at this small fraction (3\%), nevertheless their interaction leads to a slight increase {\RED in} the standard deviation, with no observable change in the mean value. We can speak interchangeably of segregation radius or of equilibrium radial position. Moreover, as the agreement is really quantitative,} we are confident in our simulation method {\RED to be used} to study other systems, such as flows on rough inclines.

\subsubsection{Trajectories in 3D tumbler}   

\begin{figure}[htbp]   
\center
\includegraphics[width=0.95\linewidth]{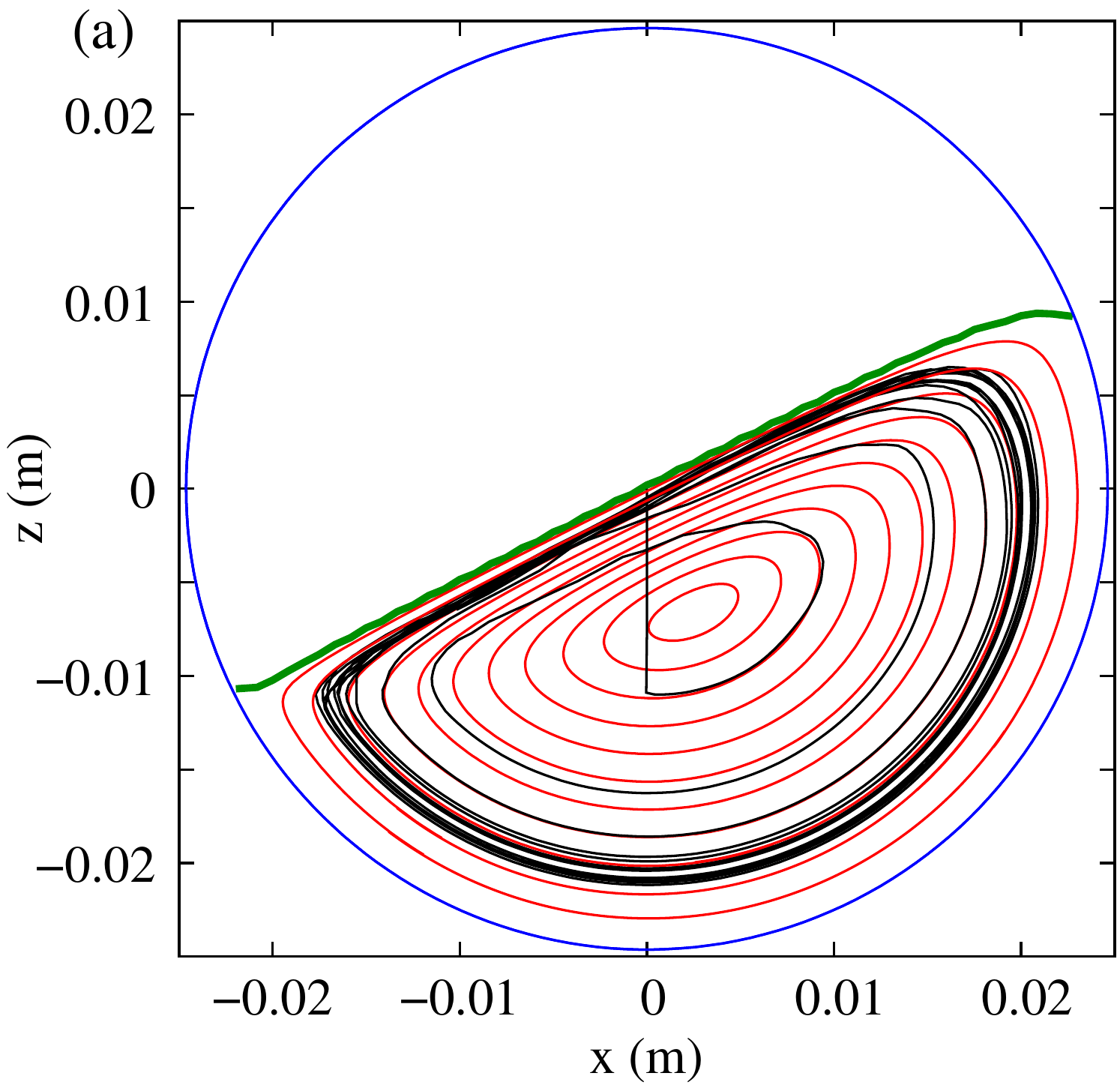} 
\includegraphics[width=0.95\linewidth]{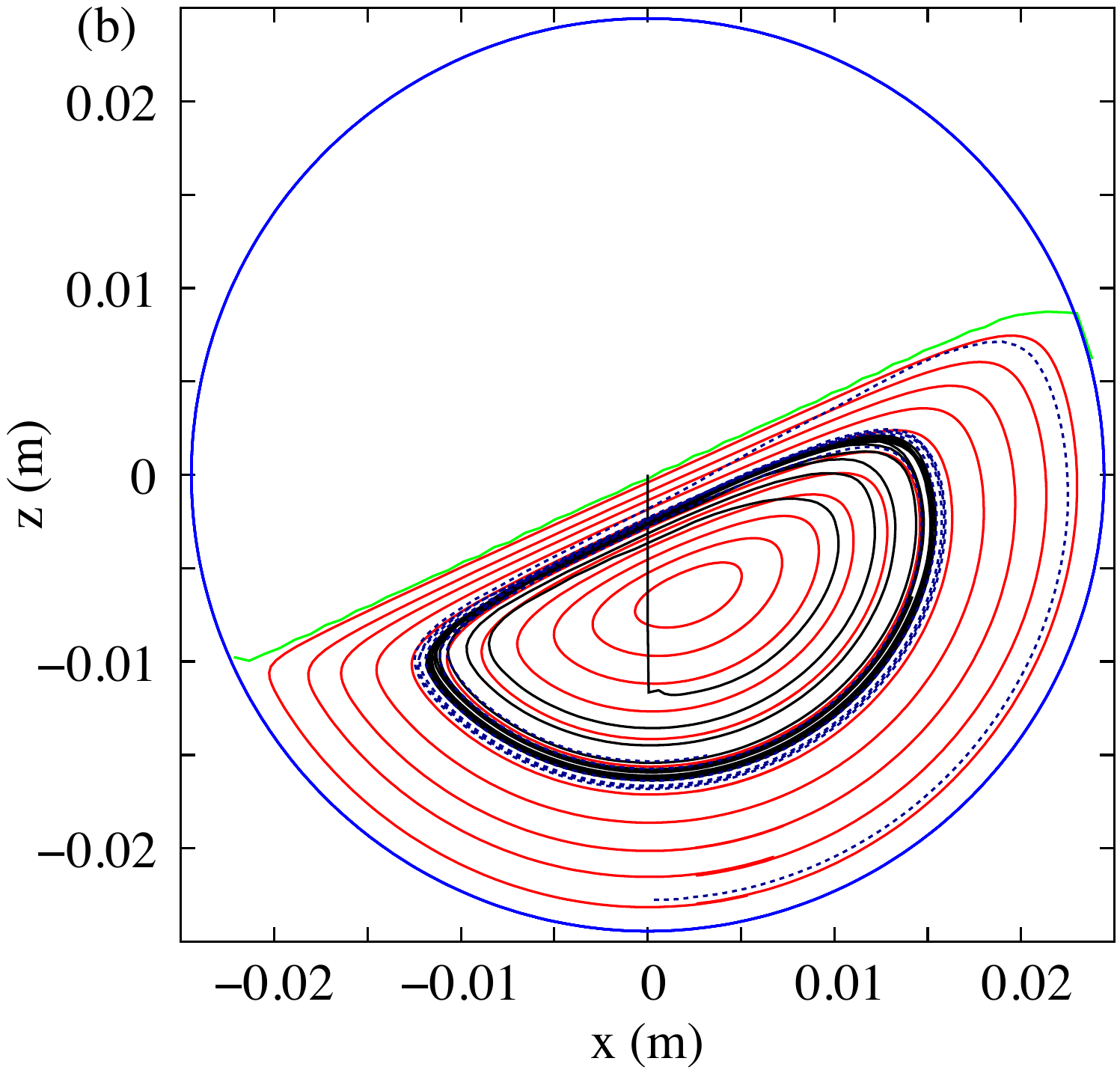}  
\caption{Trajectories of the tracer {\RED center} (black curves), and {\RED small bead} streamlines (red curves) ($D=48.5$~mm, $d_t=3$~mm). The thick green curve {\GREEN is} the free
surface. The first rotations concern the convergence, the following {\RED rotations concern} the stationary phase. (a) $d_t/d=4$, the tracer is at the limit between surface and intermediate {\GREEN positions,} its top touching the free surface, (b) $d_t/d=8$ {\GREEN {\RED  either starting from the periphery (dashed line) or from the tumbler {\RED center} (solid line)},} the tracer is at an intermediate position.}
\label{trajectory4}
\end{figure}

{\GREEN To gain a better understanding of the segregation phenomenon, tracer trajectories are studied in details.}
Fig.~\ref{trajectory4}(a) shows the trajectory of a large particle with a size ratio of~4 and the streamlines of small beads {\GREEN in a plane $x-z$.}
Two phases are distinguished: first, the unsteady stage, second a stationary trajectory when the equilibrium depth is reached. 
 
 The tracer initially falls after the tumbler has been filled (vertical line), then the rotation starts with the tracer relatively close to the stagnation point. During the first, second, third {\RED passages}, and {\RED the} first part of the fourth passage in the flowing layer, the tracer {\RED exhibits} an upward motion when compared to the small bead streamlines. It migrates towards its equilibrium position.
Accordingly,  in the static zone, from one passage to the {\RED next},
the radial position $R_{ti}$ increases. Then, after these 4 passages, the trajectory is stationary: the tracer flows along the streamlines at each passage, and presents a nearly constant radial position $R_{ti}$ with some fluctuations from turn to turn. 
This confirms the experimental observation that after 3 rotations ($\simeq 6$ passages through the flowing layer for a half-filled drum) the whole segregation process is over \cite{Thomas00}.
{\GREEN The convergence to an equilibrium depth, and consequently} the segregation process, happens mainly during flow, and is not due to processes happening during the entrance {\RED into} and/or the exit from the flowing layer. 
Here, the tracer {\RED starts} from a central position, and {\RED moves} upwards to reach its equilibrium depth. It could have been downwards if the tracer {\RED had} been released from the surface {\GREEN of the flow (or periphery in the static part).} 
An equivalent upward motion is observed for a tracer with a size ratio~8 (Fig. \ref{trajectory4}(b)), but {\RED its} amplitude {\RED is smaller,} as the starting position is closer to the equilibrium $R_t/R$ corresponding to this size ratio.
{\GREEN A more rapid downward motion toward the same equilibrium radial position is observed when the tracer is released at the periphery, probably because of the longer distance {\RED traveled} in the flowing layer.}

Once in the steady phase, the tracer trajectory and small bead streamlines are parallel
in the flowing zone. There is no relative motion  {\RED any longer, either} up or down. Plotting a circle 3~mm on the trajectory shows that the tracer {\RED with} size ratio~4 is just below the surface, and that the tracer {\RED with} size ratio~8 is on {\GREEN a mid-height} intermediate depth. Each depth in the flowing layer corresponds to one radial position in the rotating part of the tumbler. However, the tracer trajectory does not match exactly the same small bead streamline in the rotating zone and in the flowing zone. There are two small shifts between the tracer trajectory and the small bead streamlines when going in and out of the flowing layer. At the entrance (Fig.~\ref{trajectory4}(b)), the tracer starts to move after the small beads {\RED on} the same streamline (despite the shift {\RED occurring at} the previous exit), probably because its bottom is still surrounded with non-moving small beads. At the exit of the flow, the tracer stops before the small beads {\RED on} its corresponding streamline because  its lower part is touching the static curved bottom ({\GREEN note that these entrance and exit shifts are enhanced in 2D (Fig.~\ref{tambnath16_traj})).} In conclusion, these shifts are not responsible for the segregation from turn to turn. But they exist, {\RED they} probably vary with $R_t$ and might be one cause of the discrepancy between 2D and 3D. For that reason, it is not possible to easily deduce the flowing depth positions from both data of small particle streamlines and $R_t/R$. Nevertheless, the shifts are very small and the $R_t/R$ variation mainly reflects a variation {\RED in} depth within the flowing layer. 

{\RED A more accurate examination of the trajectory reveals that} the entrance in the flow induces a starting point
slightly above the equilibrium depth that the tracer will reach (Fig.
\ref{trajectory4}(a)).  {\RED Each time it passes through} the flowing layer, {\RED the} tracer exhibits a
tiny descent towards its equilibrium depth, then {\RED remains at} a constant depth to the
end of the flow, parallel to streamlines. The length at which the constant depth is reached seems to decrease with the tracer size ratio, approximately at mid-length for ratio 4, almost immediately for ratio 8
(Fig.~\ref{trajectory4}).  Segregation is so fast that slight
destabilization can be rebalanced in less than {\RED one passage in the flow}. 

In conclusion, the study of trajectory in the 3D cylindrical tumbler shows that the process responsible for the segregated radial positions of tracers is a vertical migration and stabilization of the tracer at various depths, occurring during flow. We then expect a similar phenomenon to happen during flow on an incline.

\subsubsection{Radial position and depth within the flowing layer}

One {\RED may} wonder how the different values of $R_t/R$ {\RED should be interpreted} in terms of equilibrium depth within the flowing layer of the tumbler, to anticipate conclusions {\GREEN across} the tumbler study and the {\GREEN following} incline study. In particular, the question arises whether the asymptotic small values of $R_t/R$ do correspond {\GREEN or not} to a reverse segregation {\GREEN within the flow}.
 {\GREEN A tracer} touching the bottom of the flow undergoes a small decrease {\RED in} $R_t/R$ with the size ratio, because with our protocol ($d$ decreasing) the thickness of the flowing layer slightly decreases \cite{FelixFalk07}. Numerical thickness measurements{\RED,} added with a tracer radius to obtain the tracer {\RED center} position{\RED, are} shown as a dashed line on Fig.~\ref{tambour3dfin}: it  {\GREEN would} correspond to  {\GREEN reverse} positions,  {\GREEN turning around the stagnation point.} We deduce that size ratios 20 and 25 are in reverse position, and that the small decrease between them is explained by the choice of the protocol.

In addition, in some simulations, we measure the depth of the tracer directly on its trajectory. For a ratio 8, (Fig. \ref{trajectory4}(b)) the tracer is at an intermediate depth.  For the largest size ratios (for example, ratio 20), the tracer is just touching the bottom of the flowing layer{\RED, i.e.,} the bottom of the tracer passes where the streamlines are reduced to the stagnation point, which is nowhere else than the middle point of the bottom.  {\GREEN But this method is not precise:} it is difficult to define the bottom of a flowing layer  {\GREEN near the tracer.}
The bottom of the flowing layer is defined by an averaging of small bead streamlines. The tracer passage has almost no effect on the averaging, although it probably deforms locally and during a short time the granular material  {\GREEN below and} around it when it passes ``at the bottom".  {\GREEN Consequently, the bottom of the averaged flowing layer may not be the same {\RED as} the local bottom of the flow around the tracer.} Nevertheless, we choose to call the positions of these tracers with the largest ratios {\RED "reversed"}, keeping in mind that this is somehow arbitrary. {\RED In fact, denser tracers may be found} at lower $R_t$ than the asymptotic value, {\RED probably because they more strongly} deform the bottom \cite{FelixThomas04}.
With that choice, all the {\GREEN  asymptotic} $R_t/R$ positions correspond to a tracer in a reverse bottom position within the flow. {\GREEN In conclusion, reverse depth is reached for tracers with a size ratio$\geqslant$ 20 in 3D. 
The same measurements on trajectories {\RED are} made in 2D: tracers {\RED are} found {\RED at} intermediate depth for size ratio 10, 16 and 20, and in reverse position for size ratios above 40. The reverse position can be reached both in 2D and 3D, but for greater size ratios in 2D.}    
  
\subsubsection{Differences between 2D and 3D tumblers}

{\GREEN Compared {\RED with} 3D {\RED results}, 2D results are shifted, as if} the size ratio  {\RED had} a reduced effect: the maximum of the $R_t/R$ curve occurs for a higher size ratio, the dependency is smaller, and the asymptotic value is higher (Figs.~\ref{tambour3dfin} and \ref{tambour_2DFin}).
{\GREEN We} first check that the difference {\GREEN between} 2D and 3D is not due to a {\GREEN variation
of the} thickness of the flowing {\GREEN zone. Indeed, the thicknesses} have been 
measured nearly identical for a given {\GREEN small bead size} in 2D and 3D {\GREEN tumblers.}
{\GREEN Secondly,} for a given size ratio ($d_t/d=$20) we compare the depth of {\GREEN each} tracer {\GREEN on its trajectory} within the flow: in 3D, the tracer is touching the bottom, while in 2D, 6 small beads {\GREEN are under the tracer (this flow thickness is 25$d$)}. 
The shift {\GREEN between 2D and 3D} $R_t/R$ {\GREEN curves} does correspond to a real {\GREEN difference} in depth positions {\GREEN within} the flowing layer.

{\GREEN Nevertheless, a radial position difference in 2D and 3D can also be seen for tracers at the same depth. For example,  for} {\GREEN the asymptotic values,} the largest tracers in 2D (size ratios above 40) {\GREEN and in 3D (size ratios 20 and 25)} {\RED are} {\GREEN all} measured touching the bottom of the flowing layer. {\GREEN The size of the tracer is fixed ($d_t=4.8$~mm or 4.85) and the thickness of the flowing layer is almost unchanged for these small bead sizes $d$ (a slight decrease indicated by the dashed line in Fig.~\ref{tambour3dfin}). Nevertheless,} there is a {\GREEN gap between} the 2D and 3D values of $R_t/R$ {\GREEN corresponding} to these identical reverse depths (Fig.~\ref{tambour_2DFin}). {\GREEN Note that to take into account the slight variation {\RED in} the flowing layer thickness with the small bead size, one can simply
extrapolate $R_t/R$ values in 3D up to 40 (Fig.~\ref{tambour_2DFin}), and compare radial positions exactly at the same flowing thickness: the conclusion is unchanged.}

{\GREEN Considering tracers at a same depth, the $R_t/R$ difference {\RED is} mainly {\RED due to} the larger trajectory fluctuations in 2D which shift $R_t$ to larger values in 2D when approaching the bottom (and to smaller values when {\RED nearing} the surface). It is also due to} a difference in {\RED the} entrance and exit of the flowing layer, {\GREEN which gives smaller $R_t$ values in 2D. This {\RED latter} effect pushes in the opposite direction, but is small compared to the {\RED former} one.}

{\GREEN We have seen that there is a difference between the 2D and 3D equilibrium depths 
within the flowing layer all along their evolution with the size ratio.}
To understand the cause of this {\GREEN difference}, one should compare the effective densities of the medium made {\RED up} of 
small particles. If we note $\rho$ the density of the small (or large) particles,
the effective density of the granular medium made {\RED up} of small particles is equal
to $c\,\rho$, where $c$ is the compacity. A large tracer is 
denser than a sphere/disk of the same diameter filled with a random close
packing of small particles. In 2D, such a packing gives a compacity close to
$c_{2d}\simeq 0.8$, while in 3D, $c_{3d}\simeq 0.6$. Thus, the 
density ratio between the tracer and the medium is larger in 3D than in 2D, 
leading to deeper intermediate segregation and advanced reverse segregation. This result was confirmed using tracers of decreasing densities \cite{FelixThomas04}. For tracers less dense than a random packing of small particles, only surface segregation of the tracer is observed.

 Even if the names and limits of the equilibrium depths are arguable, there are similarities but also discrepancies between 2D and 3D cases. 
In 3D, the evolution of the position shows a shift of the curve maximum toward smaller size ratios and a stronger dependency with size ratio. If the enhancement of the effect of the size ratio is due to the compacity around 0.6 (in 3D) instead of 0.8 (in 2D), we expect that it will always be present in all types of flow. Thus, {\RED care} should be taken  {\RED when extrapolating these 2D studies} to the 3D case.

\section{Rough incline}    

The experimental study of granular segregation occurring during flow down an  
incline is a difficult task to achieve in {\GREEN wide and thick} channels (3D). Indeed, in our previous 
experiments \cite{Thomas00} only the surface of the flow was visible. {\GREEN  A fraction of 10\% of large particles was used.} For large size ratios ($d_t/d\geqslant 6$), no
tracers were visible at the surface during flow, {\GREEN although for $d_t/d\leqslant 3$, large particles were at the surface.} The volume of
the deposit could be accessed after the flow {\RED had} stopped due to a slope change or
a vertical end wall. {\GREEN But the aggradation of the deposit may have modified the particle {\RED depths}.} {\GREEN Size ratio had been varied and the segregation pattern in the deposit changed {\RED according} to the size ratio: for small {\RED size} ratios, the large particles covered the surface of the deposit, while} for larger size ratios, the large particles were found inside the deposit. {\GREEN The individual positions were moderately spread inside the deposit. Nevertheless, their mean position was} at an intermediate depth, which was {\RED deeper} and deeper {\RED with} increasing size ratios. {\GREEN Because of this spread and because of the aggradation,} it was not possible to conclude {\RED on whether} these tracers were {\GREEN located at a 
well-defined intermediate} depth {\RED during flow} (corresponding to an intermediate segregation). Simulations will allow measurements during flow, in an established steady-state regime, {\GREEN with a single tracer.}

The main advantage of the incline geometry is that  {\RED the} measurements of tracer {\RED depths} within the flows are direct,
while measurements of radial positions in the tumbler {\RED involve} entrance in the flowing layer, acceleration and exit {\RED from the flowing layer.} 
Moreover, {\RED for a} solid rough incline, the bottom {\RED depth can be accurately determined}, which is not the case in a partially filled tumbler where {\RED flow passes} on loose granular matter having a curved bottom {\RED shape}. 
Another difference is that the flow thickness in the tumbler is mainly imposed by the dimensions (tumbler {\RED diameter} and small particle size). For the chosen protocol (decreasing small bead size), this layer thickness decreases with the size ratio (around $8d$ for $d_t/d= 2$; $21d$ for $d_t/d=10$; 34$d$ for $d_t/d=25$, {\GREEN  respectively: 4, 2.1 and 1.4$d_t$}), 
while for confined flows in an inclined channel the thickness of the flow can be varied independently. In the present study, the smallest thickness {\RED chosen} is comparable to those encountered
in non-confined flows on an incline (around $10d$) \cite{FelixThomas04e}, and, for this reason, the results on such thin flows {\RED are not without} interest. The thickness will be increased {\GREEN ($37d$),} to explore {\RED thickness} effects, and will reach the {\RED values for} experimental channel flows, for comparison \cite{Thomas00}.

As experimental results were obtained without following any protocol, we choose to keep the small bead size constant ($d=6$~mm), and vary the tracer size ($d_t$). Indeed, decreasing the small bead size would have {\RED resulted in} increased flow velocity and {\RED increased} calculation time for a constant flow thickness. Nevertheless, we may expect some deviations between experimental and numerical results if the tracer becomes too large compared {\RED with} the flow thickness.

\subsection{2D simulations of flows on an incline}     

\subsubsection{Intermediate segregation}

\begin{figure}[htbp]
\includegraphics[width=0.95\linewidth]{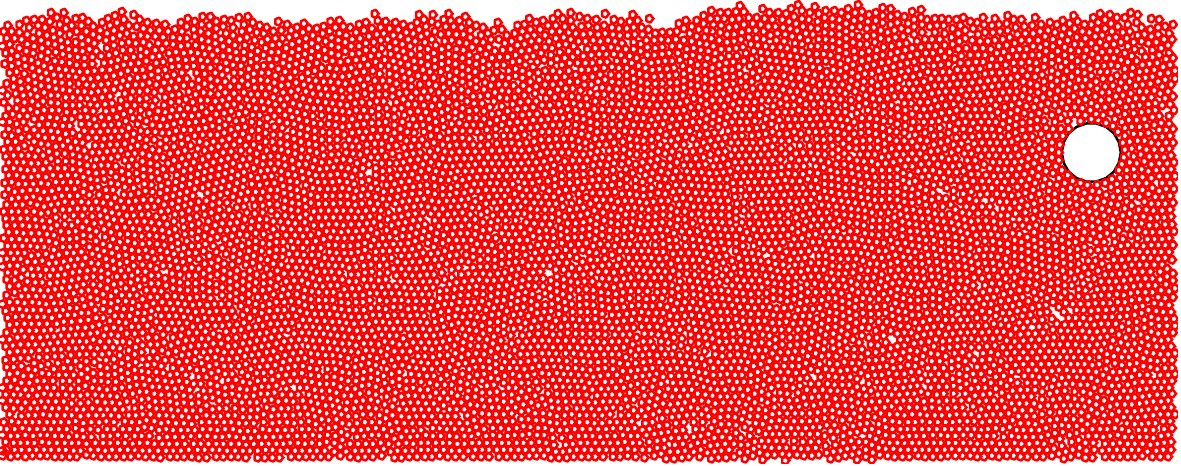}
\caption{A 2D granular flow down a rough incline with 6~mm small 
disks and a 48~mm large tracer, moving from left to right. The slope angle is 20$^\circ$, the flow thickness is 36~cm.}
\label{plan8}     
\end{figure}

Even {\RED though} quantitative agreement cannot be taken for granted, {\GREEN we first perform} 2D simulations. {\RED The} simulation domains are 160$d$ long, or 300$d$ long for the larger tracers ($d_t/d\geqslant8$).
Fig.~\ref{plan8} shows a 
48~mm diameter tracer (disk) in a granular flow made {\RED up} of 6~mm small disks flowing
down an incline. The plane slope is 20$^\circ$ and
the thickness of the flow $h_{max}$ is around 36~cm.   
The tracer with this size ratio ($d_t/d=8$) is not far from the free 
``1D surface" but remains below it, {\GREEN fluctuating around} an intermediate depth. This does not correspond to {\GREEN the behavior of a large particle during} the classical granular surface segregation of large particles, but {\RED to that of} a particle flowing inside {\GREEN the bed,} at an {\GREEN equilibrium intermediate} depth: this {\GREEN would lead to} intermediate segregation {\GREEN if several non-interacting tracers of the same size were present.}

\begin{figure}[htbp]
\includegraphics[width=0.95\linewidth]{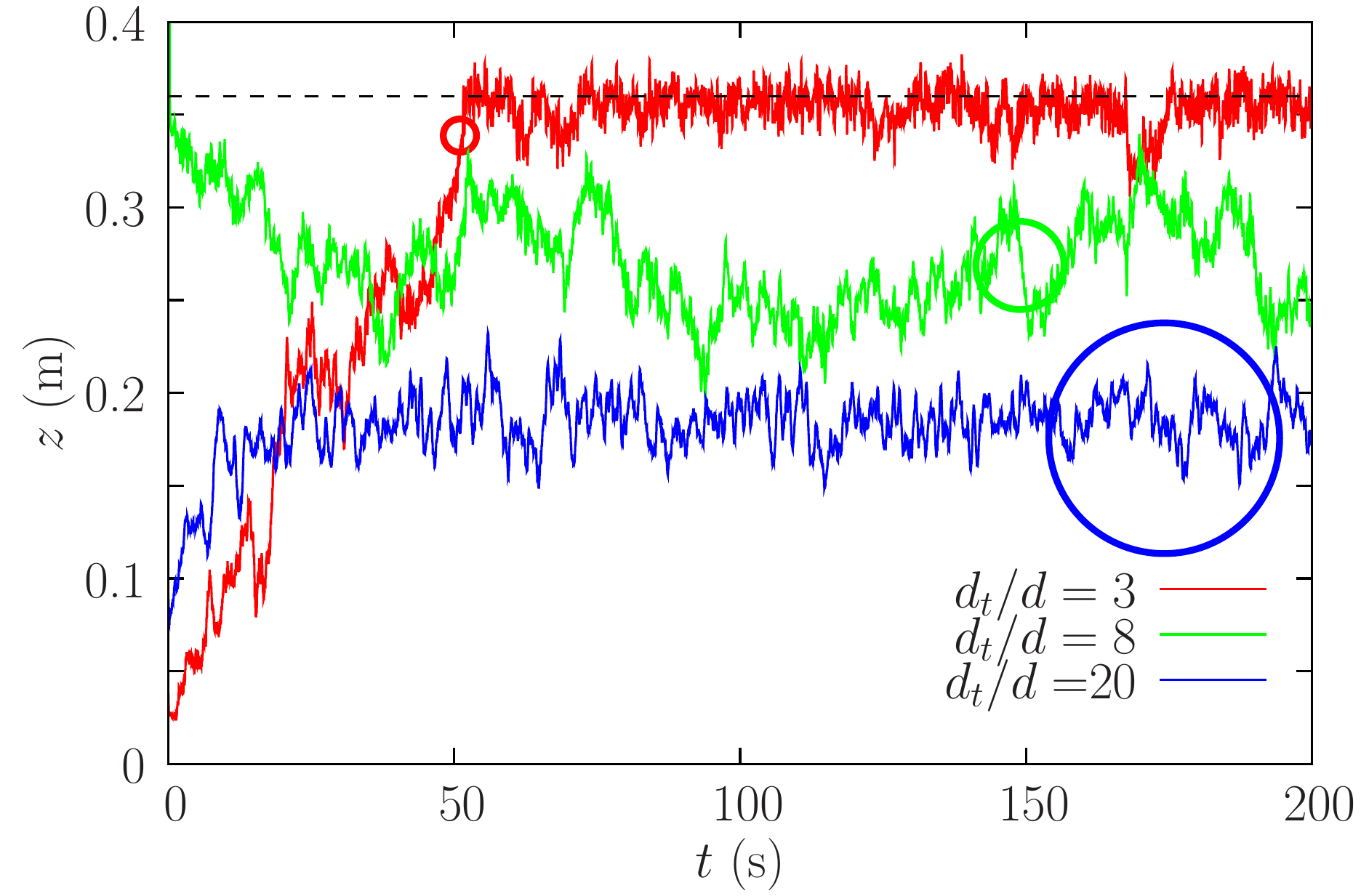}   
\caption{{\GREEN Trajectories} of the {\RED center} of 3 tracers versus time. The horizontal dashed line {\GREEN is} the free 2D ``surface". Thick circles show the {\RED sizes} of the three tracers.}
\label{plan3_8_20}
\end{figure}
Figure~\ref{plan3_8_20} shows the depth $z(t)$ of {\RED each tracer center} for three different tracer sizes versus time~$t$ (each simulation {\RED involving one} single tracer). For each 
tracer size, several initial positions at the bottom or {\RED at the} surface are tested, but only one is plotted {\RED here}. 
The steady state {\GREEN tracer depth} $h$ does not depend on the initial location {\GREEN ($h$ is the mean of $z(t)$, the initial convergence time {\RED being removed})}.
For the size ratio $d_t/d= 8$, the tracer 
almost never reaches the free ``surface" and stays at an intermediate depth. It is the  {\RED noisiest} trajectory. At intermediate depths, the trajectory is not 
stabilized by the existence of the free ``surface" or the bottom nearby.  For
 the size ratio $d_t/d=20$, the tracer reaches an equilibrium depth located
near the {\RED center} of the flow {\RED with} a layer of around 22 small particles below it. 
The $d_t/d= 3$ tracer, initially placed at the bottom,
reaches the surface, {\GREEN as in} a surface segregation {\GREEN phenomenon}. {\BLUE Stationary positions are reached after horizontal displacements of 10000$d$, 25000$d$ and 30000$d$ for tracers of size ratio 20, 8, and 3 respectively. The distance is mainly due to the gap between the initial vertical position of each tracer and its corresponding stationary position. Large values are explained by the use of a thick flow (60$d$), and by the poor efficiency of the 2D segregation.}

\begin{figure}[htbp]
\center
\includegraphics[width=0.95\linewidth]{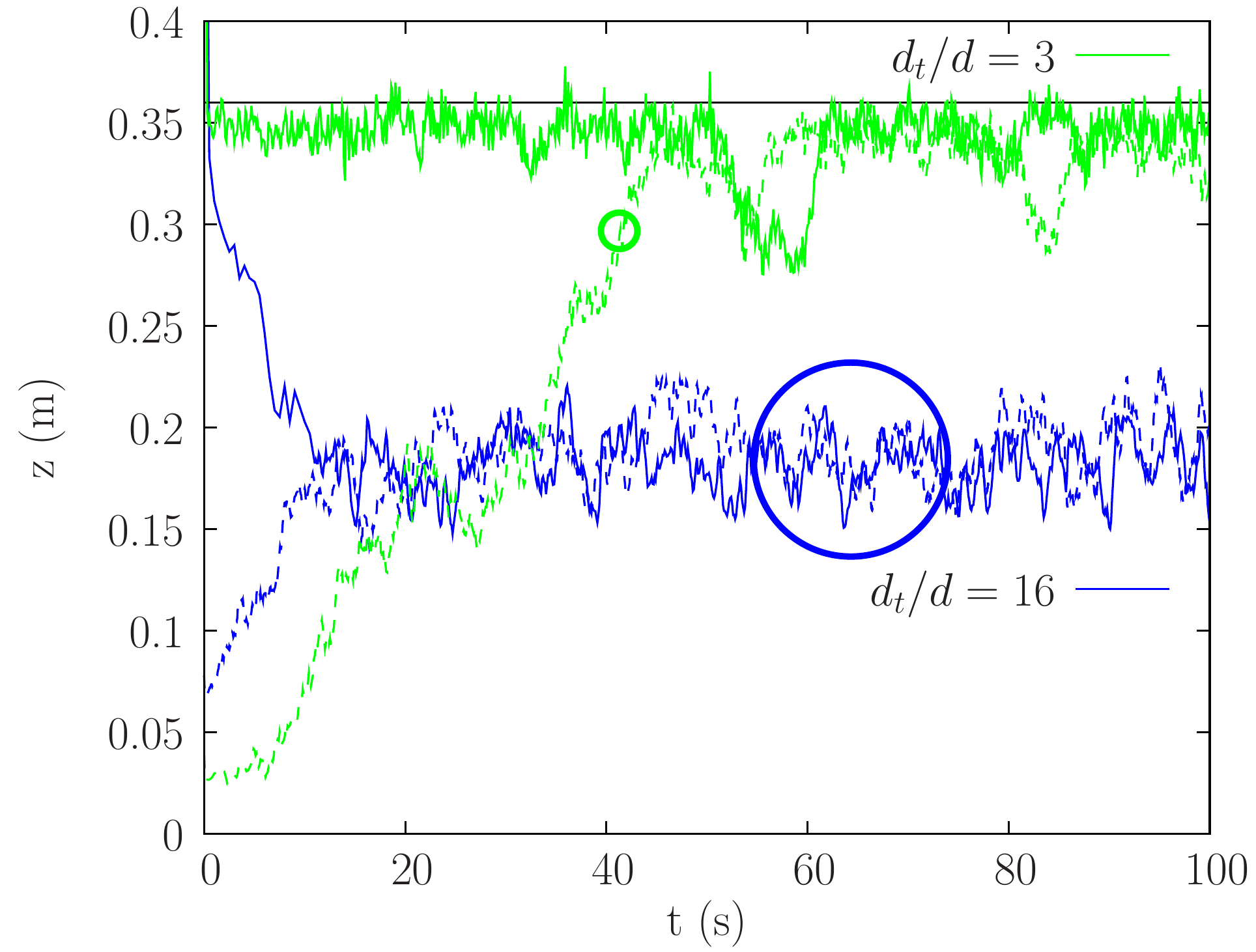}
\caption{{\GREEN Trajectories} of a tracer released at the top (solid lines)
and the bottom (dashed lines) of the flow. Both trajectories converge to the same 
equilibrium {\GREEN depth} {\GREEN(intermediate for} $d_t/d=16$, {\GREEN at surface for} $d_t/d=3$). Circles show the tracer sizes. The horizontal black line {\GREEN is} the mean position of the free {\GREEN 2D} ``surface".}
\label{plantopbottom}
\end{figure}

For a size ratio of 16, trajectories starting from the top and from the bottom reach the same equilibrium depth in about 12-15~seconds (around 4000-5000$d$) (Fig.~\ref{plantopbottom}). {\BLUE The initial gap to stationary position is the main parameter which determines the time or distance to travel along.} It takes a longer time {\BLUE (and distance)}, around 55~seconds, ({\BLUE 22000$d$}) to the tracer with size ratio 3 starting from the bottom to reach the surface: it has to move across the whole flow thickness, on a trajectory showing larger fluctuations.

\begin{figure}[htbp]
\center
\includegraphics[width=0.95\linewidth]{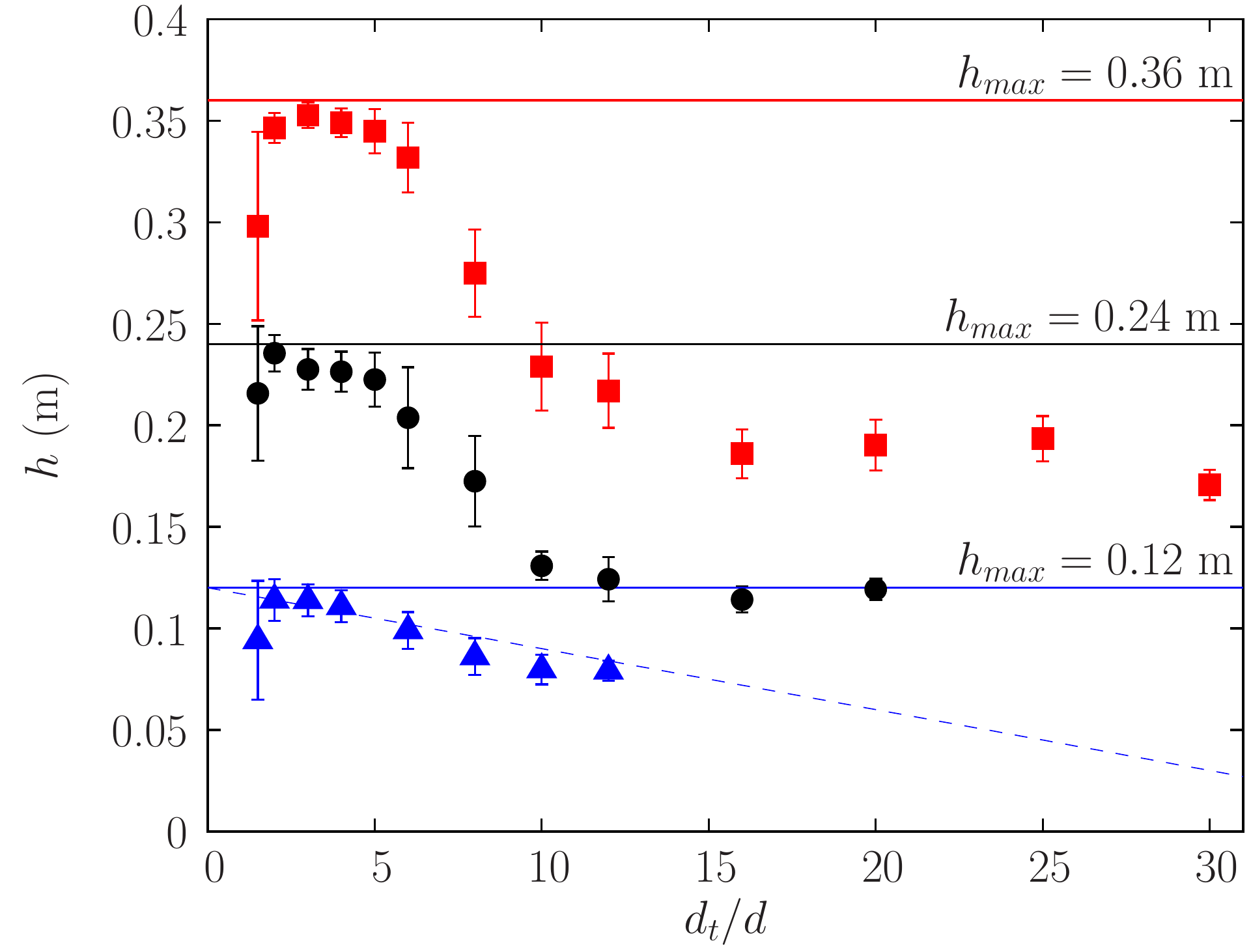}
\caption{{\GREEN Equilibrium depths of} the tracer {\RED center} in a flow down a 2D incline
versus size ratio, for 3 flow thicknesses: (blue $\blacktriangle$) 0.12~m, (black $\bullet$) 0.24~m, (red $\blacksquare$) 0.36~m. Error bars show the standard deviation. Horizontal lines show the free ``surfaces". {\RED The} oblique dashed line corresponds to the position of a tracer whose {\RED top} {\GREEN is} at the surface of the thinnest flow.}
\label{planrough} 
\end{figure}
\begin{figure}[htbp]
\center
\includegraphics[width=0.95\linewidth]{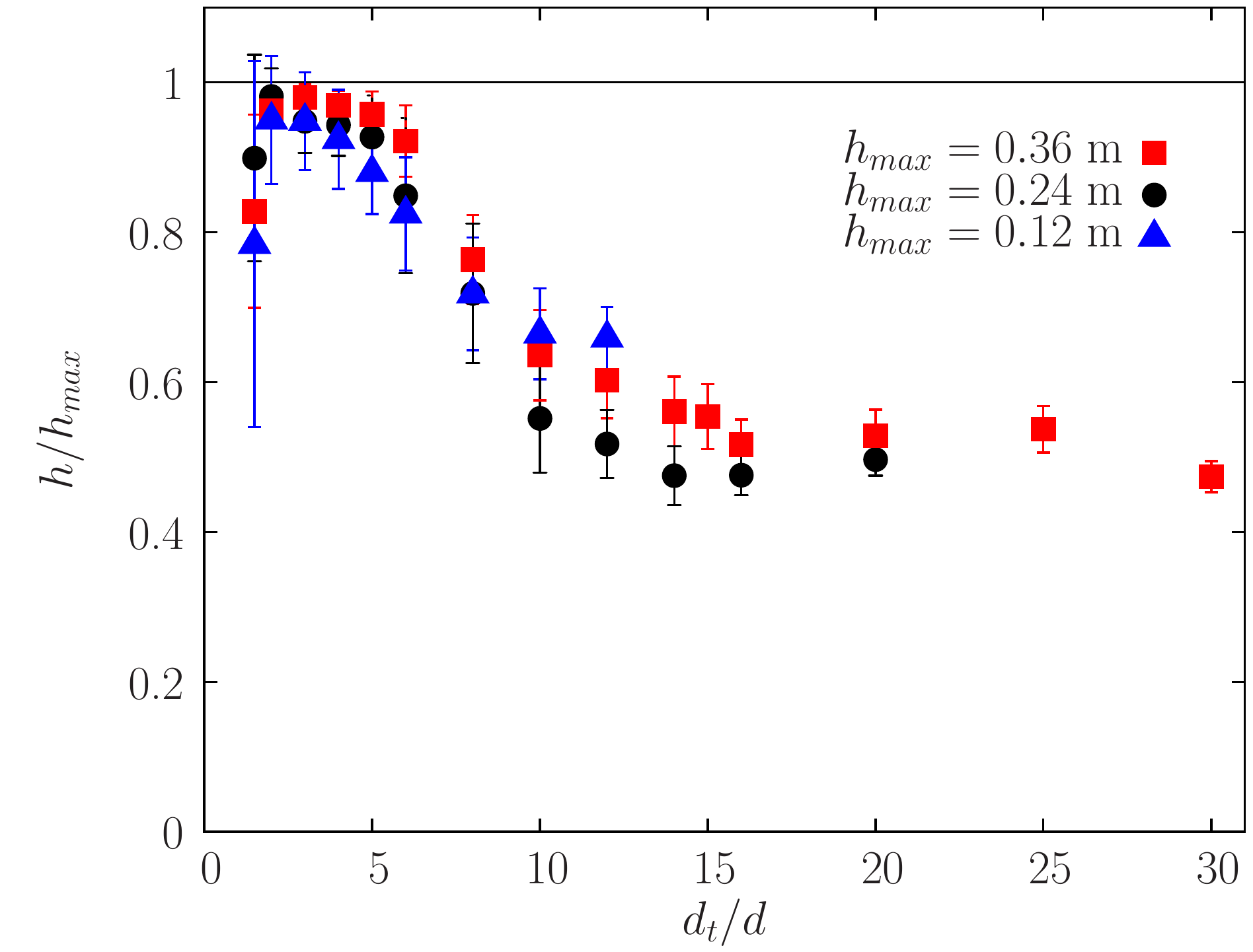}   
\caption{Relative {\GREEN equilibrium depths} of the tracer {\RED center} in the 2D flow
versus size ratio, for three flow thicknesses: (blue $\blacktriangle$) 0.12~m, (black $\bullet$) 0.24~m, (red $\blacksquare$) 0.36~m.}
\label{planrenor}  
\end{figure}

In order to study where the {\GREEN tracer stabilizes}, the mean {\GREEN depth} of one tracer in 
the stationary regime $h$ is reported for several 
size ratios $d_t/d$ ({\GREEN with} $d$ = 6~mm) and for
several thicknesses of the flow $h_{max}$ (Fig.~\ref{planrough}). {\GREEN $h$ is calculated from the flow bottom to the tracer {\RED center}.} Moderately 
large tracers ($2\leqslant d_t/d\leqslant 6$) are found at {\GREEN or near} the surface, {\GREEN and $h$ is} maximum for a size ratio between 2 and 3.  {\GREEN For these low size ratios, {\RED the} values of $h$} seem related to the distance to the free ``surface", independently of the thickness of the flow, {\GREEN showing the same curve shape relatively to the flow surface}. For larger 
size ratios ($d_t/d \geqslant 7$), the tracer position {\GREEN gets deeper and} with increasing size ratios. {\GREEN It is compatible with the $R_t/R$ vs $d_t/d$ decrease in the tumbler.} The $h$ asymptotic value for very large size ratios is close to $h_{max}/2$, and thus scales with the thickness of the flow (Fig.~\ref{planrenor}). 

The interesting result is that {\GREEN there are some} tracers which stabilize at intermediate
depths inside the flow. {\GREEN This} shows the {\RED occurrence} of intermediate segregation in a 2D flow on a rough incline, {\GREEN at least for non-interacting tracers. We can assume that small fractions of large disks would undergo intermediate segregation for these size ratios.} The small standard deviations represented as error bars {\GREEN indicate} that each tracer does not explore the whole thickness of the flow, but remains
at {\GREEN an} intermediate well-defined depth, with little randomness in its trajectory. {\GREEN These small fluctuations would correspond to a small standard deviation in the segregation of several non-interacting tracers.}

Note that for thin flows, tracers with size ratios above 10 are very large compared to the flow 
thickness and they are close to {\RED appearing} at the ``surface", although 
they interact with the bottom at the same time. The oblique dashed line shows the position of a tracer {\RED such that} its top is flush with the free ``surface" of the flow (Fig.~\ref{planrough}). It defines a boundary between surface and intermediate {\GREEN positions} (for small size ratios, {\GREEN here around 5}), and also shows a reasonable size {\GREEN limit} for a tracer in {\GREEN such} thin flow (here, $d_t/d=12$).

If the vertical position is renormalized by the thickness of the flow 
(Fig.~\ref{planrenor}), the three previous curves collapse reasonably well. 
{\GREEN Note} also that {\GREEN for} $1.5\leqslant d_t/d \leqslant6$,
rescaling like $h_{max}-h$ is a better choice as curves match well in their upper part, but {\RED they} will no longer collapse for large size ratios. {\GREEN {\RED Thus} the behavior} in a 2D flow shows two regimes: a first one ({\GREEN tracer near or at the} surface) where the equilibrium position depends on the distance to the ``surface", independently of $h_{max}$ value, and a second one where the {\GREEN equilibrium depth} is intermediate and {\GREEN tends towards $h_{max}/2$, and thus} scales with the flow thickness.

The positions $h$ show that only surface and intermediate {\GREEN depths} are obtained in 2D granular flows on an incline. {\GREEN We conclude that reverse segregation is not obtained in 2D, at least in this parameter range}. The large tracer does not reach positions below {\GREEN mid-height of the} flow, even for very large size ratios. For the thinnest flow, the depths are compatible {\GREEN both} with a reverse and an intermediate {\GREEN pattern}, considering the small number of small particles below the tracer, but for thicker flows both types of {\GREEN depths} {\RED can be differentiated}.  Equilibrium positions end up really at mid-flow for the {\GREEN largest} ratios. {\RED There are} 14 small particles below {\RED the largest tracer} in the thickest flow, {\GREEN significantly above a reverse position.}

Since in tumblers {\RED the dependency of} the position $R_t$ {\RED on} $d_t/d$ {\RED is} stronger in 3D {\GREEN than} in 2D, we expect different results for a 3D incline flow. Another point worth noting is that the {\GREEN {\RED dependency} of the} position ($h$ or $R_t/R$) {\RED on} $d_t/d$ is {\GREEN also greater} for a 2D incline than for a 2D tumbler and consequently, the asymptotic value is approached for smaller size ratios on an incline than in a tumbler flow.

\subsubsection{Slope angle}

\begin{figure}[htbp]
\center
\includegraphics[width=0.95\linewidth]{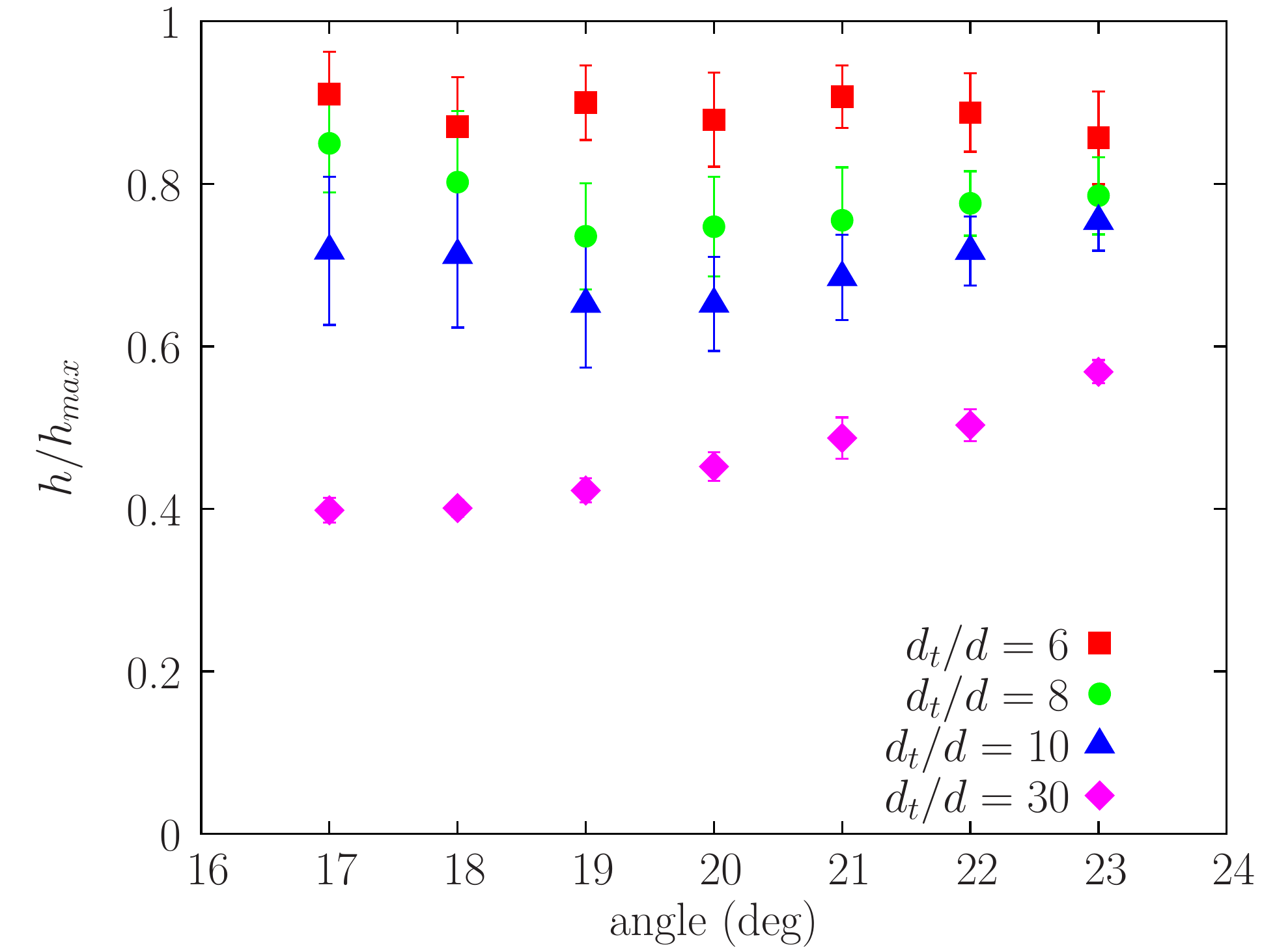}
\caption{Relative {\GREEN equilibrium depths} of the tracers in a 2D flow
versus slope angle ($\theta = 17$ to 23$^\circ$) for 4 size ratios: 
$d_t/d = 6$, 8, 10 and 30. The {\GREEN flow} thickness is $h_{max}=$ 36~cm.}
\label{planangle}
\end{figure}
In a granular flow down an incline, the easiest way to increase the shear
rate, without changing the thickness of the flow, is to increase the slope.  
Fig.~\ref{planangle} shows the relative position of four tracers, with size
ratios $d_t/d=6$, 8, 10 and 30, for several angles of the plane. Even if small evolutions are measurable, the relative vertical
 position of tracers ($d_t/d\leqslant10$) is almost unchanged for a slope change 
from 17 to 23$^\circ$ although this change induces an increase {\RED in} the mean velocity of the flow,
 and thus {\RED in} the shear rate, by a factor of 4. In the case of a $d_t/d=30$ tracer, {\GREEN a slight monotonic increase {\RED in} the tracer depth} with the slope is observed. {\GREEN For size ratio 10, the same increase is obtained but only for slopes larger than 20$^\circ$.}

\subsection{3D flows on an incline}   
A series of simulations {\RED is} conducted on the 3D {\GREEN incline}, first in a thin flow, then in thicker flows.
Even {\RED though} very large size ratios are not reachable with our computational facilities, this captures most of the phenomena and allows a comparison with the 2D case and with {\GREEN previous} experiments in a 3D channel. 

\begin{figure}[htbp]
\center
\includegraphics[width=0.95\linewidth]{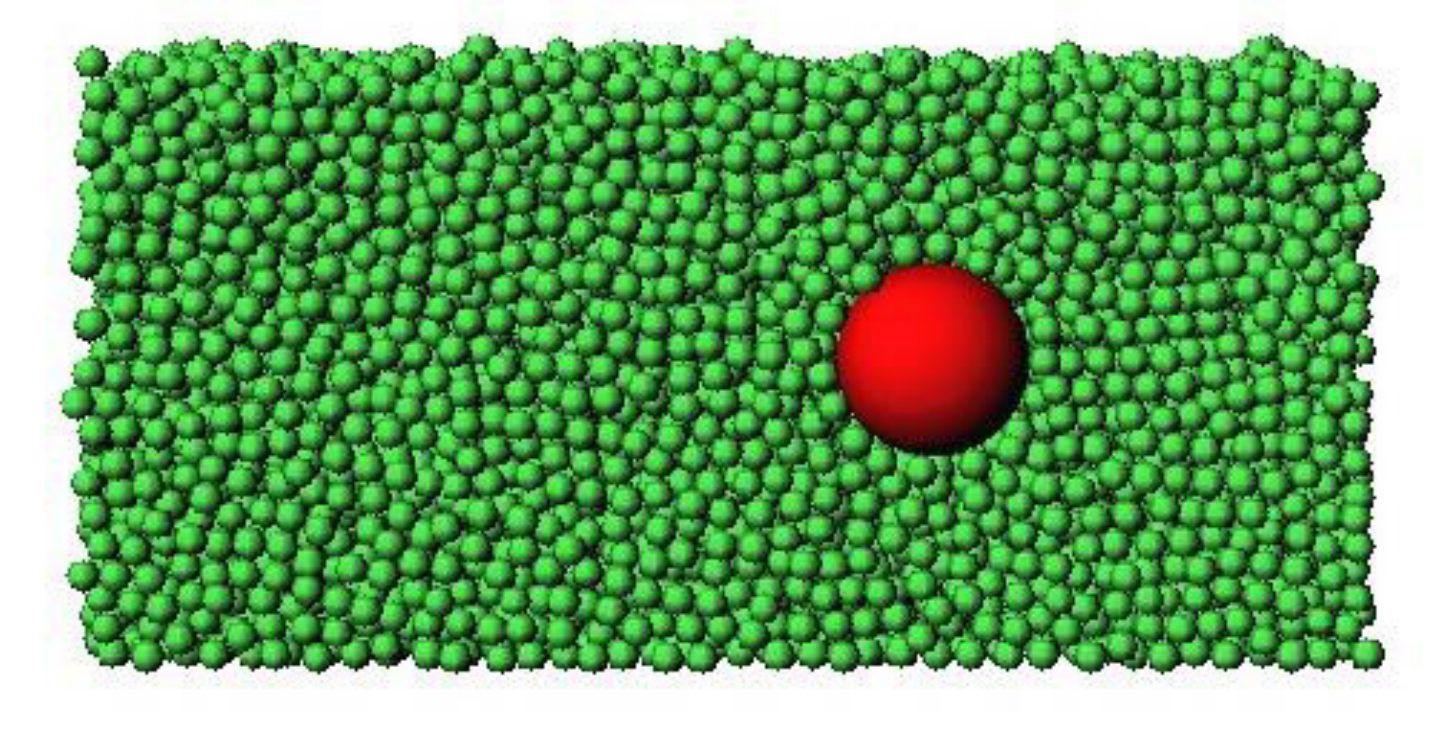}
\caption{A 3D {\GREEN incline} granular flow, with a tracer ($d_t/d=6$), {\GREEN moving} from left to right (slope is 23$^\circ$, $h_{max}=0.112$~m). Side beads have been removed to show the tracer.}
\label{plan68k}
\end{figure}

\subsubsection{Equilibrium positions}

\begin{figure}[htbp]
\center
\includegraphics[width=0.95\linewidth]{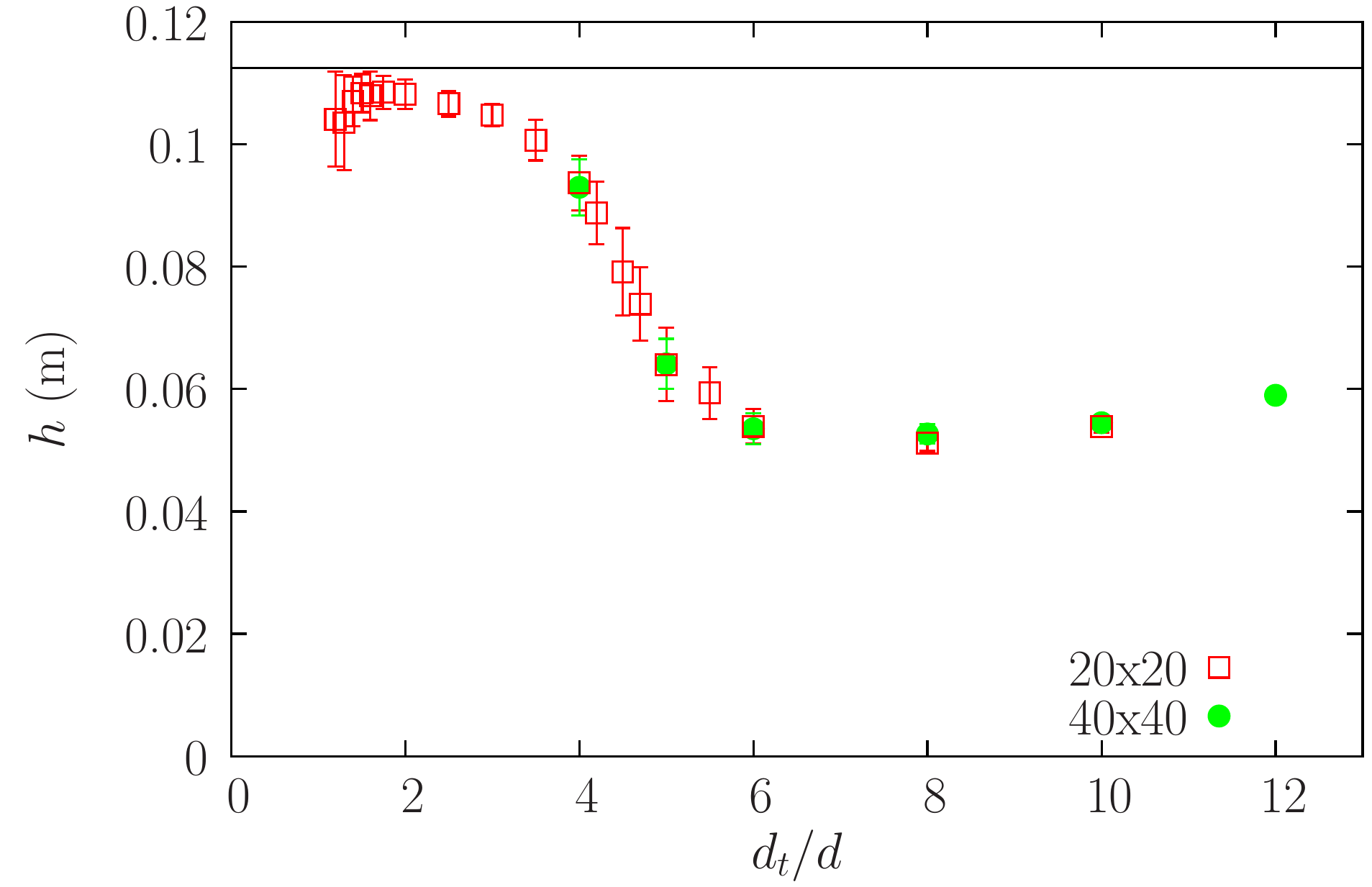}
\caption{{\GREEN Equilibrium depths} of the tracer {\RED center} in the 3D flow down an incline versus size ratio $d_t/d$.  Error bars {\GREEN show} the standard deviation. 
The horizontal line is the free surface ($h_{max}=0.112$~m, the slope is 23$^\circ$). Two {\GREEN numerical} domain sizes {\GREEN are} used: 20$d$ $\times$ 20$d$ (red~$\square$) and 40$d$ $\times$ 40$d$ 
(green~$\bullet$).}
\label{nath3d} 
\end{figure}

Figure~\ref{plan68k} shows a 3D flow, with a {\GREEN tracer having a} size ratio $d_t/d=6$ {\GREEN (small beads are $d=6$~mm)}. 
The horizontal dimensions of the simulation domain are 20$d$ $\times$ 20$d$ (0.12~m $\times$ 0.12~m) or (40$d$ $\times$ 40$d$) for the largest size ratios. Both domain sizes {\RED are} used for several
size ratios to be sure that the simulated domain is large enough (Fig. \ref{nath3d}). 
The {\GREEN flow} thickness ($h_{max}=0.112$~m $\simeq$ 18$d$) corresponds to a relatively thin flow, comparable to the flows encountered in our 3D tumbler for size ratios around 8. The tilt angle is 23$^\circ$.

For each size ratio (from 1.2 to 12), the large tracer depth $z(t)$ evolves rapidly during flow (see below Figs. \ref{nathh2} and \ref{nath12}) to stabilize finally at a constant depth $h$. Some tracers have been 
initially placed at the bottom of the flow, and some at the surface without any final difference. Once in steady state, trajectory fluctuations are small and give small standard deviation associated {\RED with} each $h$. 

The equilibrium depth $h$ depends on the size ratio between tracer and small beads. Fig. \ref{nath3d} plots the {\GREEN tracer depths} (from 7.2 to 72 ~mm in size) for size ratios ranging from $d_t/d= 1.2$ to 12. 
{\GREEN For moderately large size ratios (below 4), $h$ is near or at the surface, in accordance with the surface segregation of large beads.} As in the 3D rotating tumbler, the maximum of the curve
(i.e., tracer at the free surface) is obtained for size ratios between 1.5 and 1.8 (Fig.~\ref{nath3d}).
For size ratios approximately between 4 and 6, the tracer reaches an equilibrium 
depth inside the flow, suggesting the {\RED occurrence of} intermediate segregation in 3D flow down an incline, {\GREEN  for non-interacting tracers}.
For larger size ratios, $d_t/d>6$, the equilibrium depths reach
a saturation value near the bottom, {\GREEN  in a reverse position}. We note that the equilibrium positions are independent 
of the size of the simulation domain. The slight increase of the curve for the largest size ratios (10 and 12) is due to the increase {\RED in} the tracer size itself, {\GREEN showing that the tracer is in strong interaction with the bottom}. {\RED There are} only about 4 small beads below {\RED the tracer}. {\GREEN 
The three types of equilibrium positions (surface-intermediate-reverse) are {\RED thus} found in this thin 3D flow, suggesting that the three segregation patterns would exist for a small fraction of non-interacting tracers.}

Comparing 2D and 3D cases (Figs. \ref{planrenor} and \ref{nath3d} respectively), 
the overall behavior is the same but some differences are present. In the 
3D case, the equilibrium depth decreases more rapidly and reaches a smaller
saturation value {\RED earlier}, at a size ratio
close to $d_t/d= 6$ in 3D, instead of $d_t/d=$ 10 or 15 in 2D. 
 We also note that the standard deviations are much smaller in 3D.

\subsubsection{Thickness of the flow}

\begin{figure}[htbp]
\center
\includegraphics[width=0.95\linewidth]{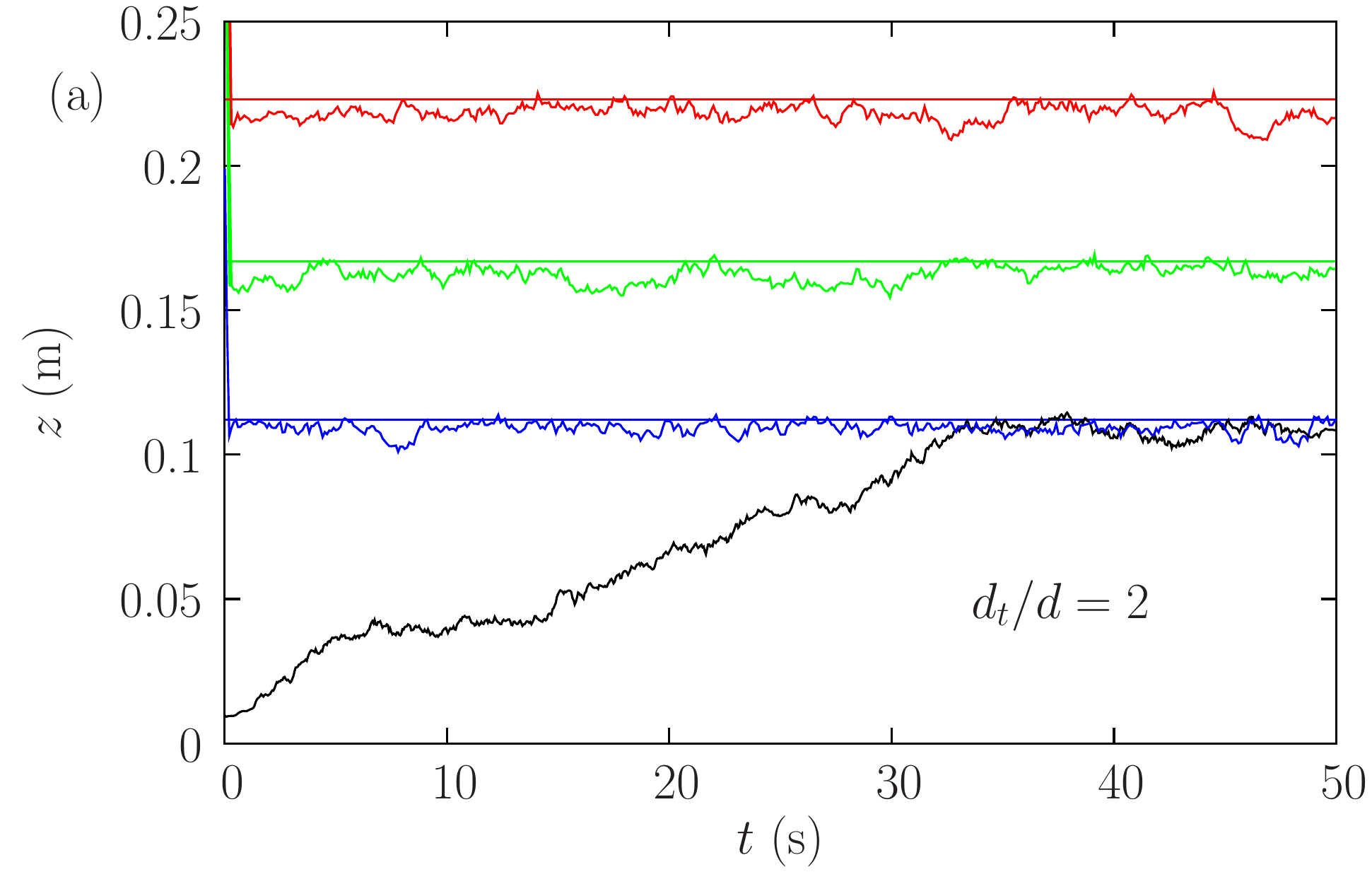}  
\includegraphics[width=0.95\linewidth]{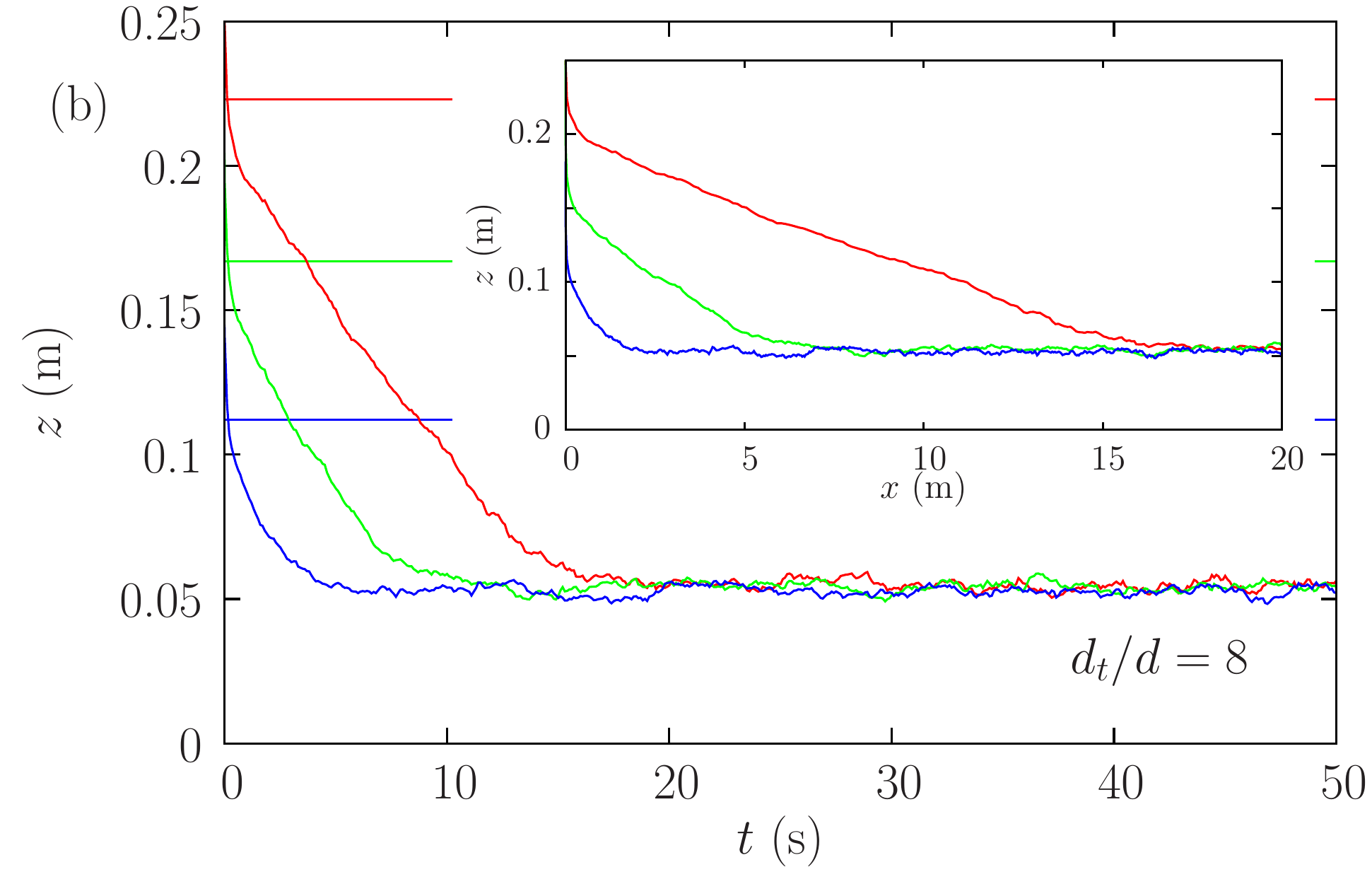}  
\caption{Trajectories of tracers in the 3D incline flow (slope is 23$^\circ$) for 3 flow thicknesses $h_{max} = 0.112$~m, 0.167~m and 0.223~m and for 2 size ratios: (a) $d_t/d=2$ and (b) $d_t/d=8$, {\BLUE insert: trajectories in $x$ coordinate}.}
\label{nathh2}
\end{figure}

\begin{figure}[htbp]
\center
\includegraphics[width=0.95\linewidth]{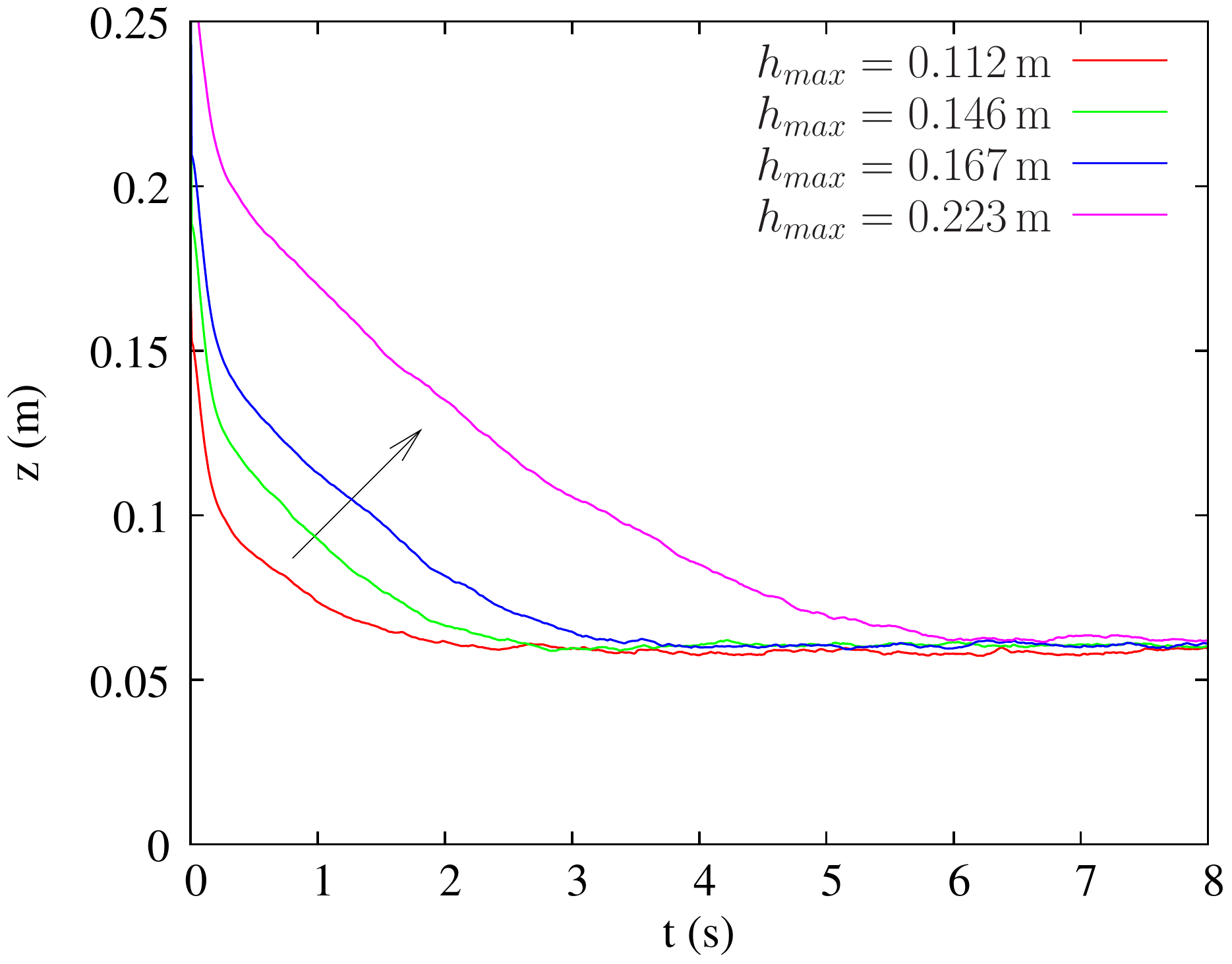}   
\caption{Trajectory of a tracer center ($d_t/d=12$), initially placed at the surface, in a 3D flow, for 4 flow thicknesses (slope is 23$^\circ$). The arrow indicates increasing thicknesses.}
\label{nath12}
\end{figure}

Figure \ref{nathh2} shows the trajectories for the first 50 seconds of 
two tracers, $d_t/d= 2$ and 8, immersed in granular flows having three different
thicknesses. The horizontal lines show {\RED the} positions of the free {\RED surfaces}
of the flows: $h_{max}= 0.112$~m, 0.167~m and 0.223~m. For the three thicknesses,
the tracer with a size ratio of 2 remains {\RED at} or goes {\RED to} the surface
of the flow showing {\GREEN the same final position {\RED as} in} a surface segregation process (only the case of the thinnest flow is shown for the ratio 2 tracer placed at the bottom). When crossing the whole thickness, the convergence is longer for this small size ratio ($d_t/d=2$) than for a larger one ($d_t/d=8$), and the trajectory presents more fluctuations.  
The large tracer ($d_t/d= 8$) 
sinks {\GREEN to reach a depth} near the bottom of the flow. {\GREEN Its} stationary depth is close
to 0.05~m, independently of the thickness of the flow. As
the tracer radius is $r_t= 0.024$~m, it does not touch the 
rough inclined plane made {\RED up} of small glued beads, but about 4 small beads remain
between the tracer and the plane. We consider this position close enough to the bottom to be called ``reverse".

When using a $t-z$ representation, parallel trajectories on Fig.~\ref{nathh2}(b) show that the sinking velocity is constant ($v_{sink}=-0.0105$m/s). For a given size ratio, the time of convergence is mainly related to the thickness of material to travel through. {\BLUE A constant sinking velocity
is an interesting feature since it can be used in theoretical models to describe granular segregation. From an experimental point of view, an $x-z$ representation (Fig.~\ref{nathh2}(b) insert) is more interesting since it gives the incline length required for an experiment. 
For a size ratio of $d_t/d=8$, changing the thickness of the flow from $h_{max}=19d$ 
to 37$d$, decreases the slope of the trajectories (compared to the rough incline) and increases the horizontal settling distance from $\simeq 400d$ to $\simeq 2500d$.
This distance increase comes from the flow thickness increase but also from 
the induced increase of the mean velocity which is about a factor 3 here.} In the case of a larger tracer
$d_t/d=12$ (Fig.~\ref{nath12}), the sinking is more rapid ($v_{sink}=-0.021$m/s), {\BLUE and the slope of the trajectories also decreases with the increase in thickness (not represented). Consequently, the settling distance
increases from $\simeq 200d$ to $\simeq 1200d$, for $h_{max}=0.112$ and 0.233~m respectively.} For a downward motion, convergence is more rapid for high size ratios (comparing Figs.~\ref{nathh2}(b), \ref{nath12} and \ref{nath10a}). Downward forces acting on tracers are stronger when tracers are larger, and consequently heavier.

\begin{figure}[htbp]
\center
\vspace{3mm}
\includegraphics[width=0.95\linewidth]{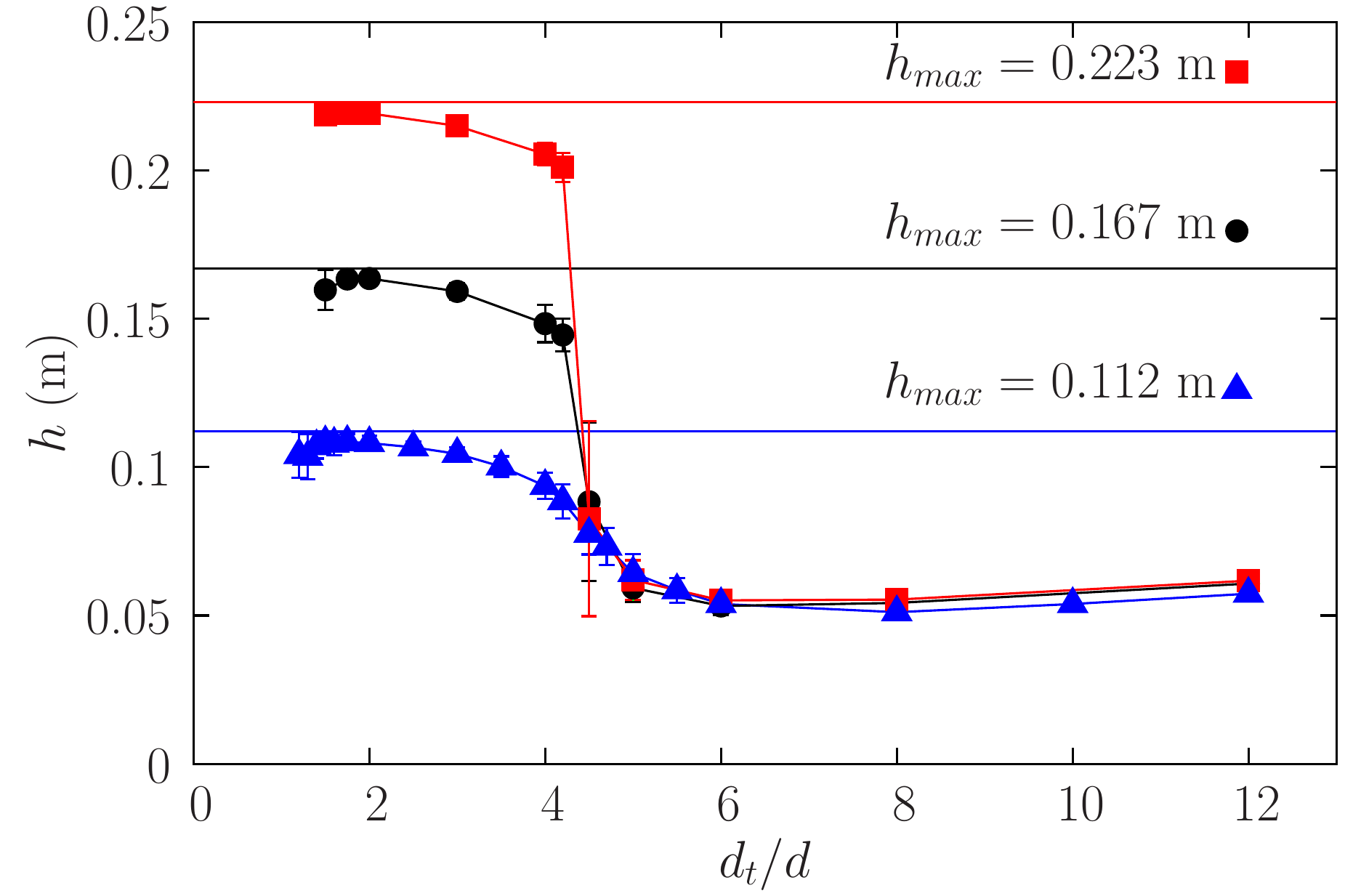}   
\caption{Equilibrium {\GREEN depths} of tracers versus size ratio {\GREEN in the 3D flow} for 3 thicknesses: (blue~$\blacktriangle$) 0.112~m, (black~$\bullet$) 0.167~m, (red~$\blacksquare$) 0.223~m.  Error bars show the standard deviation.}
\label{nathh1-2} 
\end{figure}

Figure~\ref{nathh1-2} shows the equilibrium {\GREEN depth} $h$ {\RED for} a large tracer 
for three different thicknesses {\GREEN $h_{max}$.} For size ratios up to 4, the
large tracer remains at or near the surface, independently of $h_{max}$.
 For size ratios larger than
5, the tracer sinks close to the bottom of the flow, and $h$ is independent
of $h_{max}$ (as in Figs. \ref{nathh2}(b) and \ref{nath12}). {\GREEN The slight increase with the tracer size shows the strong interaction with the bottom when in reverse position.} For the two thickest flows, 
a sharp transition between the {\GREEN surface position range}
and the {\GREEN reverse position range} appears 
for size ratios $d_t/d$ between 4.2 and 4.5, while a {\RED relatively} progressive variation is observed for the thinnest flow. The tracer {\GREEN depth} $h$ depends on the flow thickness only during the transition. Both {\GREEN parts of the curves, $h-h_{max}$} for small {\GREEN (below 4)} or {\GREEN $h$} for large size {\GREEN (above 5.5)} ratios are independent of $h_{max}$.

\begin{figure}[htbp]
\center
\vspace{3mm}
\includegraphics[width=0.98\linewidth]{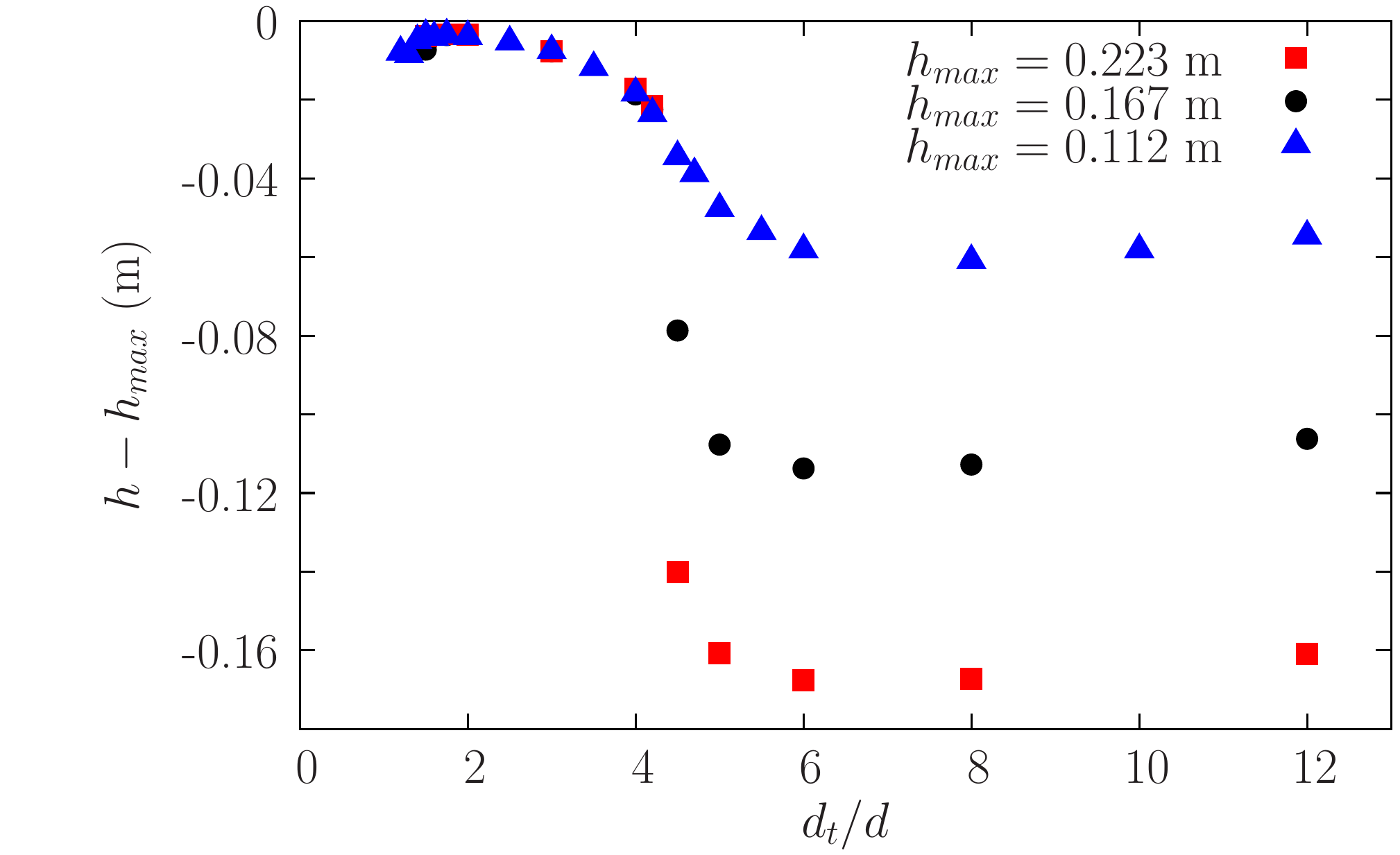}
\caption{Equilibrium {\GREEN distances from} the tracers to the surface {\GREEN $h-h_{max}$, in the 3D flow on an incline} for 3 thicknesses: (blue~$\blacktriangle$) 0.112~m, (black~$\bullet$) 0.167~m, (red~$\blacksquare$) 0.223~m.}
\label{nathh1-2hd}  
\end{figure}

Plotting $h$ (Fig.~\ref{nathh1-2}) shows that 
the distance to the bottom controls the equilibrium position for very large
tracers, independently of the thickness of the flow. In the same way, 
plotting $h-h_{max}$ (Fig.~\ref{nathh1-2hd}) shows that the distance to 
the surface is independent of the flow thickness for 
moderately large tracers ($1.5 \leqslant d_t/d \leqslant 4.2$), when {\GREEN positions near surface are reached}. It seems that two independent phenomena, one influenced by the presence of the surface and one by the bottom, determine the equilibrium {\GREEN of the tracer in each case}. {\GREEN For} a thin flow, {\RED the free surface and the bottom are close enough so that} the two phenomena interact, and the result is a progressive transition between the two influences, creating a larger range of intermediate segregation positions. In the case of a thick flow, both influences are {\GREEN almost} separated and could be studied independently.

\begin{figure}[htbp]
\center
\vspace{-1.5cm}
\hspace{-1.3cm}
\includegraphics[width=0.99\linewidth]{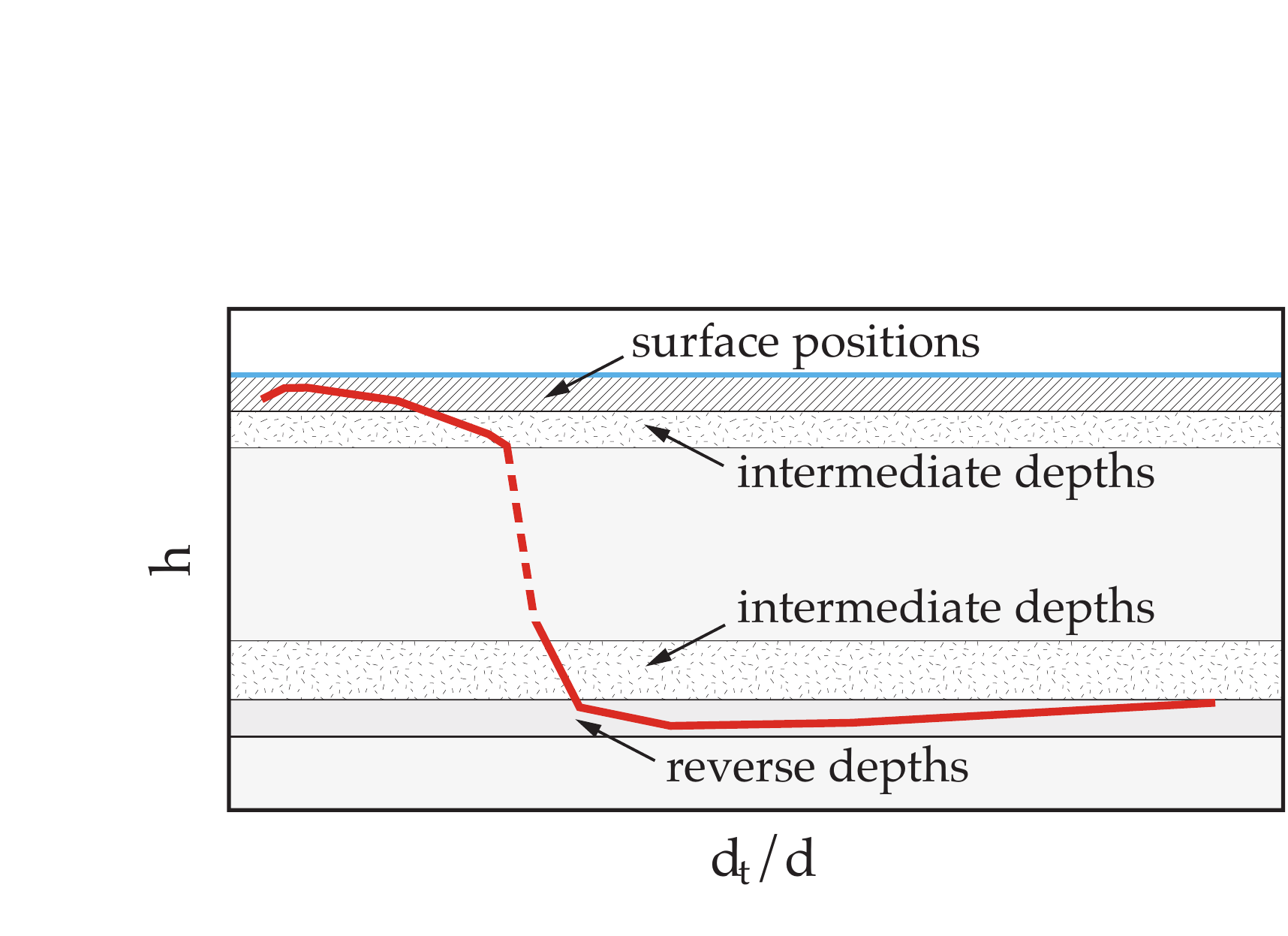} 
\caption{The {\GREEN upper part of the $h$ curve defines} two layers, {\GREEN with} surface and intermediate {\GREEN equilibrium depths}, and the {\GREEN lower part}, two layers, {\GREEN with} intermediate and reverse {\GREEN equilibrium depths} (red $h$ curve from Fig. \ref{nathh1-2}). In a thick flow {\GREEN(drawn here)}, there {\RED are} no {\GREEN equilibrium depths} in a layer {\GREEN around mid-height}. The bottom is never reached, {\RED partly} due to the tracer size {\GREEN($h\geqslant d_t/2$)} and {\RED partly} due {\RED to the presence of some} small beads {\GREEN(around 4) under} the tracer.}
\label{schema}  
\end{figure}

It could be tempting to associate the three zones {\GREEN coming from these curves (Figs.~\ref{nathh1-2} and \ref{nathh1-2hd})} {\RED with} the {\GREEN three types of equilibrium positions:} surface, intermediate and reverse ({\GREEN or the three} segregation {\GREEN types)}. But {\GREEN they do not exactly match}. For example, tracers just below the surface (as for a ratio 3.5) are not visible at the surface, whatever the flow thickness is: they should be considered in intermediate {\GREEN position}.  {\RED Symmetrically}, the tracer with a size ratio 5 is floating above the larger ones, and is in intermediate position. {\RED As for} the largest tracers ($d_t/d\geqslant8$), which show a slight increase {\RED in} their {\RED center} {\GREEN depth} due to the increase {\RED in their size}, {\RED they} are in strong interaction with the bottom plane and are {\RED thus} in reverse {\GREEN position}. The separation into three {\GREEN types} of {\GREEN equilibrium depths} (surface-intermediate-reverse) {\GREEN is convenient but} may not be representative of the phenomena happening in the granular matter. {\GREEN Only} two mechanisms {\GREEN may be the cause for equilibrium depths:} one {\GREEN due to the influence} of the surface and one {\GREEN due to the influence of} the bottom. {\GREEN Their potential combination appears or {\RED does not appear at} around mi-height {\RED of the flow}, depending on the flow thickness.} Nevertheless, {\RED in the present study} we keep the separation in the three types (surface-intermediate-reverse) {\RED that} correspond to particular positions of the tracers (and not to mechanisms). {\RED In this view,} we have to split the surface zone of the $h$ curve in two layers: one layer with surface positions {\GREEN (visible tracers)}, and one layer with intermediate depths. {\GREEN In the same} {\RED way,} we split the bottom zone of the $h$ curve in two layers: a second layer with intermediate depths and one layer with reverse depths (Fig \ref{schema}).
{\RED In this view}, thick flows have two intermediate depth layers which are separated by an empty central zone where {\GREEN there is no equilibrium depth for a tracer}. Thin flows have their two intermediate depth layers continuously connected, forming a ``thick" central layer {\GREEN of intermediate equilibrium depths}.

 \subsubsection{Comparison between the 2D and 3D flows on an incline}  

The main difference between 2D and 3D is the equilibrium depth of very large 
tracers (Figs. \ref{planrough} and \ref{nathh1-2}). The large 
tracers sink near the bottom, exhibiting a reverse {\GREEN position} in the 3D case while
they locate {\RED themselves} at intermediate {\GREEN depths} in 2D, near mid-height of the flow. For $h_{max}=0.112$ in 3D, the asymptotic equilibrium depth is also close to $h_{max}/2$ (Fig.~\ref{nath3d}), {\GREEN but this} is just a coincidence: {\GREEN other flows with different thicknesses show the same constant asymptotic value}. The fact that {\GREEN the stabilization of a large tracer} in 2D is $h_{max}$ dependent, while {\GREEN it} is independent of $h_{max}$ in 3D, shows {\GREEN that} 2D and 3D segregations of {\GREEN a few non-interacting} large tracers may be different processes.
Moreover, the transition between the {\GREEN surface and the deepest positions} is steeper in 3D than in 2D (see Figs.~\ref{planrenor} and \ref{nathh1-2}). The transition occurs between size ratios
$d_t/d = 4$ and 6 in 3D, while in 2D the whole transition occurs between 
$d_t/d = 5$ and 15. This stronger dependency in the 3D case is also observed in the tumbler. Moreover, the maximum does not occur for exactly the same size ratio in 2D and 3D. Nevertheless, similar behaviors are also noticed: for small size ratios in 2D and 3D the tracers positions are both related to the distance to the surface, independently of the $h_{max}$ value. As for the tumbler system, the 2D incline case should not be carelessly extrapolated in 3D: evolutions with the size ratio are different, {\GREEN even {\RED though} some strong similarities are observed}. {\GREEN These differences are probably linked} to the compacity difference between 2D and 3D.

\subsubsection{Comparison between tumbler and incline}  

\begin{figure}[htbp]
\center
\includegraphics[width=0.95\linewidth]{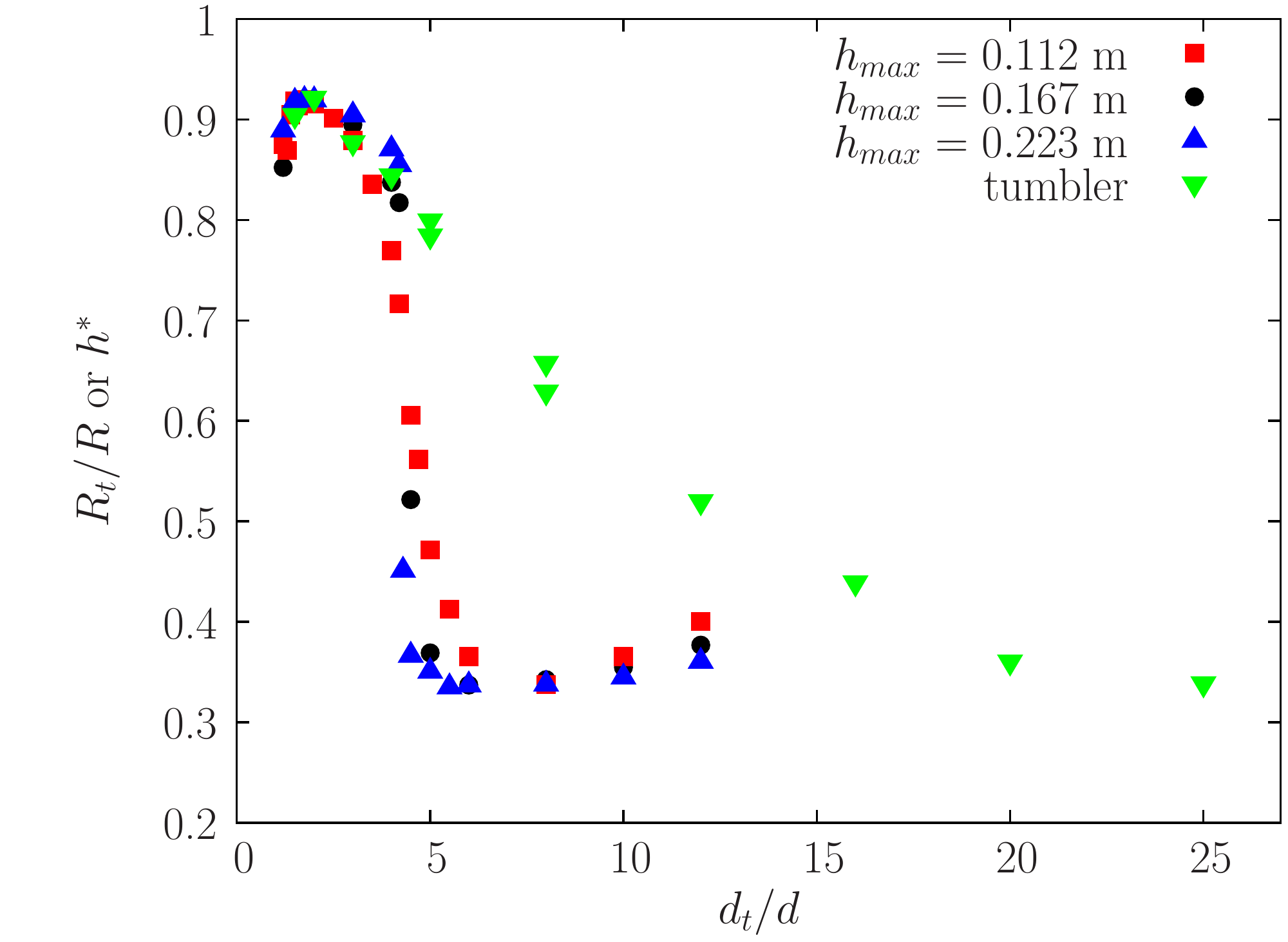}
\caption{Equilibrium positions of the tracer versus size ratio, in the 3D tumbler (green~$\blacktriangledown$) and equivalent rescaled {\GREEN(see text)} positions in the 3D incline flow, for 3 different thicknesses.}
\label{drumincline}
\end{figure}

On an incline, granular matter flows on a solid rough surface whereas in tumblers, it flows on loose curved granular material. {\GREEN The comparison} of the equilibrium position ($h$ or $R_t/R$) versus $d_t/d$ in both types of flows (incline or tumbler) gives information on {\GREEN the influence} of the structure of the flow.
Fig.~\ref{drumincline} shows normalized depth in the incline flow $h^*$ and radial positions $R_t/R$ in the tumbler at the same scale. We choose to adjust the minimal and maximal positions of $h$ to the asymptotic and maximum values of $R_t/R$, {\GREEN which corresponds to bottom and surface tracer position, respectively, within the tumbler flowing layer}. The curves match relatively well for $d_t/d\leqslant3.5$, indicating that for these small size ratios the process is mainly controlled by surface phenomena, which are quite insensitive to the substratum. For larger ratios $d_t/d\geqslant4$, curves shift with a stronger dependency in the case of rough inclines. The difference may come from the substratum. 
As {\GREEN the conclusions {\RED drawn} from} Figs. \ref{nathh1-2} and \ref{nathh1-2hd}, these data suggest that the equilibrium at a given depth comes from one phenomenon influenced by the surface, or/and one influenced by the bottom.

 Note that for 3D inclines, the tracer vertical position increases for
size ratio starting from 1, reaches a maximum for size ratios between 1.5 and
1.8, and decreases for larger values. For 3D tumblers, the maximum is obtained
for size ratios between 1.5 and 1.8 in {\RED experiments}, and between 1.5 and 2
in simulations. These non-monotonic variations are analogous to those observed
experimentally in an annular shear cell where the {\RED segregation} time and the
segregation rate both present an extremum for a size ratio of 1.6
\cite{GolickDaniels14}. This is also related to the variation of the segregation
P\'eclet number, defined as a segregation rate on a diffusive remixing, which
shows a slight maximum at 1.7 \cite{ThorntonWeinhart12}, {\GREEN  or the variation of the force acting on a tracer in 2D{\RED,} which shows a maximum at 2 \cite{Guillard16}.} 

\subsubsection{Slope angle}  

\begin{figure}[htbp]
\center
\hspace{-0.3cm}
\includegraphics[width=0.95\linewidth]{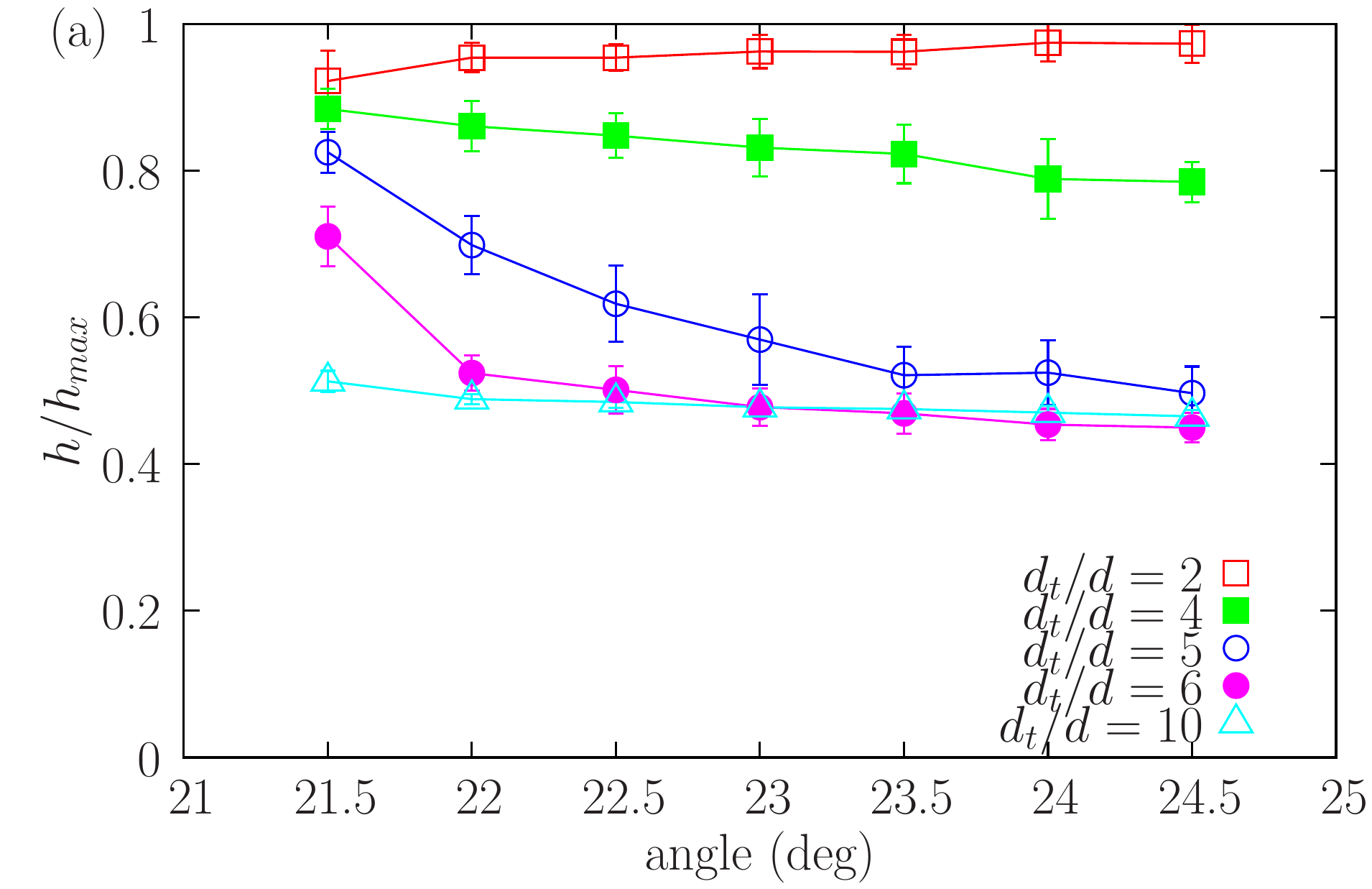}   
\includegraphics[width=0.95\linewidth]{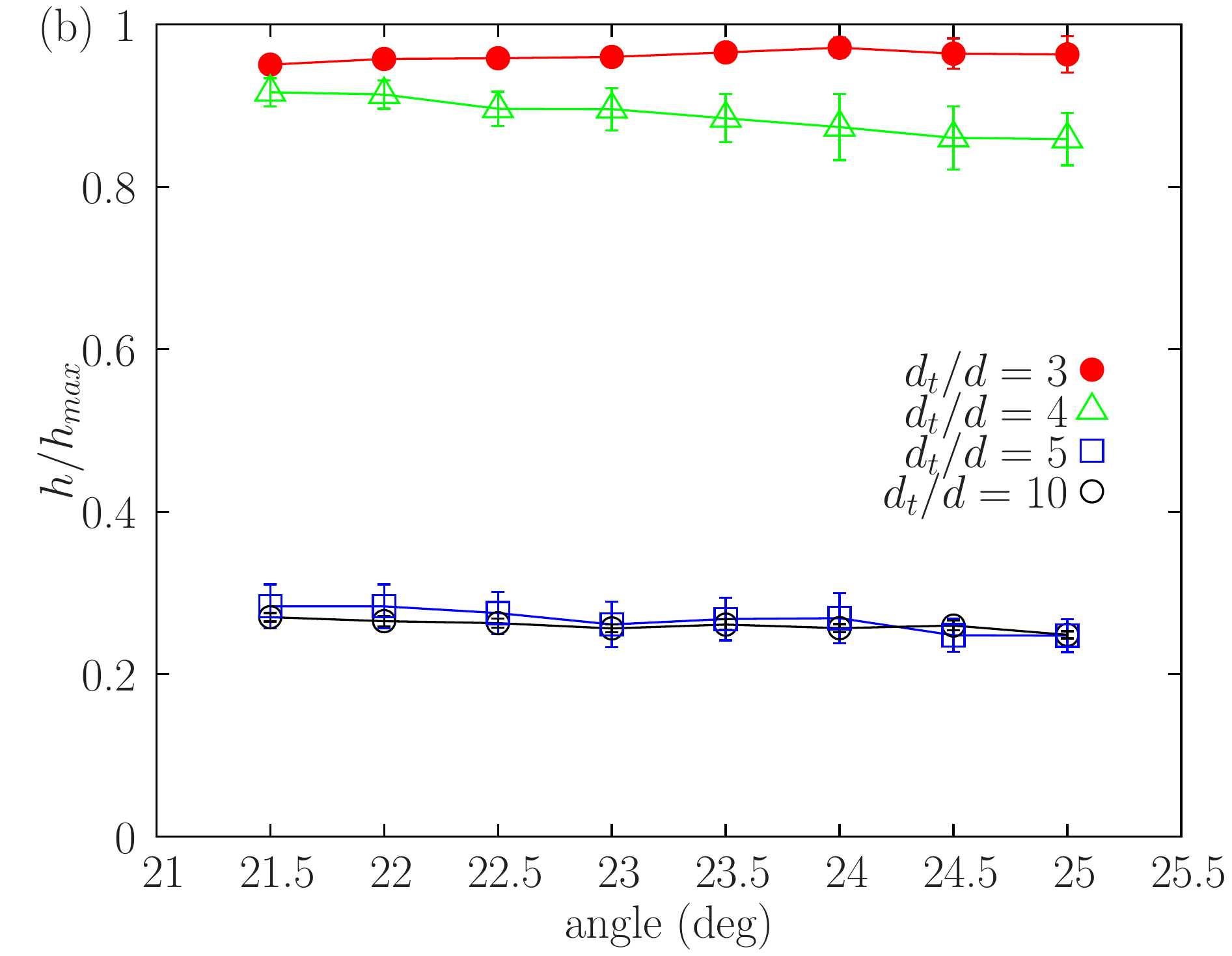}  
\caption{{\GREEN Equilibrium depths} of tracers {\GREEN in} a 3D {\GREEN flow} versus slope (size ratios $d_t/d$ from 2 to 10). {\GREEN Flow thickness} $h_{max}$= (a) 0.112~m (b) 0.223~m.
Error bars show the standard deviation.}
\label{nathangle}
\end{figure}

Several simulations {\RED are} done for different slope angles of the plane. 
{\GREEN For thin flows, positions $h$ for tracers with} small and large size ratios show no dependency {\RED on} the slope, i.e.{\RED, on} the velocity of the flow (Fig.~\ref{nathangle}(a)). 
On the contrary, when getting close to the transition between surface and reverse depths (size ratio between 4 and 6), $h$ depends on the slope. The greater the angle, the deeper 
the tracer stabilizes. This {\GREEN can be interpreted} by the fact that the flow being more rapid, it {\RED loses} cohesion and is less able to carry large and consequently heavy tracer.   

\begin{figure}[!htbp]
\center
\hspace{-0.8cm}
\includegraphics[width=0.95\linewidth]{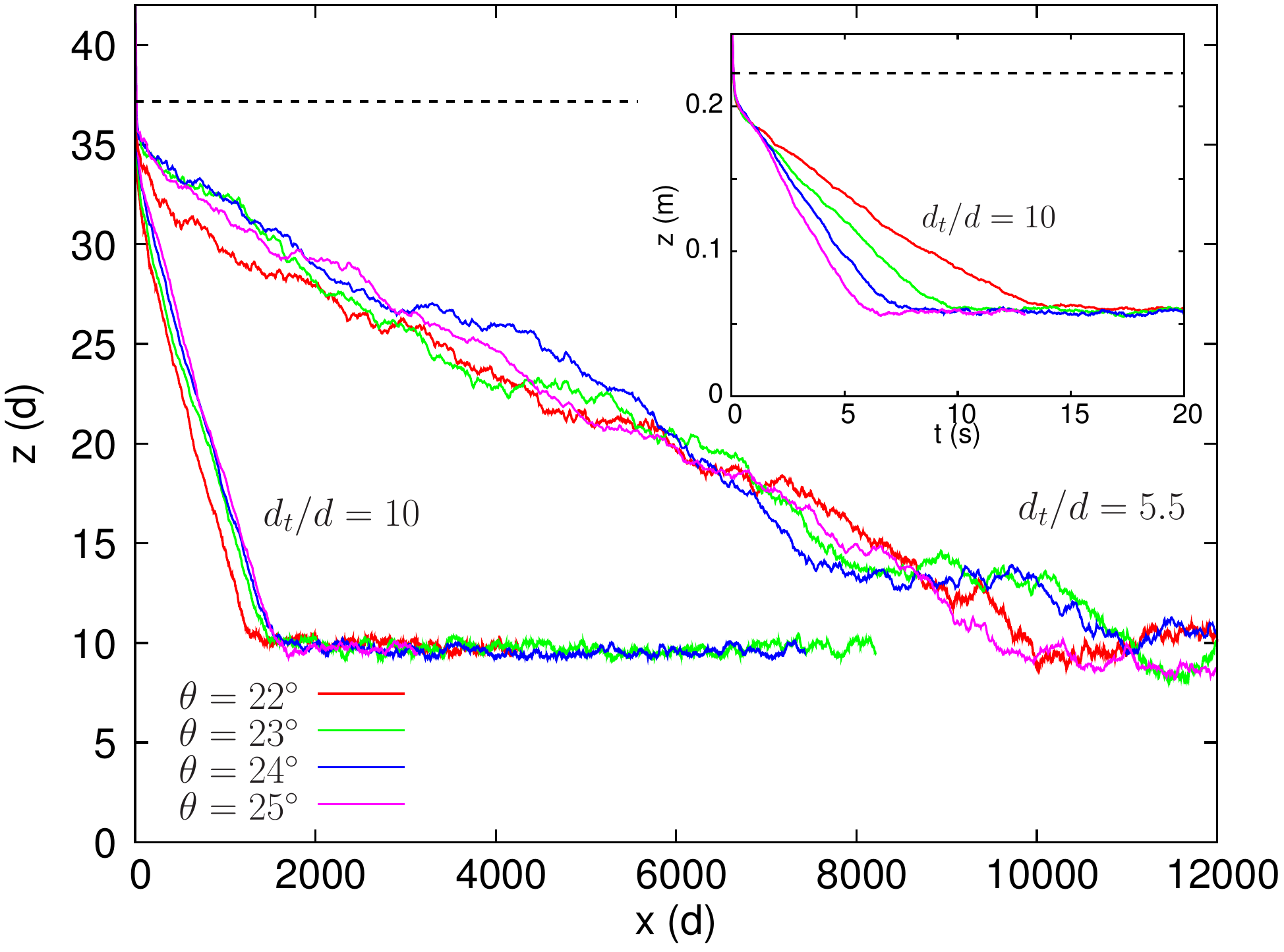}  
\caption{Tracer trajectories (x-z plane) measured in small bead diameters. $d_t/d= 10$, $h_{max}=0.223$~m (37$d$) and slope angles {\BLUE 22 to 25$^\circ$}. Dashed lines show the free surface. A set of {\RED noisier} trajectories {\GREEN are plotted} for comparison (size ratio $d_t/d=5.5$). Insert: time evolution of the tracer depth (z-t).}
\label{nath10a}
\end{figure}

In the case of a {\GREEN thicker} flow, the equilibrium depth
of the tracer shows almost no dependency {\RED on} the slope
(Fig.~\ref{nathangle}(b)). But no size ratios between $d_t/d= 4$ and 5 {\RED are} 
presented here: {\RED they do} not present the usual rapid convergence to {\RED an equilibrium} depth. Further investigations {\RED are needed} ({\GREEN ongoing study on} \cite{ThomasDOrtona18}). 

{\RED The} time evolution of a tracer depth $z$ plotted for different angles {\BLUE (22 to 25$^\circ$)}, {\RED shows that} the tracer sinks more rapidly when the slope is larger 
(Fig.~\ref{nath10a} insert). However, {\RED trajectories} (depth $z$ versus displacement along the flow $x$) {\RED for different angles} all superimpose (Fig.~\ref{nath10a}). 
This shows that the sinking of a large tracer is {\GREEN due to} successive geometrical reorganizations between {\GREEN particles}. At higher slope, flow velocity and shear rate are increased {\GREEN and} reorganizations are more frequent: the tracer sinks more rapidly {\GREEN ($z$ vs $t$)}. {\RED The trajectories considered} (in a $z-x$ space) all coincide independently of the flow rate (Fig.~\ref{nath10a}): only the number of reorganizations plays a role. {\BLUE The slope of the trajectories (compared to the rough incline) are constant with the incline angle.} 
{\BLUE The horizontal settling distances are $\simeq 1800d$ for a size ratio $d_t/d=10$ and $\simeq 10000d$ for $d_t/d=5.5$. This implies that the sinking velocities increase with the incline slope. Note that in 2D, the trajectory slope is also found constant for rough incline slopes from 17$^\circ$ to 23$^\circ$ and for $d_t/d=30$.} {\RED By contrast}, if the shear rate is increased due to an increase {\RED in} flow thickness {\GREEN (Figs.~\ref{nathh2}(b) and \ref{nath12}), the downward} tracer velocities are nearly identical ({\GREEN giving} parallel trajectories in a $z-t$ space) and the spatial trajectories {\BLUE 
do not match in a $z-x$ space (Fig.~\ref{nathh2}(b) insert)}. {\GREEN In this case,} the increase of the flow thickness  induces an increase {\RED in} the shear rate and {\RED in} the frequency of reorganizations, but also an increase {\RED in} the normal stress{\RED,} which reduces the {\GREEN  downward} velocity of the tracer. Both mechanisms compensate to induce a constant {\GREEN downward} velocity. {\BLUE Note, that the constant downward velocity is also found in 2D for $h_{max}=$36 and 24$d$ and size ratios 20 and 30, even though fluctuations are large.} {\GREEN To conclude}, an increase {\RED in} the flow velocity {\GREEN has a different} effect {\BLUE on the downward motion of the tracer} if {\GREEN coming from} a slope or from a thickness {\GREEN increase.} {\BLUE Note that neither the constant velocity, nor this type of dependence with the traveled distance has been observed when the trajectory variation is due to a change in tracer size ratio. Choosing $x$ instead of $t$ in Figs. \ref{plan3_8_20} and \ref{plantopbottom} does not have give any additional information.}

\subsubsection{Multiple tracers flows on a 3D incline}

In {\GREEN previous} experiments, 10\% volume fraction of tracers {\RED was} used \cite{Thomas00}. {\GREEN To compare simulations and experiments, the tracer
fraction is numerically varied. This will {\RED also} allow comparison {\RED between} the
segregation process and the stabilization of one single tracer}.
{\GREEN {\RED The segregated position (also labelled $h$) is the mean of the {\RED tracer} positions once {\RED the flow has} reached the stationary regime}.}

The mean flow velocity $v$ is measured for $d_t/d= 8$ and $h_{max}= 0.223$~m: it decreases by a factor 2 while the fraction increases from one tracer ($\simeq 0.8$\%) up to {\RED a} 5\% (or to 10\%) {\RED volume fraction}. As pointed
out (Figs. \ref{planrough} and \ref{nath10a}), tracer trajectory depths $z$ vs time $t$ cannot be compared for flows having non-equal velocities, only stationary depths $h$ can be compared. For a full comparison of trajectories, 
 $z$ vs $x$ displacements should be used.

\begin{figure}[!htbp]
\center
\includegraphics[width=0.95\linewidth]{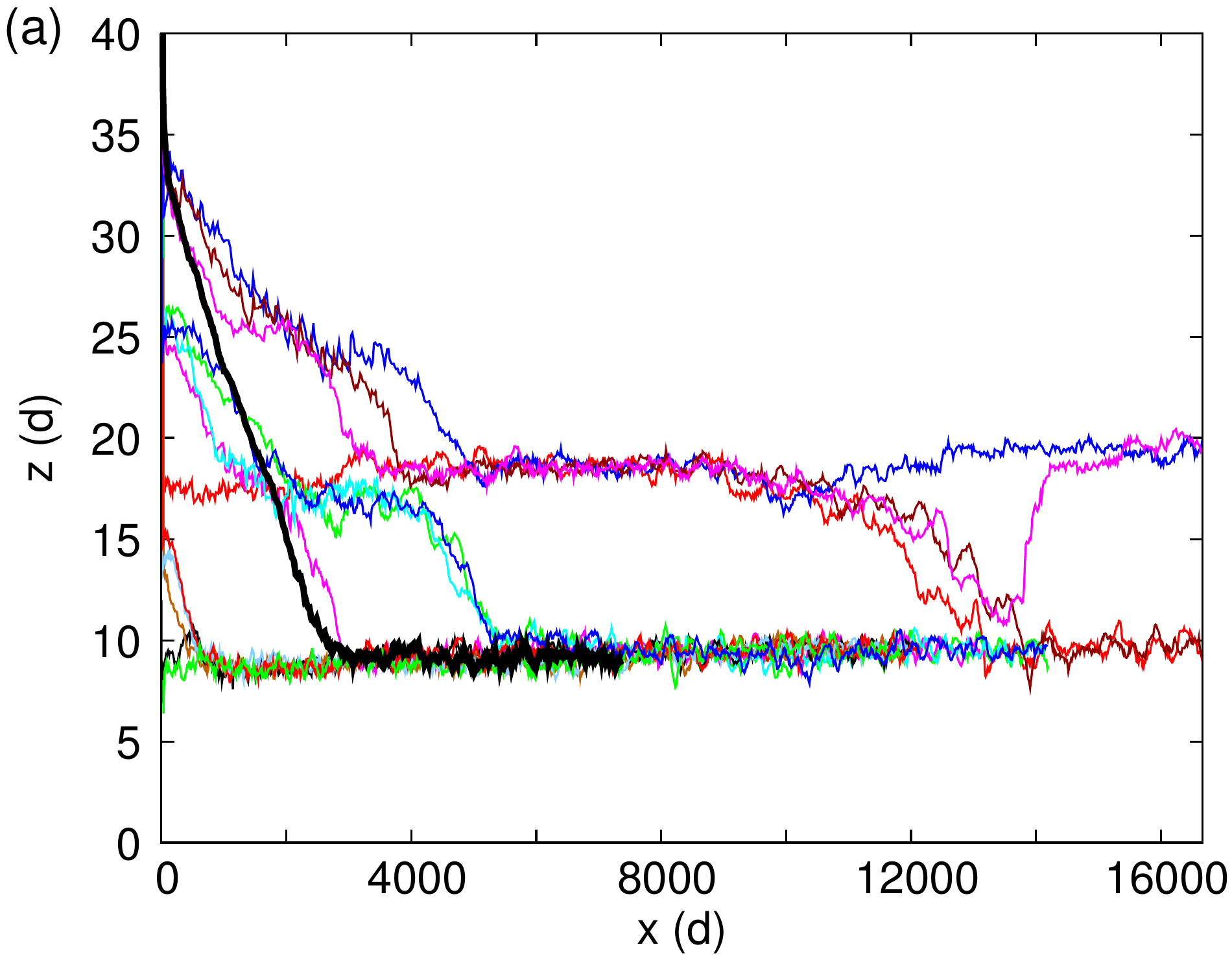}  
\includegraphics[width=0.95\linewidth]{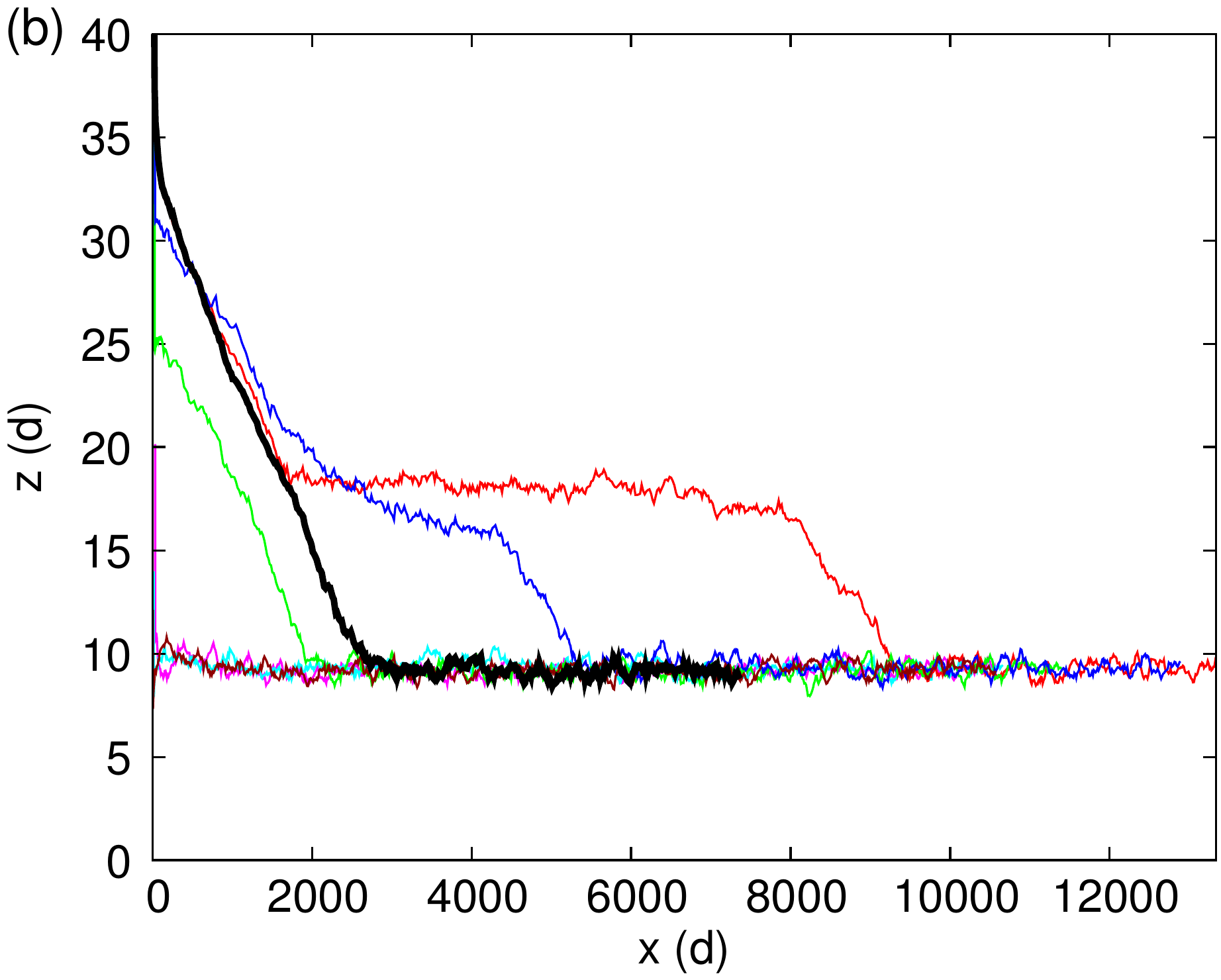}  
\caption{Tracer trajectories z-x ({\GREEN in color}) measured in small bead
diameters ($d$) in a {\GREEN 3D} flow ($h_{max}= 0.223$~m (37$d$)) with: (a) 10\%, (b) 5\% of tracers ($d_t/d=8$). 
Trajectory (thick black) of an {\GREEN identical} single tracer released at the surface.}
\label{col3710d}   
\end{figure}

Figure \ref{col3710d} compares {\RED the} trajectories of a single 
tracer and of several tracers (5\% and 10\%) in the case of a thick flow ($h_{max}= 0.223$~m $\simeq 37d$). Tracers are initially 
randomly placed. 
{\GREEN The tracer trajectories reveal a succession of displacements: horizontal displacements (the tracers cannot move downwards due to the steric exclusion effect) alternated with downward displacements (with a slope less steep than the case of a single tracer). In the case
of multiple tracers, the overall downward displacement is slower  {\RED than that of} a single tracer.} 
The depth equilibrium position of the lowest layer in the case of multiple tracers is rapidly identical to the depth of one single
tracer. For {\RED a} 10\% volume fraction, three, then two, layers of tracers form in the lower part the flow (Fig. \ref{col3710d}(a)). Successive down cascading from one layer to another corresponds to an increase {\RED in} the local fraction of the lowest layers. Very rare up-motions of tracers are observed. 
For 5\% volume fraction (Fig.~\ref{col3710d}(b)), the downward  {\RED slopes}
 of a single tracer  {\RED trajectory and}  of multiple tracer {\RED trajectories} can even be
comparable. Only one layer of tracers is present at the end, whose depth is identical to {\RED that} of a single tracer. 

First, the tracer fraction has no influence on the depth of the lowest layer. Consequently, it is possible to compare experimental and numerical data {\RED using} the lowest numerical trajectories, and the {\RED lowest} experimental tracer depths. 
The main effect {\GREEN of the increased fraction} (5 or 10\%) is the  {\RED persistence} of a second,
 {\RED possibly a third} layer above the basal layer of tracers which is full {\RED and cannot 
 include any more tracers.}  {\GREEN For this reverse segregation, there} is an asymmetric upwards spread of the {\RED tracer} positions{,} {\GREEN with a position distribution maximum at} the lowest layer depth.    

\begin{figure}[htbp]
\center
\includegraphics[width=0.95\linewidth]{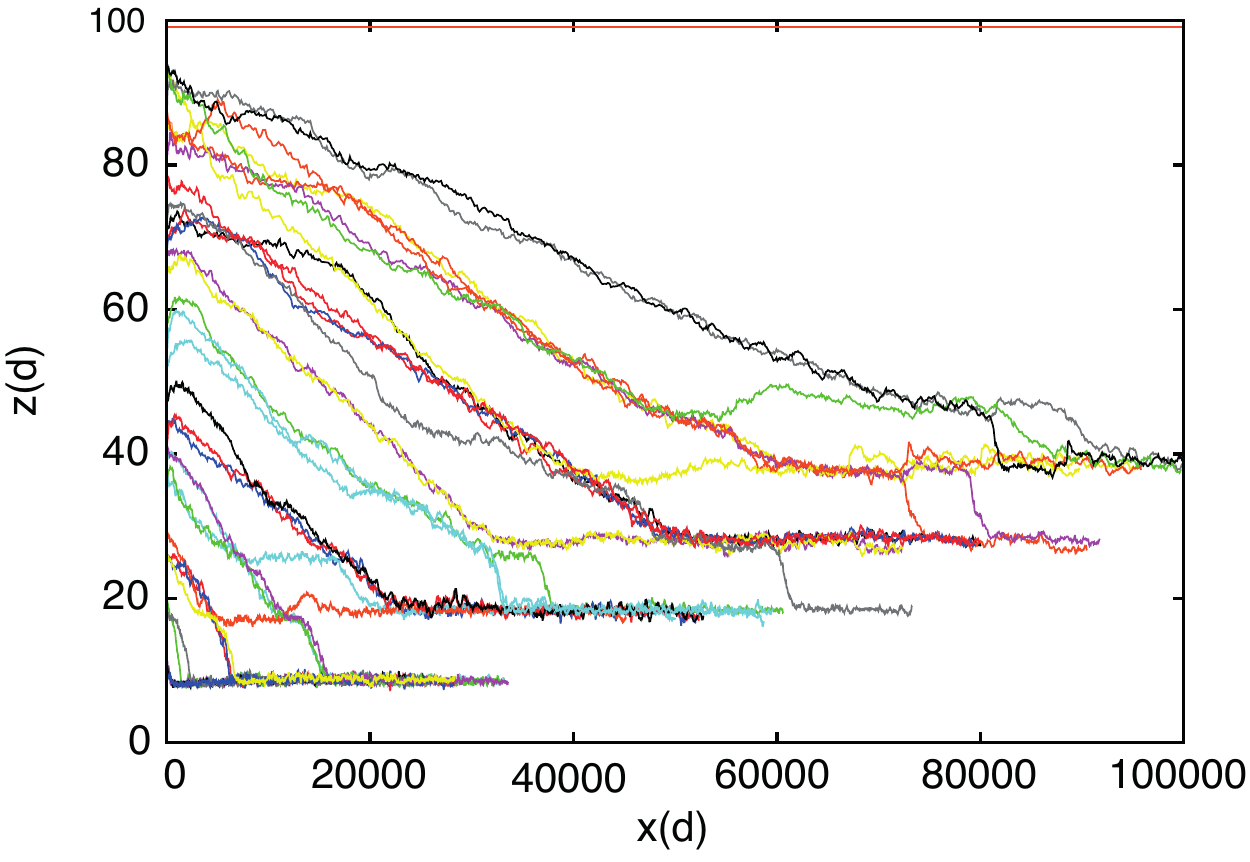}
\caption{Trajectories (z-x) of {\GREEN a few} tracers (10\%, $d_t/d=8$) in a 3D {\GREEN very} thick flow ($h_{max}=$100$d$) measured in small bead diameters ($d$).}
\label{colonnex}
\end{figure}

Secondly, the convergence to the final state of segregation is longer to establish for 5 or 10\% of tracers than {\GREEN for one single} tracer. {\GREEN Even {\RED though} some individual downward velocities are locally the same,} it takes time for tracers to move from one layer to another. Even {\RED though} the global segregation pattern is rapidly obtained (around $2000-4000d$), the distance of convergence is so long (around $10000d$) that it is not {\GREEN reachable} in usual laboratory conditions. These values are to be compared with experiments in channel with thicknesses from $28d$ to $45d$, and a surface pattern obtained at 70~cm \cite{WiederseinerAncey11}. Nevertheless, in our simulations, the depth of the lowest trajectory is rapidly defined for thick flows (Fig.~\ref{col3710d}). 
In {\GREEN our previous} experiments, flows and deposits sometimes presented a thickness larger than 37$d$. To see
how {\RED such a thickness could} affect the previous results, one simulation {\RED is} performed with $h_{max}=100d$, 10\% of tracers and $d_t/d=8$ {\GREEN corresponding to an equilibrium reverse depth} (Fig.~\ref{colonnex}). The number of basal layers increases, {\GREEN because} for a constant volume fraction the number of tracers increases with the flow thickness. As several layers of tracers develop (instead of 2 or 3) and as tracers cascade between {\RED layers}, the time and the distance needed for convergence strongly increase (Fig.~\ref{colonnex}). For experiments {\RED done} with $d=300-400$~$\mu$m particles, a distance of convergence of $100 000d$
requires a plane of 35~m. Nevertheless, the results for $h_{max}=100d$ are similar {\RED to those} {\GREEN for $37d$} (Fig.~\ref{col3710d}): reverse segregation {\RED is obtained}, the bottom layer depth at $9d$ (equal to the single tracer depth), and the formation of several layers of tracers. For larger size ratios, we may expect shorter convergence times and distances, since a single {\GREEN larger} tracer reaches its equilibrium depth {\RED faster} (Figs. \ref{nathh2}(b) and \ref{nath12}, or \ref{nath10a}).

\begin{figure}[htbp]
\center
\includegraphics[width=0.95\linewidth]{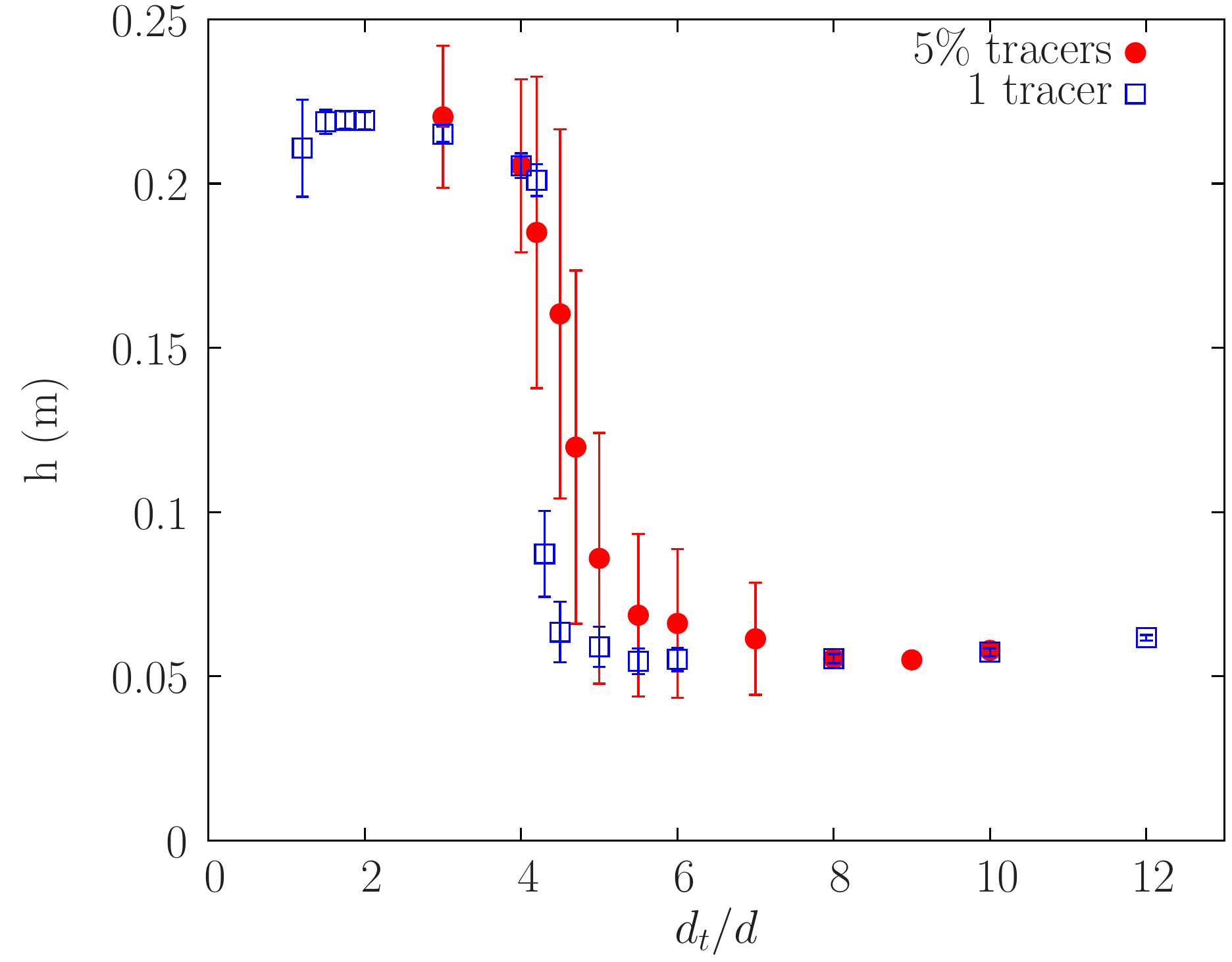}  
\caption{Equilibrium {\GREEN depths} versus size ratio in a 3D {\GREEN flow} ($h_{max}= 0.223$~m) for: a single tracer (blue~$\square$), 5\% of tracers (red~$\bullet$). Error bars show the standard deviation, but are only partly representative of the {\GREEN vertical spread} of individual tracer positions which {\GREEN can be} asymmetric, never {\RED higher} than the surface or lower than the {\GREEN reverse position} (see Fig. \ref{col3710d}).}
\label{multib}  
\end{figure}

{\GREEN Measurement of the segregated positions} {\RED is made} for 5\% of tracers, {\GREEN in a thick flow $h_{max}=0.223$~m and} for several size ratios {\GREEN in an interval} around {\GREEN the value 4.3, i.e.}{\RED,} the {\GREEN reversal} transition {\GREEN of a single tracer position} from surface to bottom (Fig.~\ref{multib}).  {\GREEN 
The segregated position of several tracers at a given moment also presents a reversal, evolving from the surface to intermediate depths, and then to reverse depths. The standard deviation is small enough to consider that segregation {\RED occurs}: tracers are not spread all through the bed, but regrouped near the mean position, especially for surface and reverse segregations.
We then quantitatively compare the results for a single tracer ($h$ is the mean on the trajectory) and for several segregated tracers, and further down with experimental results on several tracers.} Except minor {\GREEN differences}, the {\GREEN two curves are} very close.
For small size ratios ($\leqslant 4$), {\GREEN mean depths are the same, but there is} {\RED an} increase {\RED in} standard deviation for several tracers. {\GREEN The deviation includes both the trajectory fluctuations and the interaction between tracers.} For size ratios $4.2\leqslant$ $d_t/d$ $\leqslant 4.7$, larger standard deviations and an upshift of mean depths are observed for several tracers, {\GREEN giving} a smoother transition.
For larger size ratios ($\geqslant 5$), the upshift {\RED disappears}, and only larger standard deviations are observed for several tracers (note that for ratios above 8, all beads fit in the lowest layer{\RED, } giving the same standard deviations {\RED as} a single tracer). 
{\GREEN As the mean depth of a single tracer and those of several tracers are almost identical, it confirms the hypothesis that the segregation process for this low fraction is a regrouping of near non-interacting tracers at the same equilibrium depth, because this depth depends only on the size ratio. Studying a single tracer is valuable for understanding the segregation phenomena for a low fraction of tracers. With this low fraction, we observe successively surface segregation, intermediate segregation, and reverse segregation when increasing the size ratio.}  
The {\GREEN reversal} from surface to bottom happens for a size ratio {\RED (around 4.5) similar to the reversal size ratio} {\GREEN for a single tracer}(around 4.3).  
One consequence of the smoother transition {\GREEN than for a single tracer} is the disappearance of the empty central region {\GREEN where no single tracer stabilizes in a thick flow} (Fig.~\ref{schema}). For these fractions, there is a thick central {\GREEN layer} of intermediate segregation. {\GREEN In the case of multiple tracers, the segregation pattern organizes} in three layers (surface, intermediate and reverse), {\GREEN very {\RED much} like the equilibrium depths of a single tracer} in a thin flow.

\subsubsection{Comparison with experiments in channel} 

Experiments  {\RED were} performed by the sudden release of 1 kg of an initially homogeneous mixture of glass beads with 10\% of large tracers, in a 6~cm wide, 1~m long rough channel inclined with a
slope about 26.5$^\circ$ (more details in \cite{Thomas00}). Flows  {\RED were} observed
to be 1 to 2~cm thick, with deposits aggraded over 2 to 5~cm thick after the flow  {\RED had been}
stopped by a perpendicular wall, or by the change of slope to horizontal. On
cross-sections of the deposit, the segregation pattern could be separated in
three main cases, depending on the size ratio: (1) the small size ratios (1.75,
2, 2.14, 3.5) (resp. for tracer diameter 0.35, 0.7 or 3, 1.5, 0.7~mm) for which all tracers  {\RED were} at the surface with a small standard deviation, (2) the 4.3 ratio (for 3~mm tracers) for which
tracers  {\RED were} rather everywhere (surface and inside), and (3) the large ratios (5.9, 8.6,
10.4, 10.7, 15, 21.4, 44) (resp. for tracer diameter 3, 3, 0.7, 7.5, 3, 7.5, 3~mm)
for which tracers  {\RED were} found inside, with a small layer free of tracers
near the surface. {\GREEN Decrease in the mean} position of tracers and {\RED in} {\GREEN their} standard deviation  {\RED was} observed when increasing the size ratio
(for these $d_t/d\geqslant4.3$).

The three {\GREEN patterns} agree with the {\GREEN three types} of tracer {\GREEN mean depth} found in the simulations {\GREEN for several or for a single tracer} (Fig.~\ref{multib}). The upper limit of the transition between surface and reverse {\RED segregations} is experimentally found for 4.3. {\GREEN As} this transition numerically occurs between 4.2 and 4.3 for one {\GREEN tracer} and {\GREEN between 4.2 to 4.7} for 5\% of tracers, the agreement is very good. But experimental standard deviations are large, not only for ratio 4.3, but for all larger size ratios, which is not observed in simulations {\GREEN at high size ratios}. Nevertheless, standard deviations decrease with the size ratio {\GREEN in} both experiments and simulations. 

Three experiments  {\RED were done} with a wall to stop the flow at {\GREEN various} distances from the start (30, 60 and 90~cm). Tracers  {\RED were} 3~mm, and small beads  {\RED were} 300-400~$\mu$m (size ratio 8.6). Tracers  {\RED were} found inside the deposit{\RED,} with no major differences in the segregation pattern. But the mean depth of tracers in a {\GREEN cross-section taken} at the same distance from the end wall (for example at 10~cm),  {\RED showed} a slight decrease {\GREEN passing from the 30~cm to the 60~cm, and to the 90~cm long experiment}. The convergence to a final mean position {\RED was} still developing {\RED at the time when} the flow {\RED stopped}. {\GREEN We conclude that} all our {\GREEN experimental} data, established for a 90~cm {\RED traveling} distance, do not concern a perfect stationary state.
This convergence distance is compatible with the simulations, where a stationary state is not reached at $2500d$ (equivalent to 90~cm) for flows of $37d$ (equivalent to 1.3~cm) (Fig. \ref{col3710d}) or of $100d$ (equivalent to 3.5~cm) (Fig. \ref{colonnex}).
{\GREEN The fact that experimental} standard deviations {\GREEN are} larger than numerical ones can be explained by this non-fully converged state. {\GREEN It can also be explained} by the use of 10\% of tracers instead of 5\%. The decreases {\RED in} the experimental {\GREEN mean depth} and standard deviation {\GREEN with increasing} size ratios {\GREEN are compatible with} a better convergence towards the reverse position. This better convergence is compatible with the faster migration of one single tracer when increasing the size ratio (comparing Figs. \ref{nathh2}(b) and \ref{nath12}, or Fig. \ref{nath10a}). This also explains why reverse segregation is experimentally nearly reached for the ratio 44, despite a quite short channel.

\begin{figure}[htbp]
\center
\includegraphics[width=0.92\linewidth]{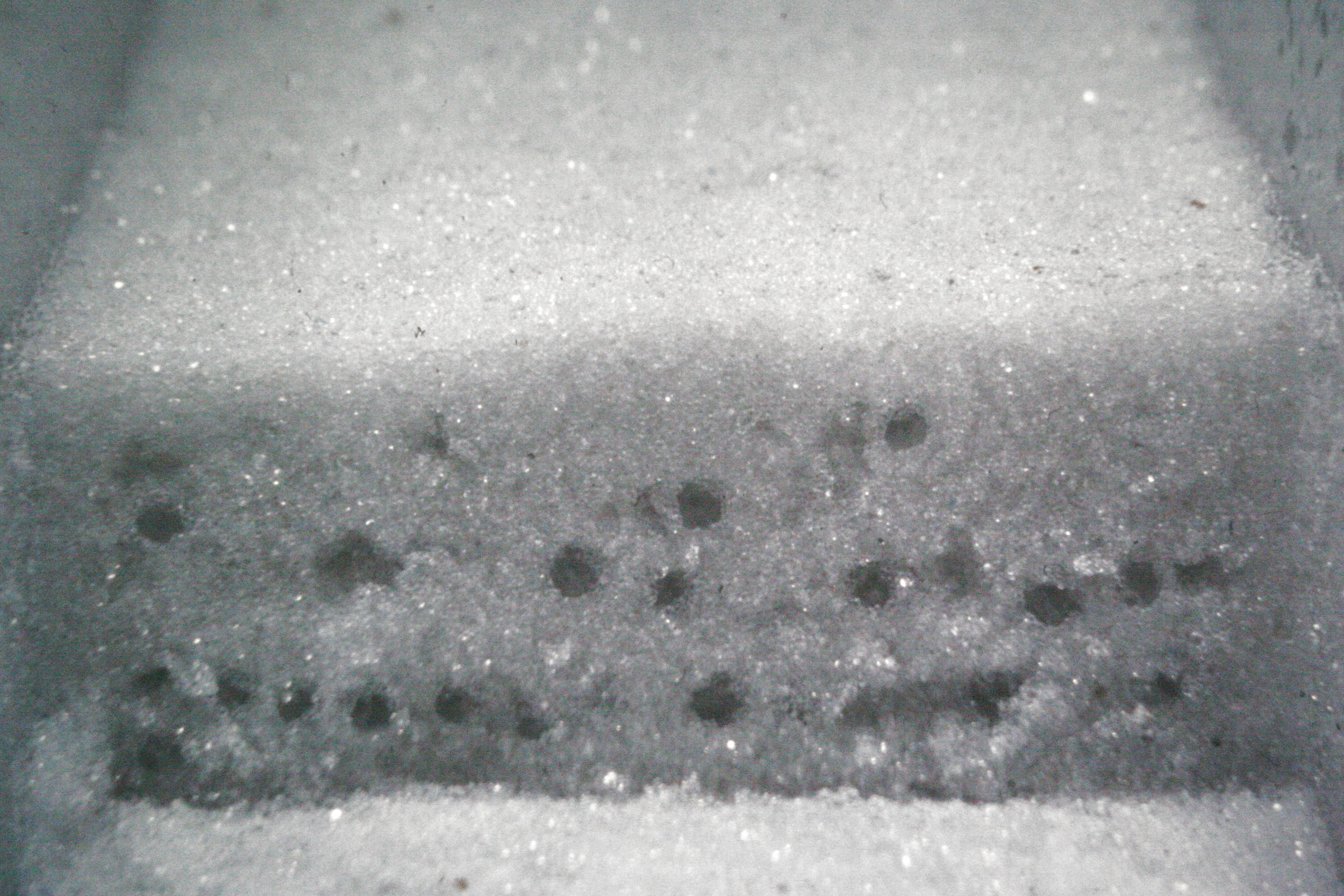}
\caption{{\GREEN Cross-section} of the deposit in a 1~m long and 6~cm wide 
chute flow experiment. The flow is composed of glass beads: 90\% of 
300-400~$\mu$m and 10\% of 3~mm.}
\label{manipcut}  
\end{figure}

In the simulations (Figs.~\ref{col3710d}(a) and \ref{colonnex}), for {\RED traveling} lengths corresponding to experiments (90~cm=2500$d$), tracers above the first layer do not organize in superimposed layers like those obtained at the end of the simulations.  For such a short flowing distance, the first bottom layer is well defined and the second layer is in formation. The other layers are still emerging and have not reached their final depth yet. Indeed, in experiments, most sections presented tracers organized in one bottom layer (Fig.~\ref{manipcut}), and sometimes in a {\GREEN blurred} second layer.
{\GREEN Experimental} measurements of this bottom layer depth were done at a distance between 15 and 30~cm from the end wall. We {\GREEN tried} to avoid perturbations due to the collision with the end wall and the possible {\GREEN local} variations of the tracer fraction near the flow front. 
For size ratios from 4.3 to 8.6, the bottom layer depth  {\RED was} {\GREEN experimentally} measured between $10$ and $11.5d$  {\GREEN with randomness} (and $13d$ for size ratio 15, above our simulation range). These values are close to the $9$ to $10.2d$ values numerically found for size ratios from 5 to 12 (Fig.~\ref{nathh1-2}). 

Even {\RED though} the stationary stage has not been reached in our experiments, experimental data reproduce well the existence of the bottom layer and its depth, the {\GREEN reversal} between surface and reverse segregations, the variation of the associated standard deviations, and the exact size 
ratio ($d_t/d=4.3$) for which the {\GREEN reversal} occurs. {\GREEN Moreover,} the numerical study {\GREEN has shown} that {\GREEN experimental} tracers settled inside ($d\geqslant 5.9$) correspond to non-converged states of reverse segregation at different degrees of convergence.

Simulation and experiment results {\GREEN both} show that the size ratio 4.3 induces intermediate segregation of the tracers, {\GREEN for which} the spread of {\RED the} experimental positions is maximum ({\GREEN in addition, it is experimentally} obtained for a non-fully converged system). The result is a nearly homogeneous mixture. This experimental spread of tracers all through the deposit is compatible with the {\GREEN numerical results obtained for a converged state}: a quite large standard deviation combined with a {\GREEN mean} position at mid-{\GREEN height}. But longer times of convergence, and effects coming from the increase {\RED in} the tracer fraction {\GREEN up to 10\%} might also be {\RED involved} in the experimental process {\GREEN for explaining the tracer spread}. For that reason, the range of size ratios around 4.3 needs further investigations. {\GREEN Nevertheless, combining an intermediate segregation and an appropriate tracer fraction could be a means to prevent any segregation during a flow.}

\section{Conclusion}

In 3D granular flows, the {\GREEN selection of an equilibrium depth} of a large tracer depends mainly on the size ratio between the tracer and the small beads, and {\GREEN to a lesser extent} on the nature of the flow. {\GREEN Comparison between depths of single tracers and mean depths of several tracers (3 to 10\%) shows that the stabilization of one tracer and the segregation process select identical equilibrium depths. In that case, the segregation is the regrouping of non-interacting large tracers at the same equilibrium depth.} 

 In a tumbler, a precise study of trajectories has shown that the {\GREEN depth} is established during the flowing phase and is recorded in the static rotating part. The flow substratum is a loose granular material whose boundary with the flowing layer is difficult to define, but trajectories of the largest size ratios seem to place tracers at the bottom of the flow. {\RED Thus} surface {\GREEN positions}, intermediate {\GREEN positions} with deeper and deeper {\GREEN depths toward} reverse {\GREEN positions} are observed when increasing the size ratio between the tracer and the small particles. The transition between surface and reverse {\GREEN depths} is progressive, and a large range of tracer size ratios is found to be at intermediate {\GREEN depths}. 

For all 3D flows down a rough incline, {\GREEN the reversal also happens when increasing the size ratio.} Two {\GREEN cases} are clear: {\GREEN positions} near the surface, which corresponds to tracers at the surface, or {\GREEN to non visible tracers,} just {\GREEN under} the surface (surface and intermediate depths), and {\GREEN positions of} tracers floating {\GREEN at or very close to} the bottom (intermediate and reverse depths). {\RED The} existence of intermediate {\GREEN positions} near half-{\GREEN height} depends on the flow {\GREEN thickness.}
For thick flows, the {\GREEN reversal} between {\GREEN surface and bottom positions} is sharp, with no tracers stabilized around mid-{\GREEN height}.
For thin flows, the {\GREEN reversal} is progressive and tracers stabilize at every intermediate {\GREEN depth} inside the flow. We conclude that, in 3D, the tracer position {\GREEN is also} determined by the type of substratum (solid or loose) and {\GREEN  the flow thickness}.  

For multiple tracer flows on 3D incline {\GREEN (5 to 10\%),} the three segregation patterns {\GREEN (surface, intermediate, reverse)} are observed {\GREEN when increasing the size ratio,} {\GREEN corresponding to the three types of depth stabilization of a single tracer}. The transition is smoother {\GREEN and happens at} the same size ratio (around 4.3) for simulations and experiments, {\GREEN corresponding to single tracer reversal}. But reverse segregation is long to establish, and a large spread in the positions remains {\RED over} a long {\RED traveling} distance. During this travel, the flows (with size ratio above 5) can be considered as nearly homogeneous, except near their surface where only small particles are present. {\GREEN The intermediate case (size ratio 4.3) remains almost homogeneous}. The choice of {\GREEN reverse and especially intermediate positions} could be an opportunity to {\GREEN maintain} {\RED a} homogeneous mixture for usual industrial transfers. Further studies are needed to set the precise parameter range where these processes can be used.

The case of 2D flows has been studied in tumblers and inclines. For small size ratios, the position of tracers relative to the free surface behaves similarly in 2D and 3D, although {\GREEN deeper} positions are found for large size ratios in {\GREEN  3D}. The dependency of the {\GREEN stabilized depth {\RED on} the size ratio} is similar but {\GREEN weaker} in 2D {\GREEN both in tumbler and on incline}. {\GREEN Moreover} for the largest size {\RED ratio} tracers on 2D inclines, the reverse positions do not exist. Tracers stabilize at intermediate positions near mid-{\GREEN height and their} position scales with the flow thickness contrarily to the 3D case. In 2D tumblers, the equilibrium position evolution {\GREEN with size ratio} is also {\GREEN weaker}, with a shifted maximum, and leads to an intermediate radial {\GREEN position of equilibrium} for the largest tracers. However, the position of these largest tracers does correspond to a reverse depth comparable to {\RED that of} the 3D {\GREEN case}, but {\GREEN this depth is} obtained for larger size ratios {\GREEN than in 3D.} The {\GREEN difference} between 2D and 3D, probably due to a granular packing compacity difference, does not {\GREEN emerge in} all processes in the same manner.
The highest care should be taken before extrapolating results of studies between 2D and 3D cases.  

\begin{acknowledgments}
The authors thank F. Schwander, F. Smith, J. Favier and D. Martinand for carefully rereading this 
manuscript.
This work was granted access to the HPC resources of Aix-Marseille Universit\'e financed by the project Equip@Meso (ANR-10-EQPX-29-01) of the program ``Investissements d'Avenir" supervised by the Agence Nationale de la Recherche.
\end{acknowledgments}

\end{document}